%% file: main.tex
\documentclass[twocolumn,tighten]{aastex631}
\usepackage{CJKutf8}
\usepackage{amsmath}
\usepackage{enumitem}
\usepackage{booktabs,pbox}
\usepackage{bm}
\usepackage{siunitx}
\usepackage{float}

\graphicspath{{./figures/}{../figures/}}

\DeclareRobustCommand{\ion}[2]{%
\relax\ifmmode
\ifx\testbx\f@series
{\mathbf{#1\,\mathsc{#2}}}\else
{\mathrm{#1\,\mathsc{#2}}}\fi
\else\textup{#1\,{\mdseries\textsc{#2}}}%
\fi}

\newcommand{\NII}  {[\ion{N}{ii}]} 

\newcommand{\m}    {$\mu$m}

\shorttitle{Embedded Clusters}
\shortauthors{Graham/PHANGS}

\begin{document}


\title{PAH Marks the Spot: Digging for Buried Clusters in Nearby Star-forming Galaxies}


\input{authors}


\begin{abstract}
\input{00_abstract}
\end{abstract}


\keywords{}


\section{Introduction} \label{sec:intro}
\input{01_introduction.tex}


\section{Sample and Data} \label{sec:sample_data}
\input{02_sample_data.tex}


\section{Visual Cluster Identification} \label{sec:by_eye}
\input{03_byeye.tex}

\section{Machine Learning for Identification} \label{sec:ML}
\input{04_ML.tex}


\section{Results} \label{sec:results}
\input{05_results.tex}


\section{Discussion} \label{sec:discussion}
\input{06_discussion}


\section{Conclusions} \label{sec:conclusions}
\input{07_conclusions}


\input{08_acknowledge}

\appendix
\section{Appendix A - Ideal Embedded Cluster Candidates} \label{sec:ECC_RGBs}
\input{09_appendixA}
\section{Appendix B - \NII\ Contribution to H$\alpha$} \label{sec:NII_Ha}
\input{10_appendixB}

\pagebreak
\bibliography{main.bib}


\end{document}

%% file: authors.tex
\newcommand{\UWyoming}{\affiliation{Department of Physics and Astronomy, University of Wyoming, Laramie, WY 82071, USA}}
\newcommand{\STScI}{\affiliation{Space Telescope Science Institute, 3700 San Martin Drive, Baltimore, MD 21218, USA}}
\newcommand{\NOIRLab}{\affiliation{International Gemini Observatory/NSF NOIRLab, 950 N. Cherry Avenue, Tucson, AZ 85719, USA}}
\newcommand{\UToledo}{\affiliation{Ritter Astrophysical Research Center, University of Toledo, Toledo, OH 43606, USA}}
\newcommand{\JHU}{\affiliation{Department of Physics and Astronomy, The Johns Hopkins University, Baltimore, MD 21218 USA}}
\newcommand{\OSU}{\affiliation{Department of Astronomy, The Ohio State University, 140 West 18th Ave., Columbus, OH 43210, USA}}
\newcommand{\MPIA}{\affiliation{Max Planck Institut f\"ur Astronomie, K\"onigstuhl 17, 69117 Heidelberg, Germany}}
\newcommand{\UCSD}{\affiliation{Department of Astronomy \& Astrophysics, University of California San Diego, 9500 Gilman Drive, La Jolla, CA 92093, USA}}
\newcommand{\ANU}{\affiliation{Research School of Astronomy and Astrophysics, Australian National University, Canberra, ACT 2611, Australia}}
\newcommand{\nrao}{\affiliation{National Radio Astronomy Observatory, 520 Edgemont Road, Charlottesville, VA 22903, USA}}
\newcommand{\MPE}{\affiliation{Max-Planck-Institut f\"ur Extraterrestrische Physik (MPE), Giessenbachstr. 1, D-85748 Garching, Germany}}
\newcommand{\Oxford}{\affil{Sub-department of Astrophysics, Department of Physics, University of Oxford, Keble Road, Oxford OX1 3RH, UK}}
\newcommand{\STScIESA}{\affiliation{AURA for the European Space Agency (ESA), Space Telescope Science Institute, 3700 San Martin Drive, Baltimore, MD 21218, USA}}
\newcommand{\UConn}{\affiliation{Department of Physics, University of Connecticut, Storrs, CT 06269, USA}}
\newcommand{\ESO}{\affiliation{European Southern Observatory, Karl-Schwarzschild Stra{\ss}e 2, D-85748 Garching bei M\"{u}nchen, Germany}}
\newcommand{\Plata}{\affiliation{Instituto de Astrofísica de La Plata, CONICET--UNLP, Paseo del Bosque S/N, B1900FWA La Plata, Argentina}}
\newcommand{\BrynMawr}{\affiliation{Department of Physics and Astronomy, Bryn Mawr College, Bryn Mawr, PA 19010, USA}}
\newcommand{\Virginia}{\affiliation{Department of Astronomy, University of Virginia, Charlottesville, VA 22904, USA}}
\newcommand{\Utah}{\affiliation{Department of Physics and Astronomy, University of Utah, Salt Lake City, UT 84112, USA}}
\newcommand{\Rice}{\affiliation{Department of Physics and Astronomy, Rice University, Houston, TX 77005, USA}}
\newcommand{\CPP}{\affiliation{Department of Physics and Astronomy, California State Polytechnic University Pomona, Pomona, CA 91768, USA}}
\newcommand{\Lycoming}{\affiliation{Department of Physics and Astronomy, Lycoming College, Williamsport, PA, 17701, USA}}
\newcommand{\CPH}{\affiliation{Department of Physics and Astronomy, California State Polytechnic University Humboldt, Arcata, CA 95521, USA
}}
\newcommand{\kipac}{\affiliation{Kavli Institute for Particle Astrophysics \& Cosmology (KIPAC), Stanford University, CA 94305, USA}}
\newcommand{\Whitman}{\affiliation{Whitman College, 345 Boyer Avenue, Walla Walla, WA 99362, USA}}
\newcommand{\UniCA}{\affiliation{Université Côte d'Azur, Observatoire de la Côte d'Azur, CNRS, Laboratoire Lagrange, 06000, Nice, France}}
\newcommand{\UOA}{\affiliation{Department of Physics, University of Arkansas, 226 Physics Building, 825 West Dickson Street, Fayetteville, AR 72701, USA}}
\newcommand{\Michigan}{\affiliation{Department of Astronomy, University of Michigan, Ann Arbor, MI 48109, USA}}
\newcommand{\UAlberta}{\affiliation{Department of Physics, University of Alberta, 4-183 CCIS, Edmonton, Alberta, T6G 2E1, Canada}}
\newcommand{\Ensenada}{\affiliation{Instituto de Astronom\'ia, Universidad Nacional Aut\'onoma de M\'exico, Unidad Acad\'emica en Ensenada, Km 103 Carr. Tijuana—Ensenada, Ensenada, B.C., C.P. 22860, M\'exico}}
\newcommand{\UNAM}{\affiliation{Instituto de Astronom\'{\i}a, Universidad Nacional Aut\'onoma de M\'exico, Ap. 70-264, 04510 CDMX, Mexico}}

\correspondingauthor{Gabrielle~B.~Graham}
\email{gbgraham19@gmail.com}

\author[0009-0003-0926-8791]{Gabrielle~B.~Graham}
\UWyoming
\author[0000-0002-5782-9093]{Daniel~A.~Dale}
\UWyoming
\author[0009-0000-8325-9736]{Chase~L.~Smith}
\UWyoming
\author[0009-0003-8440-5813]{Elisabeth~Brann}
\BrynMawr
\author[0009-0006-9568-7360]{Kaycee~D.~Conder}
\UWyoming
\author[0009-0005-0394-3754]{Samuel~Crowe}
\Virginia
\author[0009-0009-9022-8757]{Sumitra~Dhileepkumar}
\Utah
\author[0009-0009-7062-7283]{Nicole~A.~Imming}
\Rice
\author[0009-0005-9024-0212]{Emilio~Mendez}
\CPP
\author[0009-0003-3269-4561]{Zachary~Pleska}
\Lycoming
\author[0009-0008-7485-8583]{Kelsey~Sako}
\CPH
\author[0000-0002-8553-1964]{Amirnezam Amiri}
\UOA
\author[0000-0003-0410-4504]{Ashley~T.~Barnes}
\ESO
\author[0000-0003-0946-6176]{Médéric Boquien}
\UniCA
\author[0000-0003-0085-4623]{Rupali~Chandar}
\UToledo
\author[0000-0001-8241-7704]{Ryan~Chown}
\OSU
\author[0000-0001-9852-9954]{Oleg~Y.~Gnedin}
\Michigan
\author[0000-0002-3247-5321]{Kathryn Grasha}
\ANU
\author[0000-0001-9628-8958]{Stephen~Hannon}
\MPIA
\author[0000-0002-8806-6308]{Hamid~Hassani}
\UAlberta
\author[0000-0002-4663-6827]{R\'emy~Indebetouw}
\nrao
\Virginia
\author[0000-0003-4770-688X]{Hwihyun~Kim}
\NOIRLab
\author[0000-0002-0432-6847]{Jaeyeon~Kim}
\kipac
\author[0009-0001-5949-1524]{Hannah~Koziol}
\UCSD
\author[0000-0003-3917-6460]{Kirsten~L.~Larson}
\STScIESA
\author[0000-0003-0946-6176]{Janice~C.~Lee}
\STScI
\author[0000-0002-2545-1700]{Adam~K.~Leroy}
\OSU
\author[0000-0002-0119-1115]{Elias~K.~Oakes}
\UConn
\author[0000-0002-0579-6613]{M.~Jimena~Rodríguez}
\STScI
\Plata
\author[0000-0002-5204-2259]{Erik~Rosolowsky}
\UAlberta
\author[0000-0002-4378-8534]{Karin~Sandstrom}
\UCSD
\author[0000-0002-3933-7677]{Eva~Schinnerer}
\MPIA
\author[0000-0002-9183-8102]{Jessica Sutter}
\Whitman
\author[0000-0002-8528-7340]{David~A.~Thilker}
\JHU
\author[0000-0003-1242-505X]{Leonardo~Ubeda}
\STScI
\author[0000-0002-3784-7032]{Bradley~C.~Whitmore}
\STScI
\author[0009-0005-8923-558X]{Tony~D.~Weinbeck}
\UWyoming
\author[0000-0002-0012-2142]{Thomas~G.~Williams}
\Oxford
\author[0000-0001-8289-3428]{Aida Wofford}
\Ensenada
\author[0000-0002-6972-6411]{J. Eduardo M\'endez-Delgado}
\UNAM
\author[0009-0009-9148-2159]{Qiushi~Chris~Tian}
\JHU
\author{the PHANGS team}


%% file: 00_abstract.tex
The joint capabilities of the Hubble Space Telescope (HST) and JWST allow for an unparalleled look at the early lives of star clusters at near- and mid-infrared wavelengths. We present here a multiband analysis of embedded young stellar clusters in 11 nearby, star-forming galaxies, using the PHANGS-JWST and PHANGS-HST datasets. We use the {\it Zooniverse} citizen science platform to conduct an initial by-eye search for embedded clusters in near-UV/optical/near-infrared images that trace stellar continuum emission, the Paschen$\alpha$ and H$\alpha$ recombination lines, and the 3.3~$\mu$m polycyclic aromatic hydrocarbon feature and its underlying continuum. With this approach, we identify 292 embedded cluster candidates for which we characterize their ages, masses, and levels of line-of-sight extinction by comparing the photometric data to predictions from stellar population models.  The embedded cluster candidates have a median age of 4.5~Myr and an average line-of-sight extinction $\left< A_V \right> = 6.0$~mag. We determine lower limits on source stellar masses, resulting in a median stellar mass of $10^3$~$M_{\sun}$. We use this sample of embedded cluster candidates to train multiple convolutional neural network models to carry out deep transfer learning-based searches for embedded clusters. With the aim of optimizing models for future catalog production, we compare results for four variations of training data using two neural networks. Confusion matrices for all eight model configurations, as well as inter-model identification trends, are presented. With refinement of the training sample, we determine that optimized models could serve as a pathway for future embedded cluster identification beyond our 11 galaxy sample.

%% file: 01_introduction.tex
Many stars that form in giant molecular clouds originate in stellar clusters \citep{Lada&Lada2003} that are deeply embedded in their natal environments for the first few million years of their existence \citep{Kruijssen2012,Krumholtz2019,kim2023,rodriguez2023,Sun2024}.  During this embedded early stage of cluster evolution, ultraviolet and optical emission is highly extinguished, limiting the possibility of robust multiwavelength spectral energy distribution (SED) fitting to estimate the stellar characteristics of the clusters \citep{deshmukh2024,Hannon2023,hoyer2023}.  After this earliest stage in a stellar cluster's life, feedback from stellar winds, thermal pressure due to photoionization heating, and radiation pressure work to clear away the surrounding gas and dust, resulting in a partially exposed system.  After several million years of cluster evolution, supernovae additionally push away the interstellar material, resulting in a fully exposed stellar cluster.  Such feedback processes can halt or trigger new waves of star formation and enhance the metallicity of the surrounding environment.  A comprehensive understanding of the evolution of stellar clusters in nearby galaxies, including their earliest, most embedded phases, will refine our view of how feedback processes shape the morphology, dynamics, and composition of the interstellar medium \citep{klessen2016,rodriguez2024,schinnerer2024,Krumholtz2019}.

Nearby galaxies are ideal laboratories for studying the diversity of embedded cluster formation. Their proximity enables high-resolution views of individual massive stars and star clusters. The combination of HST and JWST has been especially groundbreaking in recovering large cluster populations in galaxies as far as 20~Mpc, especially embedded clusters with high levels of optical extinction (\cite{pedrini2024, rodriguez2024, Whitmore2023, hassani2023}).  The Physics at High Angular resolution in Nearby GalaxieS (PHANGS) project is collecting a rich database of ultraviolet, optical, infrared, and millimeter data to study 75 galaxies at distances between $\sim$5 and 20~Mpc \citep{leroy2021,emsellem2022,lee2022,lee2023}.  One of the overarching goals of the project is to connect the processes underlying star formation on small scales to the global properties of the galaxies.  The PHANGS-HST portion of the project has produced a comprehensive catalog of optically identified stellar clusters \citep{maschmann2024}.  Similar to the PHANGS-HST program, most studies of stellar clusters in nearby galaxies have leveraged optical observations, which by nature are not optimized for detecting deeply embedded sources.  Until the advent of JWST, any infrared campaigns focused on stellar clusters suffered from relatively poor sensitivity and spatial resolution \citep[e.g.,][]{bastian2006,Whitmore2014,Corbelli2014,adamo2017,Jones2017}.  

Recent JWST-based efforts have leveraged hydrogen recombination lines coupled with polycyclic aromatic hydrocarbon (PAH) emission features to study the earliest phases of star cluster formation.  To understand the complete lifecycle of stellar clusters, it is necessary to also study the onset of clustered star formation, a phase that is presumably heavily obscured by dust.  \cite{linden2024} utilize variations in 3.3~\m\ emission in combination with Pa$\alpha$ line strength to show that young massive clusters in NGC~3256 indicate short PAH-clearing timescales, $<\sim$3--4 Myr.  The Feedback in Emerging Extragalactic Star Clusters (FEAST) survey shows a tight correlation between 3.3~\m\ PAH surface density luminosity and Br$\alpha$ star formation rate (SFR) on $\sim$7~pc scales in NGC~628 \citep{Gregg2024}.  

In this study, we likewise aim to identify the most deeply embedded stellar clusters at their earliest phases of their evolution.  Accordingly, we utilize the 3.3~\m\ PAH feature along with the Pa$\alpha$ line as key components in our identification scheme.  PAHs indicate the presence of interstellar dust, and the 3.3~\m\ PAH feature is particularly sensitive to the ultraviolet light from young stellar populations \citep[e.g., Figure~16 of][]{draine2021}.  Moreover, the $\sim$0\farcs1 angular resolution at 3.3~\m, corresponding to $\sim$5--10~pc for our galaxy distances, helps determine if the dust emission is compact and thus potentially indicative of an embedded star cluster.  We additionally rely on an elevated Pa$\alpha$/H$\alpha$ ratio to indicate conspicuously buried systems.  A third criterion for identifying embedded stellar cluster candidates is the lack of any discernible stellar continuum emission, because of dust extinction, in broadband filters spanning near-ultraviolet and optical wavelengths as seen by HST.

This pairing of infrared JWST imaging with HST near-ultraviolet/optical imaging allows for a unique look into the very earliest stages of stellar cluster formation and evolution \citep{whitmore2021,Whitmore2023,levy2024}.  We initially target 11 nearby star-forming galaxies from the PHANGS-JWST Cycle~2 survey for which we have a full complement of HST and JWST NIRCam imaging observations. Harnessing these data, we first identify embedded cluster candidates with a human by-eye search. We then use this sample to conduct supervised machine learning because the by-eye process is labor-intensive. We provide a catalog of embedded cluster candidates for the Cycle~2 program, along with measurements of key physical characteristics such as their luminosities, masses, ages, and extinctions.

Section~\ref{sec:sample_data} provides an overview of the sample and observational data. Sections~\ref{sec:by_eye} and \ref{sec:ML} review our approaches to human identification and to machine learning identification, respectively. Section~\ref{sec:results} presents the results, \S~\ref{sec:discussion} discusses the results, and \S~\ref{sec:conclusions} summarizes the main findings.

%% file: 02_sample_data.tex
We focus our initial analysis on 11 nearby star-forming galaxies from the PHANGS-JWST Cycle~2 treasury survey (ID: 3707; PI: A. Leroy; \citealt{chown2025}).  Program observations encompass imaging from the NIRCam F150W, F187N\footnote{Some PHANGS-JWST Cycle~2 galaxies were not observed with the F187N filter as their redshift places the Pa$\alpha$ wavelength beyond the filter profile.}, F300M, and F335M, and MIRI F770W and F2100W filters.  In this work, we utilize all available NIRCam imaging as well as the MIRI F770W data.  At low redshifts, the F150W and F300M filters largely trace the stellar continuum along with some contributions from a hot dust continuum, the F335M and F770W filters capture emission features from polycyclic aromatic hydrocarbons and an underlying mix of stellar and dust continuum emission, and the F187N filter is centered on the 1.87~\m\ Paschen$\alpha$ (Pa$\alpha$) hydrogen recombination line.  The PHANGS data processing pipeline (\texttt{pjpipe}) is detailed in \cite{williams2024}, including pertinent updates made to the pipeline for the PHANGS Cycle~2 data.  These updates include improving the anchoring of flux levels, optimizing relative and absolute astrometry, background matching of the individual mosaic tiles, refining the removal of NIRCam instrumental artifacts, and background subtraction in MIRI images.  The production of the Pa$\alpha$ line and continuum images is outlined in T. D. Weinbeck et al.\ (2025, in preparation).  Additionally, we include in our analysis near-ultraviolet and optical imaging from the PHANGS-HST program (F275W, F336W, F435W/F438W, F555W, and F814W; \citealt{lee2022}), as well as continuum-subtracted HST F657N/F658N maps \citep{chandar2025,barnes2025}.  The 11 galaxies selected for this analysis are those for which both HST H$\alpha$ and Cycle~2 JWST Pa$\alpha$ were available. These galaxies and their relevant properties are outlined in Table \ref{tab:sample}. We adopt distances, inclinations, and star formation rates from \cite{leroy2021}.

\setlength{\tabcolsep}{3pt}
\input{galaxy_sample.tex}
\setlength{\tabcolsep}{1pt}

\begin{figure*}[!ht]
\centering
 \includegraphics[width=0.9\textwidth]{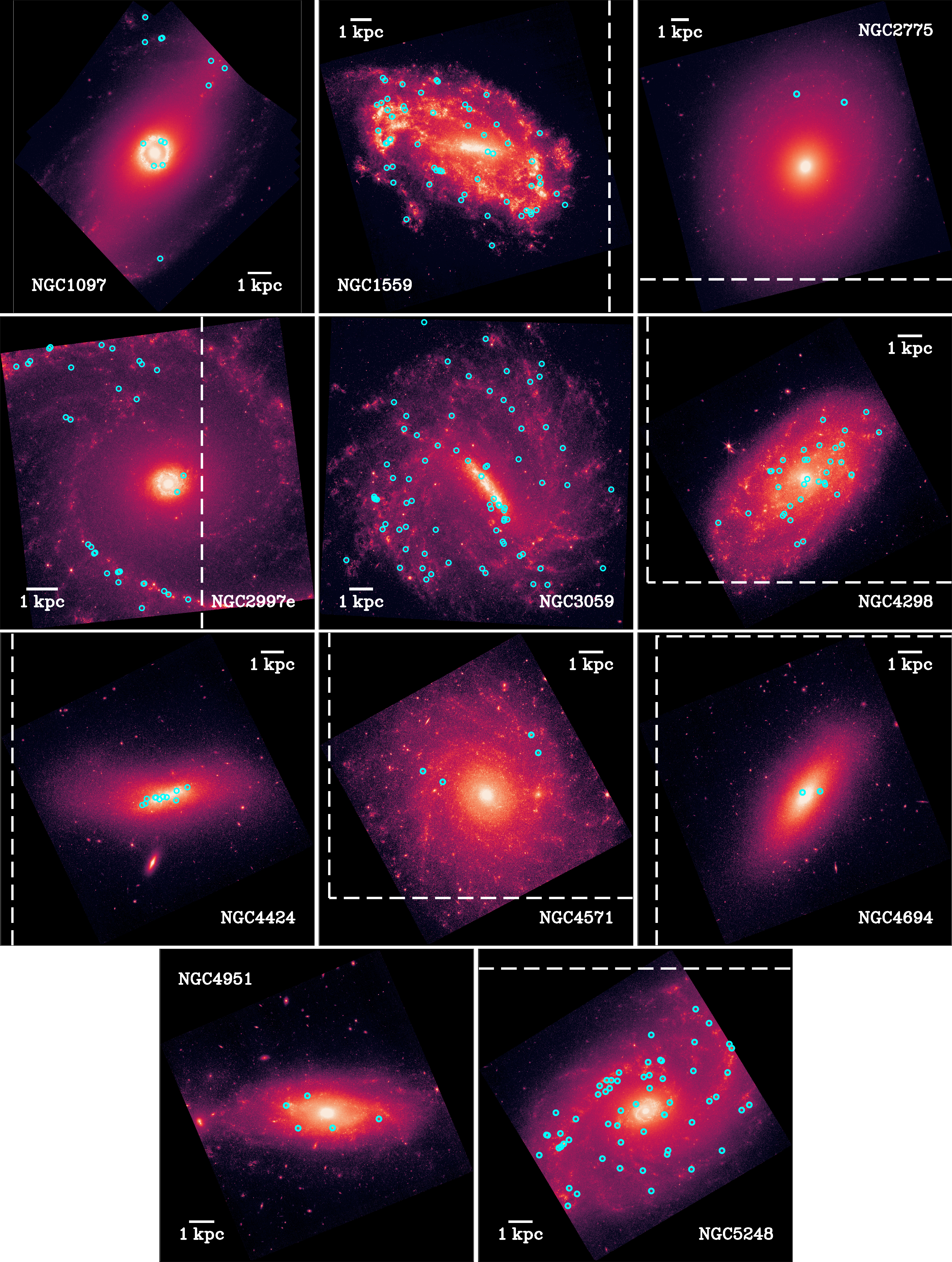}
 \caption{F335M images of our galaxy sample and the embedded cluster candidates identified in this study. The cyan circles mark the locations of the human-identified embedded cluster candidates in our final catalog.  As would be expected for young, dusty clusters, our sample of embedded cluster candidates traces the dusty spiral arms, bars, and rings of their host galaxies. The white, dashed lines indicate the edges of the fields of view for the HST H$\alpha$ imaging.}
 \label{fig:composite_image}
\end{figure*}

Figure~\ref{fig:composite_image} displays each galaxy in the JWST F335M band with overlays of our final 292 human-identified embedded cluster candidates (see Section~\ref{sec:by_eye} for more details on the by-eye identifications).

\subsection{Image Regridding and Alignment}
\label{sec:ImgProcessing}
Many facets of our identification and classification of these embedded clusters rely on accurate astrometric alignment and pixel scale agreement across multiple images. There are two main steps for creating the maps we utilize throughout this work: regridding and alignment. We first resample all data to the NIRCam F335M grid using the \textit{reproject\_exact} function from \texttt{Astropy}.  On the F335M astrometric grid, the pixel scale no longer oversamples the F150W and F187N images but this does not impact our analysis.  We then correct for sub-pixel offsets using point source catalog matching to align each of the images to a reference image. To circumvent the issue of the Pa$\alpha$ line data lacking compact point sources, we use the matching continuum map as our alignment reference, as the line and continuum data inherently aligned. As our primary interest is spatial correlation, we only convolve filters to a common resolution for the purpose of continuum subtraction. All other data remain at native resolution for source identification. Section \ref{sec:contsub} outlines the necessary convolutions we perform to produce continuum-subtracted 3.3~$\mu$m PAH maps and Pa$\alpha$ line maps.

\subsection{Continuum Subtraction}
\label{sec:contsub}
A pure $\mathrm{F335M-F300M}$ subtraction image has limitations in probing 3.3~$\mu$m PAH feature enhancement, noticeably in areas of faint F335M emission, where continuum noise is most problematic, and in regions where high stellar surface density dominates the F335M band \citep{rodriguez2023, sandstrom2023}.  Utilizing a new method from Koziol et al. (2025, in preparation), we are able to improve our subtraction of the underlying continuum emission and extract 3.3~$\mu$m PAH feature emission flux in dim areas. This improved subtraction allows us to more accurately trace the presence of this particular PAH feature. 

To perform the subtraction, we first convolve the regridded F150W and F300M data to the same resolution as the F335M data (0\farcs11). We then subtract an underlying dust continuum using:
\begin{equation}
    \mathrm{F335M_{PAH} = F335M-(0.77\cdot F300M)-(0.057\cdot F150W)}
\end{equation}
assuming image units in MJy~sr$^{-1}$ \citep[see also Koziol et al., in preparation;][]{sandstrom2023}.
We determine the point-source sensitivity of the F335M$_{\mathrm{PAH}}$ maps, as the 3.3~\m\ PAH emission is a key selection criterion for our sample. To acquire a first-order estimation of the detection limit, we implement a method similar to \citet{lee2023}. We perform aperture photometry for a minimum of 50, 0\farcs248 diameter circular apertures in `empty' regions across the F335M$_{\mathrm{PAH}}$ maps. We adopt this diameter as it matches what we use for source photometry in Section~\ref{sec:source_properties}. The noise is estimated as the standard deviation of the background-subtracted medians for these `empty' apertures, where the 3$\sigma$-clipped median of a 0\farcs124 thick annulus is used as the local background. We take the point-source sensitivity as the estimated noise times a factor of 5 for a `significant' detection. The resulting point-source sensitivities for the F335M$_{\mathrm{PAH}}$ maps ($5\sigma_{\mathrm{PAH}}$) are in Table~\ref{tab:sample}.

Additionally, the F300M and F150W are used to extract the Pa$\alpha$ nebular emission from the F187N data. For embedded cluster selection and training machine learning models, we utilize the Pa$\alpha$ maps from Weinbeck et al.\ (2025, in preparation). However, for deriving cluster properties, we choose to perform our own continuum subtraction on the Level 3 F187N data. The steps we implement here are similar to those of Tian et al.\ (2025, in preparation) and those which \cite{chandar2025} employ to produce the HST H$\alpha$ continuum-subtracted maps we utilize in this work.

The F150W and F187N data are first regridded and convolved (in units of MJy~sr$^{-1}$) to the grid and resolution of the F300M data (0\farcs1). 
Then, weights are determined for each flanking filter, $W_{\mathrm{F150W}}$ and $W_{\mathrm{F300M}}$, using the distance between the central wavelengths of each filter to the F187N filter $\left(\lambda_{\mathrm{F150W}},\lambda_{\mathrm{F300M}},\ \mathrm{and}\ \lambda_{\mathrm{F187N}}\right)$ such that

\begin{equation}
    \begin{split}
    W_{\mathrm{F150W}}=\frac{|\lambda_{\mathrm{F300M}}-\lambda_{\mathrm{F187N}}|}{|\lambda_{\mathrm{F150W}}-\lambda_{\mathrm{F300M}}|}
    \end{split}
\end{equation}
and
\begin{equation}
    \begin{split}
    W_{\mathrm{F300M}}=\frac{|\lambda_{\mathrm{F150W}}-\lambda_{\mathrm{F187N}}|}{|\lambda_{\mathrm{F150W}}-\lambda_{\mathrm{F300M}}|}.
    \end{split}
\end{equation}

Using these weights, the underlying stellar continuum flux $\left(F_{\mathrm{cont}}\right)$ is calculated in log-space as:
\begin{equation}
    \begin{split}
    \mathrm{log_{10}}(F_{\mathrm{cont}})&=W_{\mathrm{F150W}}\mathrm{log_{10}}(F_{\mathrm{F150W}})\\
    &+W_{\mathrm{F300M}}\mathrm{log_{10}}(F_{\mathrm{F300M}}).
    \end{split}
\end{equation}

The $\mathrm{Pa}\alpha$ line flux $\left(F_{\mathrm{Pa}\alpha}\right)$ is computed in linear space, in units of MJy~sr$^{-1}$, as
\begin{equation}
    F_{\mathrm{Pa}\alpha}=F_{\mathrm{F187N}}-F_{\mathrm{cont}}.
\end{equation}
Lastly, the Pa$\alpha$ line data are converted to units of erg~s$^{-1}$~cm$^{-2}$. We note that for each of these subtractions, we do not correct for Pa$\beta$ contribution to the F150W band. We discuss the effects of this contribution to our results in Section \ref{sec:discussion}.

%% file: galaxy_sample.tex
\begin{deluxetable*}{lccrrccccc}[!ht]
\tablecaption{Sample of PHANGS-JWST Galaxies Used in This Analysis}
\tablewidth{0pc}
\setlength{\tabcolsep}{7pt}
\label{tab:sample}
\tablehead{
\colhead{Galaxy}&\colhead{R.A.} &\colhead{Dec.} &\colhead{$D$}&\colhead{$i$}&\colhead{log$_{10}$SFR}&\colhead{log$_{10}M_*$}&\colhead{\NII/H$\alpha$}&\colhead{$r_{25}$}&\colhead{${5\sigma_{\mathrm{PAH}}}$}\\
\colhead{}     &\colhead{(J2000)}&\colhead{(J2000)}&\colhead{(Mpc)}&\colhead{(deg)}&\colhead{($M_\odot\;\mathrm{yr^{-1}}$)} &\colhead{($M_\odot$)}  &  &\colhead{(arcsec)}&\colhead{(${\rm \mu Jy}$)}
}
\startdata
NGC~1097$\dagger$ & 02:46:18.9 & $-$30:16:29  & 13.58${\;\pm\;2.05}$ & 48.6${\;\pm\;6.0}$ &$+$0.68 &10.8 & 0.52& 314.1&{0.0147}\\ 
NGC~1559          & 04:17:36.6 & $-$62:47:00  & 19.44${\;\pm\;0.45}$ & 65.4${\;\pm\;8.4}$ &$+$0.60 &10.4 & 0.45& 125.1&{0.0214}\\ 
NGC~2775          & 09:10:20.1 & $+$07:02:17  & 23.15${\;\pm\;3.49}$ & 41.2${\;\pm\;0.6}$&$-$0.06 &11.1 & 0.57& 128.0&{0.0126}\\ 
NGC~2997          & 09:45:38.8 & $-$31:11:28  & 14.06${\;\pm\;2.80}$ & 33.0${\;\pm\;9.0}$ &$+$0.64 &10.7 & 0.50& 307.0&{0.0215}\\ 
NGC~3059          & 09:50:08.2 & $-$73:55:20  & 20.23${\;\pm\;4.04}$ & 29.4${\;\pm\;11.0}$ &$+$0.38 &10.4 & 0.45& 114.1&{0.0113}\\ 
NGC~4298          & 12:21:32.8 & $+$14:36:22  & 14.92${\;\pm\;1.36}$ & 59.2${\;\pm\;0.8}$ &$-$0.34 &10.0 & 0.38& 75.4&{0.0109}\\ 
NGC~4424          & 12:27:11.6 & $+$09:25:14  & 16.20${\;\pm\;0.69}$ & 58.2${\;\pm\;6.0}$ &$-$0.53 &~9.9 & 0.37& 90.6&{0.0120}\\ 
NGC~4571          & 12:36:56.4 & $+$14:13:02  & 14.90${\;\pm\;1.07}$ & 32.7${\;\pm\;2.1}$ &$-$0.54 &10.1 & 0.40& 106.4&{0.0099}\\ 
NGC~4694          & 12:48:15.0 & $+$10:59:01  & 15.76${\;\pm\;2.38}$ & 60.7${\;\pm\;6.0}$ &$-$0.81 &~9.9 & 0.37& 59.9&{0.0106}\\ 
NGC~4951          & 13:05:07.7 & $-$06:29:38  & 15.00${\;\pm\;4.19}$ & 70.2${\;\pm\;2.2}$ &$-$0.46 &~9.8 & 0.35 & 94.9&{0.0096}\\ 
NGC~5248          & 13:37:32.0 & $+$08:53:07  & 14.87${\;\pm\;1.32}$ & 47.4${\;\pm\;16.3}$ &$+$0.36 &10.4 & 0.45& 122.2&{0.0133}\\ 
\enddata
\tablenotetext{}{Values are from \cite{leroy2021} and references therein.  $\dagger$ Indicates a Seyfert classification from \cite{veroncetty2010}, though NGC~1097 is listed as a broad-lined LINER. A description of how we derive values of \NII/H$\alpha$ is found in Appendix~\ref{sec:NII_Ha}.}
\end{deluxetable*}

%% file: 03_byeye.tex
Similar to an abundance of other large-scale data and machine learning projects \citep[e.g.,][]{Johnson2015,Zevin2017,Mahabal2019, Frissell2024}, we utilize the citizen science platform \textit{Zooniverse} to streamline our `by-eye’ identification process.  The primary features we implement in our \textit{Zooniverse} project include multi-image viewing, region marking within images, and subject retirement counts. The latter feature is the number of people who must view a set of images (a single subject) for it to become inactive (i.e., no more people can classify on that image set). We create separate identification workflows for each of the Cycle~2 galaxies within our project (11 workflows in total), allowing for ease of data management during postprocessing. As a pilot, we keep the \textit{Zooniverse} project internal to a team of 16 experts; however we note that it could easily be publicly released for large-scale citizen science.  

\subsection{Selection Criteria}
We aim to identify embedded sources in our sample of Cycle~2 galaxies, specifically by tracing sources that are detectable in the infrared but undetectable in the UV-optical regime, aside from H$\alpha$ emission. We narrow this broad selection requirement to four primary selection criteria:
\begin{enumerate}
    \item compact F335M PAH emission indicative of dust-embedded, clustered star formation,
    \item visible F300M source to help with continuum subtraction for the F335M emission,
    \item a high Pa$\alpha$/H$\alpha$ ratio that points to high extinction,
    \item no detectable source in any of the HST broadband filters.
\end{enumerate}
To limit the number of images necessary for identification in \textit{Zooniverse}, we produce an HST `white-light’ image to display the last criterion, which is a summation of the five UVIS/WFC3 broadband images. The regridding and alignment of our imaging data, as well as the continuum subtraction of the 3.3~$\mu$m PAH, Pa$\alpha$, and the H$\alpha$ images are described in Section~\ref{sec:ImgProcessing}. 

Several recent efforts focused on nearby galaxies have utilized hydrogen recombination line ratios to trace regions of embedded star formation, such as Pa$\beta$/H$\alpha$ to identify sources in NGC~1313 \citep{Messa2021} and Pa$\alpha$/H$\alpha$ to characterize clusters in NGC~628 \citep{calzetti2024} and M51 \citep{calzetti2025}.  We use here the Pa$\alpha$/H$\alpha$ ratio to trace areas of high extinction.  We exclude pixels in the Pa$\alpha$ image with a signal-to-noise ratio (S/N) of S/N~$ < 5$ but do not place the same constraint on the H$\alpha$ data.  By not requiring secure detections in H$\alpha$, we retain regions in the earliest stages of cluster formation for which H$\alpha$ may be completely extinguished and thus indicate exceptionally high levels of inferred extinction \citep[see also][]{Messa2021,McQuaid2024}.  

\subsection{{\it Zooniverse} Identification} 
\label{sec:zooID}
\begin{figure}[!ht]
 \centering
 \includegraphics[width=\columnwidth]{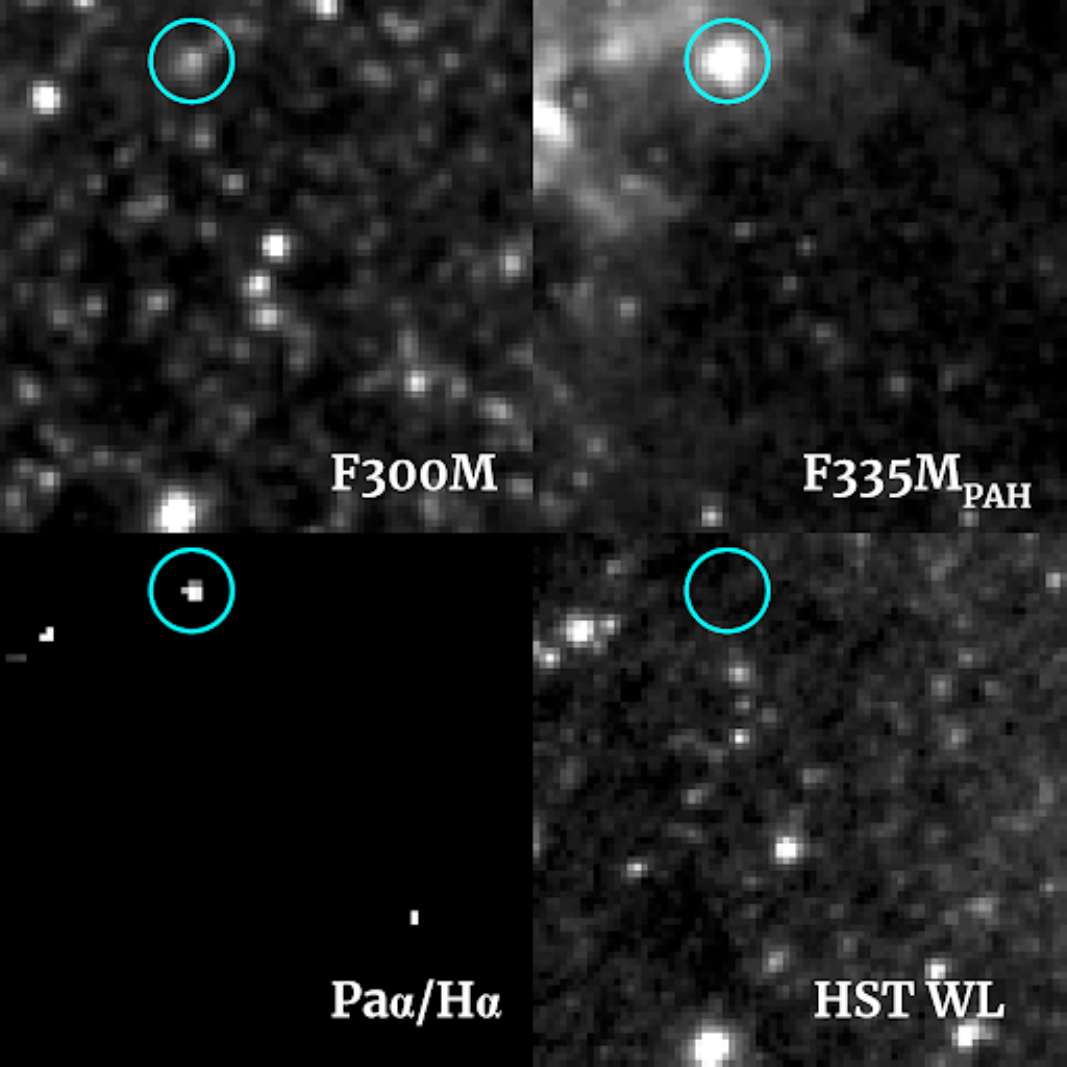}
 \caption{An example of an embedded cluster candidate in NGC~3059, denoted by the cyan circle.  This source displays point-like emission in F300M and F335M, significant P$\alpha$/H$\alpha$ excess indicating high levels of extinction, and no visible emission in our HST white-light image. Ideal candidate examples from four galaxies are shown in Appendix~\ref{sec:ECC_RGBs}}.
 \label{fig:idCase}
\end{figure}

Our by-eye identification process relies on the inspection of many separate small image cutouts from each image.  To produce these image cutouts, we populate a grid of $5\arcsec \times 5\arcsec$ boxes across each galaxy with box centers separated by 4\farcs5.  This process results in a 0\farcs5 overlap between boxes, helping ensure that no embedded cluster candidates are missed because they lie on a box edge.  A region is removed from the by-eye inspection process if it lies outside the observing footprint in any of the four identification images.  For viewing purposes within the {\it Zooniverse} platform, we apply a uniform linear normalization using the 0.5 and 99.5 percentile pixel values to the images and scale the Pa$\alpha$/H$\alpha$ ratio cutouts to display values between 0 and 1. This corresponds to an $A_V$ of approximately 3.8~mag for our upper scale limit of 1 using Equation~\ref{eq:extinction} in Section~\ref{sec:cluster_extinctions}, meaning our selection will be biased toward sources with $A_V\geq3.8$ mag. For ease of visualization, the cutout array dimensions were scaled up by 12$\times$ along each axis using the \textit{repeat} function from \texttt{NumPy}.  We save the image cutouts and include the lower-left WCS coordinate of the data in the filename for ease of extracting the source coordinates in postprocessing.  We then generate a manifest file, which informs \textit{Zooniverse} which images to display for a single boxed region.  In total, for the 11 Cycle~2 galaxies studied here, there are 2674 boxed regions with imaging data across all four images.  Figure~\ref{fig:idCase} displays an example of the images for a single boxed region. The cyan circles denote an embedded cluster candidate.

To tag a source as an embedded cluster candidate, users utilize the circle drawing tool within {\it Zooniverse} to mark the source center in the F300M image.  Each circle center is saved as an individual tagging, so users are able to tag more than one source in the same boxed region.  Once a boxed region reaches a retirement count of four separate viewers, it is no longer available in the workflow for inspection.  Once all boxed regions for a galaxy reach the retirement count, we export our tagged objects.  In total, we generated 3402 tags across our 2674 boxed regions for the 11 Cycle~2 galaxies in our sample. 

\subsection{Final Catalog Creation}
\label{sec:proc_Zooniverse}
The \textit{Zooniverse} workflow output contains $x$- and $y$-pixel coordinates and the cutout name for each tagged object.  From this information, we extract the right ascension and declination for each tagged object.  We tabulate the world coordinate system (WCS) positions alongside the username associated with each tagging.

To create our final embedded cluster candidate catalog, we group the catalog of tagged objects by cross-matching the positions and usernames of each object.  If at least three out of four people have drawn circles with centroids that agree to within five NIRCam long-wavelength pixels, or 0\farcs3, those tags are grouped to represent a single source.  Tagged objects at the edges of the image cutouts often had multiple identifications by the same user due to the 0\farcs5 edge overlaps between cutouts. To confirm that at least three different people tagged such a source, we also compare the usernames of the tags. If a duplicate tagging occurs, the first tag position for that user remains in the group of tags for that source, and the others are removed. Source groups that have duplicate tags account for $\sim 40\%$ of our initial sample. Another result of the 0\farcs5 overlap is sources receiving more than four unique tags (final group size greater than four). This was a more rare occurrence, with only four sources in this category. 

The preliminary coordinates for each embedded cluster candidate are derived from the average right ascension and declination for each tagging in its group. This results in an initial catalog of 361 objects.  Following initial cataloging, we refine our candidate catalog through visual examination of the clusters in the context of the entire galaxy.  We use the JWST F770W to `clean' our catalog by ensuring our candidates lie within the dust extent of the galaxy and that there is no discernible emission at the location of each candidate in the HST F814W data.  Though the emission near 21~$\mu$m can trace embedded star formation, the F2100W imaging data were not used to support this task due to its poorer angular resolution.  This additional inspection removes 70 candidates from our candidate list, and provides the added benefit of culling higher-redshift line emitters that contaminate the F187N filter. With the final list of embedded cluster candidates, we produce centroided F335M coordinates using the \textit{centroid\_2dg} from \textsc{Photutils} which we use for the remainder of our analysis. Utilizing $\it{Zooniverse}$ and following catalog cleaning, we identify 292 embedded cluster candidates across the 11 available PHANGS Cycle~2 galaxies. The by-eye detection rate varies significantly between galaxies, and we address possible reasons for this in Section~\ref{sec:Var_Num}.

\subsubsection{Reliability of the Final Sample}
Human classification of objects has been carried out by a single classifier, all the way up to large-scale citizen science projects. Our project was conducted by a group of 16 trained identifiers who developed the selection criteria. While we do not gain the extremely large statistics of a public citizen science project, keeping the project internal produced a cleaner initial sample with few sporadic taggings due to substantial training on the selection process. \citet{whitmore2021} refer to this as identifiers having similar `internal weighting systems.' We do not carry out an in-depth completeness test via artificial source injection. However, we do employ multiple steps, pre- and postidentification, to understand the quality and reliability of our final sample. In addition to catalog cleaning in Section~\ref{sec:proc_Zooniverse}, we compute concentration index values for our by-eye sample to test if they are consistent with what we expect for compact cluster objects (see Section~\ref{sec:F150wCI}), as this can be difficult to discern when visually identifying. We also ensure that all of our final sources have 3.3~$\mu$m PAH peak values greater than the 5$\sigma_{\mathrm{PAH}}$ detection limit. Moreover, in Section~\ref{sec:ModelTrends} we compare the location of our sample of embedded clusters used for machine learning on a color-color diagram to embedded sources from \citet{rodriguez2024} and the empirical templates of \citet{whitmore2025}.

%% file: 04_ML.tex
The usage of machine learning for stellar cluster detection, classification, and characterization has become an increasingly common means to accommodate growing astronomical datasets \citep{Wei2020, whitmore2021, Hannon2023}. In particualr, the ability of JWST to spatially resolve embedded cluster candidates motivates our goal to produce a method which can accelerate and systematize the process of cataloging these objects as new data become available \citep{whitmore2021, thilker2022}.

Using our sample of `by-eye' sources, we train convolutional neural network (CNN) models for the purpose of identifying embedded clusters within the PHANGS sample of nearby galaxies. Due to the sample size of our by-eye catalog, we opt to use a binary system for training and testing models. Our two labels include our catalog of embedded cluster candidates (Label~1) and an encompassing class of other F335M$_{\rm PAH}$ `bright’ sources (Label~0). To mitigate bias in the objects in our machine learning sample, we used standard source-finding techniques (described below) to create a general catalog of all F335$_{\mathrm{PAH}}$ peaks. From this general catalog, we populate our Label~1 and Label~0 objects as follows. If an embedded cluster candidate from the by-eye catalog has a matching peak in the general catalog within our source match tolerance, it is included in the Label~1 sample. We set a 0\farcs 2 distance (3.2 long-wavelength NIRCam pixels) and with these parameters, we retain $\sim80\%$ of the embedded clusters in our `by-eye' catalog for machine learning. Given that all of our by-eye selected sources meet the detection limits in Table~\ref{tab:sample}, we postulate that the $\sim20\%$ loss is likely a combination of the small separation threshold and the limited number of 3.3~\m\ PAH `bright' sources extracted in this step.

To maintain a set ratio of Label~1 to Label~0 objects in the final catalog, we randomly select other sources from each general catalog such that the ratio of other PAH-bright sources to embedded cluster candidates for each galaxy is 20:1. To reduce crowding, we exclude any sources between 0\farcs2 and 0\farcs5 of an embedded cluster candidate from selection for the Label~0 source catalog. In the central-most regions of galaxies, starlight dominates the F335M band \citep{sandstrom2023}, leading to a higher detection rate of sources in the central region in comparison to the outer regions. When we perform source extraction on the entire F335M$_{\rm PAH}$ image, this selection bias manifests, leading to a spatial sample bias for the catalog of Label~0 objects. As such, we perform source extraction for the galaxy center and outside regions separately on the F335M$_{\rm PAH}$ maps. Elliptical regions are defined by visually inspecting the nonsubtracted F335M map, since the central stellar-dominated region is more apparent in these images. 

Across all galaxies, we produce a sample of 4779 other PAH-bright objects and 233 human-identified embedded clusters for the purpose of training and testing machine learning models. We note that the ratio of 20:1 is chosen due to the small sample size of our embedded cluster candidate (Label~1) catalog. While a balanced distribution of sources is optimal for training classification models \citep{dablain2022}, this would limit our catalog to 466 objects for both training and testing models, which may result in over-fitting of the training data \citep{Raiaan2024, Soekhoe2016, Pasupa2016}. The impact of imbalanced datasets on model training is further discussed in Section \ref{sec:MLdiscussion}.

With the aim of employing these models for embedded cluster classification in the Cycle~1 sample of galaxies, as well as future datasets, we seek to understand how varying the data we include in our training alters model classification. Our choice in image sets reflects all filters available for the Cycle~2 data as well as the overlap between the Cycle~2 and Cycle~1 samples (i.e. excludes F150W and Pa$\alpha$). We also vary the inclusion of the F770W data in our image sets to determine its impact on model performance.  In total, we produce four variations of training data using the regridded and aligned JWST and HST images from \S~\ref{sec:sample_data}. All available HST filters are included, and the JWST filter inclusion is varied to reflect data availability for the Cycle~2 and Cycle~1 PHANGS sample. Table \ref{tab:MLfilt_summary} highlights the JWST bands in each data configuration as well as the total filters when accounting for the HST data.

\setlength{\tabcolsep}{12pt}
\input{ML_filters.tex}
\setlength{\tabcolsep}{1pt}

We create multiextension fits files (MEFs) for all objects in our source catalog for each data variation. To produce MEFs, we create $299\times299$ pixel cutouts in each filter using the object's F335M pixel center as the location of the cutout. We then stack the data cutouts into a single $N\times299\times299$ array, where $N$ is the number of filters we include in the MEF. We include the galaxy, ID number, known label, and F335M X-Y pixel center of each object in the MEF header. From these, we select an 80\% subset of objects from each label to use for model training and reserve the remaining 20\% for model testing. Table \ref{tab:MLobjects} summarizes the distribution of sources in our ML catalog.

\setlength{\tabcolsep}{6pt}
\input{ML_catalog.tex}
\setlength{\tabcolsep}{1pt}

\subsection{Convolutional Neural Network Training}
\label{ml_training}
To employ embedded cluster identification with deep transfer learning, following the work of \cite{Wei2020}, we utilize two CNNs: \texttt{VGG19-bn} \citep{Simonyan&Zisserman2015} and \texttt{ResNet18} \citep{he2015}. \texttt{VGG19-bn} and \texttt{ResNet18} are both deep convolutional neural networks, meaning they contain multiple convolution and resampling layers to reduce multidimensional input data to a one-dimensional feature vector, which the neural network then maps to an output classification. An overview of the network architectures is available at \cite{Simonyan&Zisserman2015} and \cite{he2015}, respectively. Model classification weights are pretrained on real-world $299\times299$ pixel RGB images, meaning they possess the prior optimization necessary to learn elementary shapes, contours, and patterns within images. We implement pretrained classification weights from {\fontfamily{pcr}\selectfont ImageNet} (`IMAGENET1K\_V1' in \texttt{PyTorch}; \citealt{Paszke2019}), an image database with over 14 million images \citep{Deng2015}. By altering the input training data, we are able to reoptimize the final weight of the neural network -- which is responsible for object classification -- for the purpose of embedded cluster candidate identification.

We direct the reader to \S~4.2 of \cite{Hannon2023} for an in-depth discussion of the training procedure, but we outline necessary training features here. We begin training by randomly selecting objects from our training set to populate a batch size of $B$ sources. When loading the input data, we first crop the original MEF to its central $50\times50$ pixels to reduce image crowding, then resize back to a shape of $299\times299$ using \textit{resize} from \texttt{skimage}. \cite{Tammina2019} explored incorporating rotation on their training data and found there was a decrease in accuracy of $\approx15\%$, but the model was considered to be more robust in handling variation in real-world data sets. As such, to produce variation in the orientation of our training data, we apply a random rotation of the data array between $0^{\circ}$ and $360^{\circ}$ and alternate flipping the data along the vertical axis. For a single training batch, the $B\times N\times299\times299$ data array becomes the input for classification, and the known object labels become the target output.

A single CNN model classifies on RGB data, meaning it possesses three channels for input data of a single object. Our image sets contain 8, 9, 10, and 11 data layers, so we train one model for every three layers in an image set.  As in \cite{Wei2020}, any remaining unfilled data layers in the last model are set to a constant zero array. We then concatenate these models to produce one larger identification model, which uses information from all filters for classification. Prediction labels for each object in the training batch are output as a 1D tensor, which we use to compute the cross entropy loss between the prediction and target labels using \textit{CrossEntropyLoss} from \texttt{Pytorch}. The gradient of the loss then determines the change necessary in the parameters to minimize the loss. This, in combination with the learning rate, determines by how much the parameters change for each batch of training. A validation test is run with the new parameters, and if the loss of the validation set decreases from the prior batch, the optimizer updates with the new model parameters (classification weights). For our work, we use an \textit{Adam} optimizer \citep{kingma2017} from \texttt{PyTorch} and fix our learning rate to $10^{-4}$. We use a batch size of 32 objects and train for 4000 batches for \texttt{ResNet18} models, and reduce the batch size to 14 and execute 3000 batches of training for \texttt{VGG19-bn} models. We do not include weights to account for the imbalance of objects in each class in our training sample, and we discuss the impact on results in Section~\ref{sec:C2_machine_learning}. We utilize the MedicineBow Compute Environment within the Advanced Research Computing Center at the University of Wyoming, as the available GPUs allow for parallelization of computations and reduce the time necessary for training. In total, we train 10 models for all eight model configurations (four data variations for each of the two CNNs). Hyperparameters (number of batches, batch size, learning rate, etc.) are kept consistent across all ten models for a given data configuration. The results of testing the final models are presented in \S~\ref{sec:C2_machine_learning}.

%% file: ML_filters.tex
\begingroup
\begin{deluxetable}{ccccc}[ht!]
\tablecaption{Data Variations for Machine Learning}
\label{tab:MLfilt_summary}
\tablewidth{\columnwidth}
\tablehead{\colhead{} &  \multicolumn{4}{c}{Data Configuration}\\
\cline{2-5}
\colhead{JWST Filter} & \colhead{1} & \colhead{2} & \colhead{3} & \colhead{4}}

\startdata
F150W & x & x &   &   \\ %
F300M & x & x & x & x \\ %
F335M & x & x & x & x \\ %
F770W &   & x &   & x \\ %
Pa$\alpha$ & x & x & & \\ %
\hline
Total Filters & 10 & 11 & 8 & 9\\%
\enddata

\tablenotetext{}{JWST data inclusion for the four configurations we use to train and test CNN models. All HST data (F275W, F336W, F435W/F438W, F555W, F814W, and H$\alpha$) are included in each configuration, as these data are available for both the Cycle~2 and Cycle~1 PHANGS galaxies.}
\end{deluxetable}
\endgroup

%% file: ML_catalog.tex
\begin{deluxetable}{ccrrr}[!ht]
\tablecaption{Machine Learning Objects}
\label{tab:MLobjects}
\tablehead{\colhead{Object} & \colhead{Label} & \colhead{Total} & \colhead{Training}  & \colhead{Testing}}
\startdata
Cluster & 1 & 233 & 187 & 46 \\ %
Non-Cluster & 0 & 4779 & 3824 & 955 \\ %
\enddata
\tablenotetext{}{Summary of the object dataset used for machine learning.  The non-cluster objects include all peak detections from our ``find peaks'' algorithm that do not correspond to a human-identified embedded cluster candidate.  Also listed are the number of sources used for training and testing, split into 80\% and 20\% subsets of the total sample, respectively.}
\end{deluxetable}

%% file: 05_results.tex
\subsection{Embedded Cluster Catalog vs Global Properties}\label{sec:globalprops}
\begin{figure*}[!ht]
   \centering
   \includegraphics[width=0.8\linewidth]{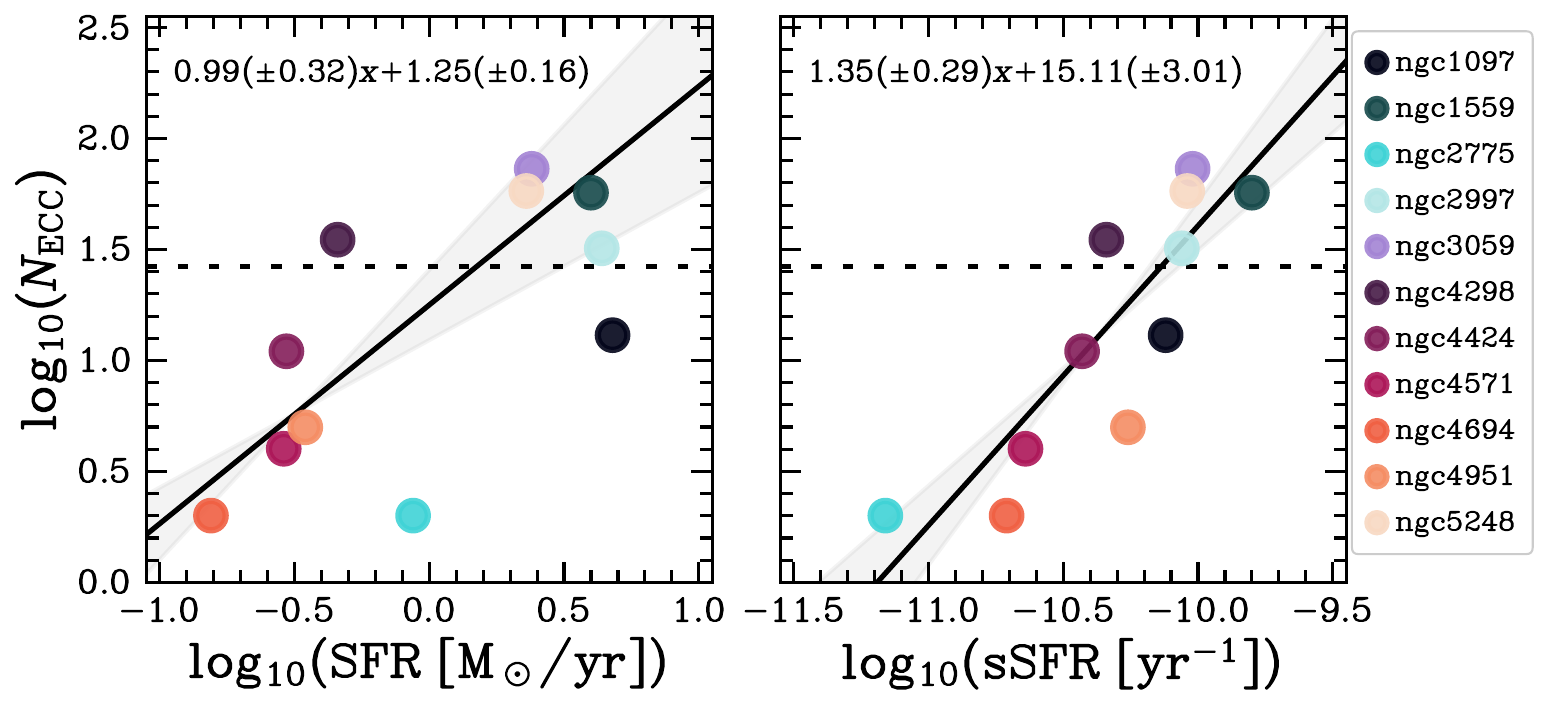}
   \caption{Number of embedded cluster candidates versus global star-formation properties. Values for each galaxy are in Table~\ref{tab:sample} and embedded cluster candidate counts from this study are in Table~\ref{tab:phot_summary}. Left: $\mathrm{log_{10}(SFR})$ versus Number of embedded cluster candidates. Right: $\mathrm{log_{10}(sSFR})$ versus Number of embedded cluster candidates. The black line in each plot show the linear fit to the data with the $\pm 1\sigma$ region of the fit in light gray. We exclude NGC2997 and NGC1097 in fitting due to lack of wide coverage. The number of embedded cluster candidates positively correlates with each of these properties.}
   \label{fig:Nclus_props}
\end{figure*}

The number of sources we identify per galaxy spans from 2 to 73 embedded cluster candidates, and we expect this number to scale with global star-formation activity indicators \citep{Adamo2020}. For each galaxy, we compare the number of human-identified embedded cluster candidates ($N_{\mathrm{ECC}}$) to the SFR and specific star formation rate (sSFR). We calculate sSFR by taking the ratio of the SFR to the stellar mass $(M_*)$ for each galaxy (columns (6) and (7) in Table~\ref{tab:sample}). Figure~\ref{fig:Nclus_props} (left) displays $N_{\mathrm{ECC}}$ versus SFR, and Figure~\ref{fig:Nclus_props} (right) displays $N_{\mathrm{ECC}}$ against sSFR for our sample. Color indicates the galaxy, and the horizontal dashed line represents the mean number of embedded cluster candidates for our sample $(\left<N_{\mathrm{ECC}}\right> =26.5)$ per galaxy. The linear fit for each dataset is shown in black and the 1$\sigma$ uncertainty envelope for the fit in light gray. The numeric fit parameters and errors are also included in each plot. We exclude NGC~2997 in fitting, as the HST data only covers the eastern portion of the galaxy, resulting in artificially lower $N_{\mathrm{ECC}}$ than what we may expect with larger coverage for identification. We also exclude NGC~1097 from fitting, as the JWST F335M data coverage extends less than 0.5$r_{25}$ from the galaxy center.

\begin{figure}[!ht]
    \centering
    \includegraphics[width=0.85\linewidth]{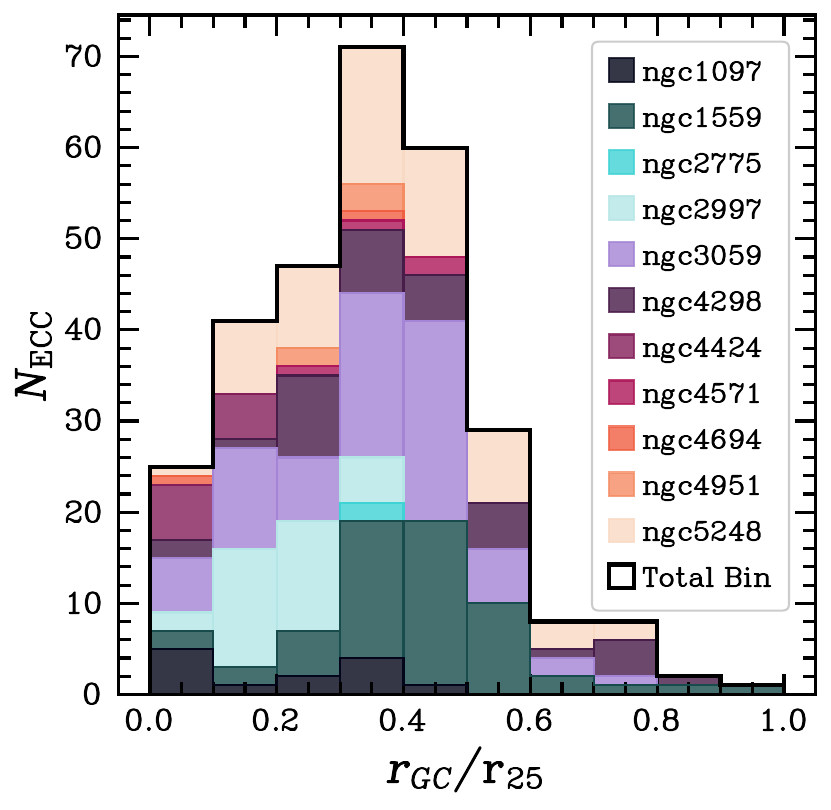}
    \caption{Histogram of de-projected galacto-centric radial distances ($r_{\rm GC}$) as a fraction of $r_{25}$ for all 292 embedded cluster candidates. Table~\ref{tab:sample} gives $r_{25}$ for each of the galaxies in our sample. The black outline represents the total bin count and each color denotes the number of clusters from each galaxy.}
    \label{fig:fR25}
\end{figure}

Our sample size limits the conclusiveness of these relations, but we do find a positive correlation between the number of embedded cluster candidates in a galaxy and both the SFR ($r=0.76$) and the sSFR ($r=0.87$). \cite{rodriguez2024} find a similar trend between the SFR and the number of 3.3~$\mu$m PAH emitters in PHANGS Cycle~1 galaxies. We additionally show here that the number of these compact, dusty sources closely relates to the rate of new star formation activity in a galaxy.
We also assess the distribution of sources for each galaxy by calculating the deprojected galactocentric radii. To derive this value, we first compute the on-sky angular distance and position angle of our source F335M centers to the respective galaxy center. We then deproject this distance by correcting for galaxy inclination effects along each axis of the galaxy and normalize to $r_{25}$. The galaxy center coordinates and values of $r_{25}$ for our sample are in Table~\ref{tab:sample}. Figure~\ref{fig:fR25} shows the histogram of deprojected galactocentric distances for our 292 embedded cluster candidates in bins of 0.1$r_{25}$. We find that a significant fraction of our sources lie within 0.5$r_{25}$ (83.6\%). However, the distribution of sources is limited by the spatial coverage for each galaxy (see Figure~\ref{fig:composite_image}). The two galaxies limited most by spatial coverage are NGC~1097 and NGC~2997. The JWST F335M data for each of these galaxies covers less than 0.5$r_{25}$, and as such, our by-eye source detection was limited to the innermost regions of these galaxies. When we exclude sources from these two galaxies, the fraction of sources within 0.5$r_{25}$ drops slightly to 80.6\%.

\subsection{Embedded Cluster Candidate Properties}\label{sec:source_properties}
We conduct aperture photometry for sources in our `by-eye' sample in the JWST NIRCam filters F150W, continuum-subtracted F187N (Pa$\alpha$), F300M, and F335M, as well as the HST-WFC3 continuum-subtracted F657N/F658N (H$\alpha$).  We choose a primary aperture with $r_{\rm ap}=0$\farcs124 and a sky annulus from $r_{\rm ann}=0$\farcs248 to $r_{\rm ann}=0$\farcs372, following choices previously adopted for PHANGS near-infrared stellar cluster photometry \citep{rodriguez2023,rodriguez2024}.

For each medium- and wide-band image, we correct for background contribution in the aperture using the $\sigma$-clipped median (using five iterations with $\sigma=3$) in the sky annulus.  For the continuum-subtracted narrow-band images (F187N and F657N/F658N), we do not perform local background subtraction. For the H$\alpha$ data, we correct for \NII\ contribution according to
\begin{equation}
     F_{\mathrm{H}\alpha} =\frac{ F_{\mathrm{H}\alpha+\mathrm{[NII]}}}{1+\left([\mathrm{NII]}/{\mathrm{H}\alpha}\right)}.
\end{equation}
To derive values for \NII/H$\alpha$ for our sample, we fit a linear relation to \NII/H$\alpha$ versus log$_{10}(M_*)$ for galaxies from \cite{leroy2021} with \NII/H$\alpha$ values available in \cite{kennicutt2008}. The details of the fitting procedure are in Appendix~\ref{sec:NII_Ha}, and the resulting values of \NII/H$\alpha$ for the galaxies in our sample are in Table~\ref{tab:sample}.

We rely on the F150W data to derive the concentration index and masses of sources in our sample. Since a prominent F150W source is not part of our selection criteria, we conduct source matching between the F150W data and the F335M centers of our embedded cluster candidates. To source match, we model a background using the \textit{Background2D} function from \textsc{Photutils}. We use SExtractorBackground, which estimates the background of an $N\times N$ box as $\mathrm{(2.5\cdot median) -(1.5\cdot mean)}$. In instances where $\mathrm{(median-mean)/std>0.3}$, the straight median value becomes the background value. We apply a 3$\sigma$ clip to our background estimation using 15 iterations, rather than the default 5. We adopt a box size of 85~pc from \cite{rodriguez2024}, who determined this is an optimal size for their sample of PHANGS galaxies at distances ranging from $\sim$5~Mpc to $\sim$20~Mpc. 

We create a 3$\sigma$ threshold map from the 2D background data using \textit{detect\_threshold}. Lastly, we extract F150W sources using \textit{find\_peaks} and their centroids with \textit{centroid\_2dg}. We crossmatch our embedded cluster candidate F335M centroids to the F150W peaks catalog using \textit{match\_to\_catalog\_sky}. We choose this function over \textit{search\_around\_sky} for source matching because it ensures only one match is made for each source, whereas the latter function could return multiple matches for a given source. We set our separation threshold to 0\farcs063. Table~\ref{tab:phot_summary} includes the F150W pixel-to-pc scale for each galaxy, along with the number of F335M sources with a matching F150W peak from this process.  We conduct aperture photometry using the F150W centroids.  We use the photometry to derive masses and concentration indices for sources with `good' matching F150W photometry, i.e., background-subtracted flux greater than 0, and the brightest pixel has S/N$>5$.

\setlength{\tabcolsep}{1.5pt}
\input{analysis_summary}
\setlength{\tabcolsep}{1pt}

\subsubsection{Pa\texorpdfstring{$\alpha$}{alpha} Equivalent-Width Age Dating}
\label{sec:pa-alpha_ew_age_dating}
By design, our embedded cluster sample is invisible in the broadband optical filters, and thus we are unable to estimate cluster ages utilizing optical or optical/near-infrared SED fitting \citep[e.g.,][]{thilker2024,henny2025}.  Thus, we determine the ages of the embedded clusters via their Pa$\alpha$ equivalent widths.  We utilize Starburst99 \citep{leitherer1999,leitherer2014} to convert between Pa$\alpha$ equivalent widths and the age of a stellar cluster.  We model a stellar population using a Kroupa initial mass function \citep[IMF;][]{Kroupa2002} with an upper-mass limit of 100$~M_\odot$.  We input Geneva solar metallicity tracks with standard mass loss \citep{Meynet1994}, and construct our model from 0.1~Myr to 1~Gyr using a timestep of $10^4$ yr. 

The Starburst99 model output provides Pa$\beta$ but not Pa$\alpha$ equivalent widths.  We utilize the modeled Pa$\beta$ luminosities (in $\rm{erg\ s^{-1}}$) and Pa$\alpha$ continuum flux (in $\rm{erg\ s^{-1}\ \text{\AA}^{-1}}$) to determine the modeled Pa$\alpha$ equivalent width via
\begin{equation}\label{eq:pa_equivalent_width_model}
EW_{{\rm Pa}\alpha,{\rm modeled}} \text{(\AA)} = 1.97\times\frac{L_{{\rm Pa}\beta}}{F_{{\rm c,Pa}\alpha}},
\end{equation}
where $L_{{\rm Pa}\beta}$ is the luminosity of the Pa$\beta$ line, $F_{{\rm c,Pa}\alpha}$ is the underlying continuum flux at the Pa$\alpha$ line wavelength ($\lambda=1.87~\mu\rm{m}$), and 1.97 is the intrinsic Pa$\alpha$/Pa$\beta$ line ratio adopted from CLOUDY \citep{ferland2017,reddy2023}.

For our catalog of human-identified embedded cluster candidates, we compute the observational $\mathrm{EW_ {Pa\alpha}}$ as
\begin{equation}\label{eq:pa_equivalent_width}
 EW_{{\rm Pa}\alpha,{\rm observed}}\text{(\AA)} = \frac{F_{{\rm l,Pa}\alpha}}{F_{{\rm c,Pa}\alpha}}\times \Delta\lambda_{\mathrm{F187N}},
\end{equation}
with $F_{{\rm l,Pa}\alpha}$ and $F_{{\rm c,Pa}\alpha}$ being the photometric flux of the line emission and continuum emission, respectively, and $\Delta\lambda_{\mathrm{F187N}}=240$~\AA\ being the bandwidth of the NIRCam F187N filter.

\begin{figure}[!ht]
 \centering
 \includegraphics[width=\columnwidth]{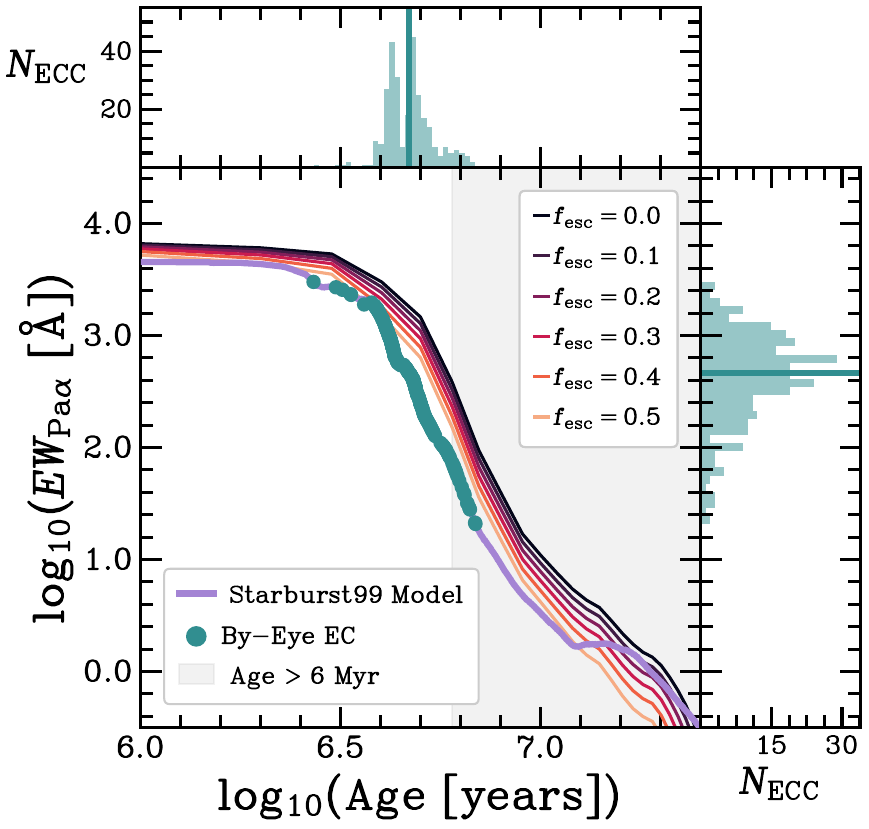}
 \caption{Cluster candidate ages using photometric Pa$\alpha$ equivalent widths (see \ref{sec:pa-alpha_ew_age_dating} for details). The cyan circles show the distribution of 291 sources with good age estimates using the Starburst99 model (purple). The histograms show the distribution of these sources on each axis with the medians indicated by a vertical line. A majority of our embedded cluster candidates are less than 6~Myr old, with a significant portion less than 5~Myr old (73\%). For comparison, six CIGALE models are shown with color indicating the escape fraction modeled. These models show that the age derived will vary significantly at lower equivalent widths and less so for high equivalent widths. The young ages derived via Pa$\alpha$ equivalent widths indicate that it is likely that the first supernovae have yet to unbind the natal shroud of gas and dust \citep{deshmukh2024, linden2023}.}
 \label{fig:EWage}
\end{figure}

To derive ages for the embedded cluster candidates, we take the observed Pa$\alpha$ equivalent width and linearly interpolate among the modeled Pa$\alpha$ equivalent widths and their corresponding modeled ages.  In Figure~\ref{fig:EWage}, we denote these ages with cyan points and compare them to the model values (purple line). Note that we impose a minimum age of 1~Myr due to the presence of degeneracies in the model below this age; therefore, all sources with an equivalent width higher than that corresponding to an age of 1~Myr in the model are assigned an upper-limit age of 1~Myr. None of our embedded cluster candidates fall onto this upper limit, but a small sample of HST Class 1+2 clusters from \cite{maschmann2024} do when we use this method for age dating in Section \ref{sec:F150wCI}. The gray shaded region highlights ages beyond 6~Myr, the approximate age at which supernovae feedback begins to dominate mechanical feedback mechanisms for stellar clusters \citep{schinnerer2024}.

Starburst99 assumes Case-B recombination, meaning all Lyman photons are absorbed and reemitted as either free-free or free-bound emission by the surrounding gas \citep{leitherer1999}. To visualize the impact of this assumption on our ages, we model Pa$\alpha$ equivalent width using the Code for Investigating GALaxy Emission \citep[\texttt{CIGALE;}][]{boquien2019}. We use a Chabrier \citep{chabrier2003} IMF and vary the escape fraction $(f_\mathrm{esc})$ between 0.0 and 1.0, with an $f_\mathrm{esc}$ of zero representing absorption and reemission of all Lyman photons by the surrounding dust and gas, i.e., no ionizing-photon `leakage' into the surrounding interstellar medium. In Figure \ref{fig:EWage}, the gradient lines show models for five values of $f_\mathrm{esc}$ from 1 to 30~Myr. There is an average equivalent width ratio of 0.73 for ages of 1--3~Myr, 0.44 for ages of 4--6~Myr, and 0.38 for ages of 7--10~Myr between models with $f_\mathrm{esc}=0.5$ and $f_\mathrm{esc}=0$. Discussion of the impact of $f_\mathrm{esc}$, as well as other model assumptions, on the age estimates is found in Section \ref{sec:discussion}.

We derive a very sharp distribution of ages for our sources, with a significant fraction with ages between 4 and 5~Myr. There is only one embedded cluster candidate in our sample that receives an age estimate of less than 3~Myr. Based on stellar population age distributions from other nearby-galaxy studies, we would expect a broader range of ages, specifically in the 1--3~Myr range. This is thought to be a result of the interplay between the orbital timescale for a burst of star-formation and the disruption timescale of clusters \citep{Whitmore2007, Fall2009}. We postulate that the dearth of young clusters stems from the method we use for age dating. By performing no extinction correction on the nebular or continuum Pa$\alpha$ data, we are inherently assuming equal extinction for these values. However, nebular extinction can exceed stellar extinction by a factor of 2 \citep{Calzetti1994}. For our heavily embedded sources, performing such a correction would increase the derived Pa$\alpha$ equivalent widths and result in younger source ages.

\subsubsection{Visual Extinction}\label{sec:cluster_extinctions}
\begin{figure}
    \centering
    \includegraphics[width=\columnwidth]{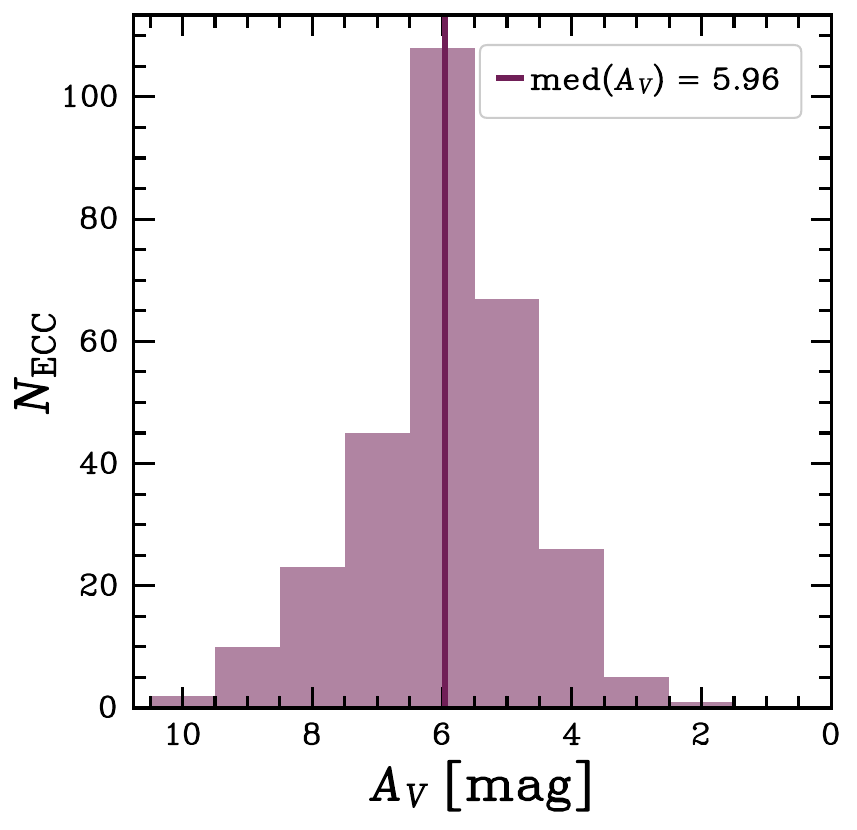}
    \caption{Distribution of $A_V$ for the 291 visually-identified embedded cluster candidates with good age estimates. A large portion ($\sim$98\%) of our embedded cluster candidates exhibit $A_V$ values greater than 3~mag, which most likely is a reflection of our by-eye selection criteria.}
    \label{fig:a_v_distribution}
\end{figure}

We estimate the visual extinction ($A_V$) by comparing the measured Pa$\alpha$/H$\alpha$ line ratio for each embedded cluster candidate to the intrinsic Pa$\alpha$/H$\alpha$ line ratio of 0.1093 \citep{ferland2017,reddy2023}. This yields the visual extinction as
\begin{equation}\label{eq:extinction}
A_V = \frac{-2.5}{A_{{\rm Pa}\alpha}/A_V-A_{{\rm H}\alpha}/A_V} \mathrm{log_{10}}\left(\frac{{\rm Pa}\alpha_{\rm obs}/{\rm H}\alpha_{\rm obs}}{0.1093}\right)
\end{equation}
We determine the values for $A_{{\rm H}\alpha}/A_V$ and $A_{{\rm Pa}\alpha}/A_V$ from the standard Milky Way ($R_V=3.1$) dust models of \citet{draine2003}.  Figure~\ref{fig:a_v_distribution} shows the distribution of $A_V$ across the embedded cluster candidate sample. We note a peak in the distribution at $\sim A_V \approx 6$~mag. The distribution is subject to bias due to our selection criteria of a `high' Pa$\alpha$/H$\alpha$ lower limit corresponding to $\sim3.8$~mag. However, our result is in good agreement with the visual extinction of other embedded clusters in nearby galaxies \citep{prescott2007, linden2023, linden2024}.

\subsubsection{Concentration Index}\label{sec:F150wCI}
The concentration index (CI) is a way of quantifying the central compactness of a source by measuring the ratio of aperture fluxes from two different radii. Similar works \citep{Whitmore2023, rodriguez2024} use the concentration index as a metric to distinguish between stars and stellar clusters in nearby galaxies. \cite{Whitmore2023} adopt an inner aperture size of $r_{\rm apt}=1\ \mathrm{pix}$ and outer aperture size of $r_{\rm apt}=4\ \mathrm{pix}$ for the JWST F200W band, corresponding to an outer aperture size of 0\farcs12.  We implement the same aperture sizes to compute the F150W concentration index ($CI_{\rm F150W}$) for our subsample of 212 embedded cluster candidates with a matching F150W source.

Table~\ref{tab:phot_summary} gives the mean $CI_{\rm F150W}$ of the embedded cluster candidates from each galaxy. For comparison, we compute the concentration index for the available catalogs of PHANGS-HST compact stellar clusters (Class~1+2 in \citealt{maschmann2024}) and AGB stars selected from the PHANGS-HST DOLPHOT catalogs. Details on the detection, selection, and usage of these sources are found in \citet{thilker2022}, \citet{lee2023}, and \citet{williams2024}.

\begin{figure*}[!ht]
    \centering
    \includegraphics[width=0.95\linewidth]{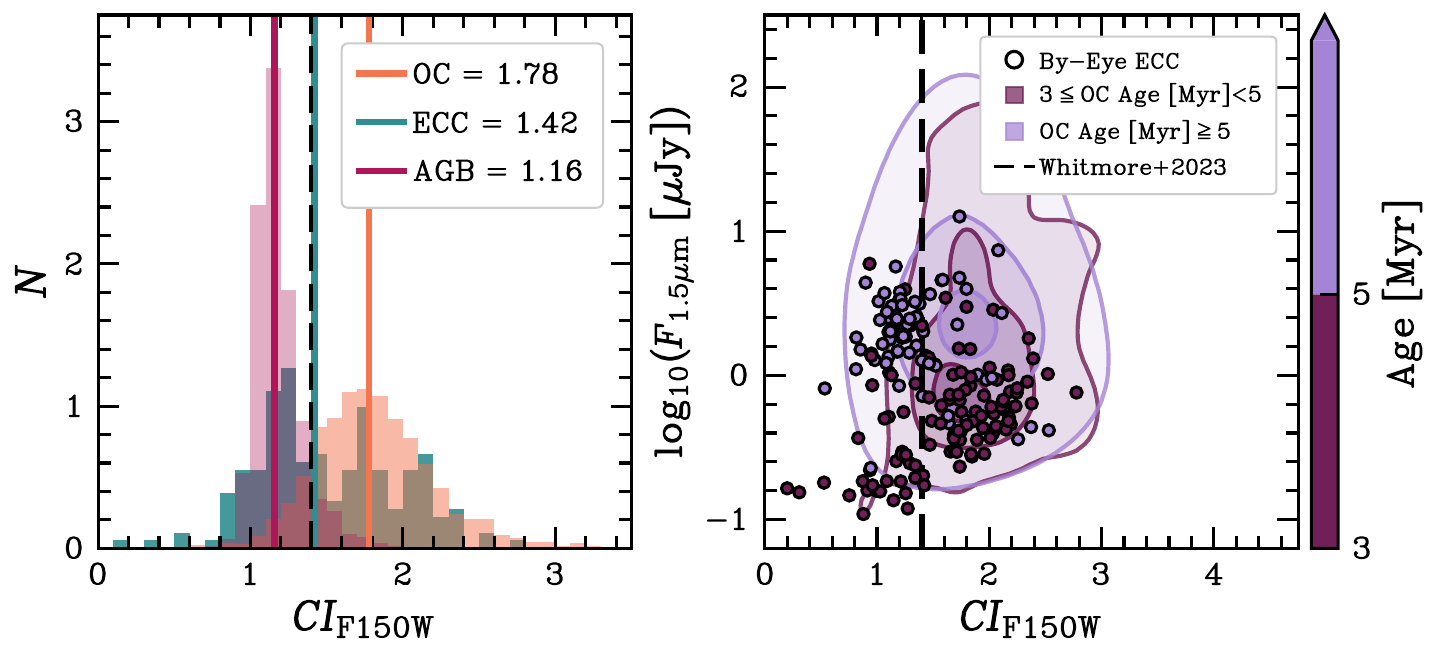}
    \caption{Left: Normalized histogram of 181 of our embedded cluster candidates (ECC, cyan), 1117 AGB stars (AGB, pink), and 2308 HST Class 1+2 optical clusters (OC, orange; \citealt{maschmann2024}) with F150W concentration index $CI_{\rm F150W}$ greater than zero and a peak $S/N>5$. Bins are normalized to the total number of objects and the bin width. The solid vertical lines indicate the median for each distribution of objects.  The bimodal distribution of the embedded cluster candidates may be indicative of evolved stars present within our by-eye sample. Right: F150W surface brightness against $CI_{\rm F150W}$ where color indicates the age derived via Pa$\alpha$ equivalent width age-dating. The scatter shows our embedded cluster candidates and the contours denote the $\mathrm{10^{th},\ 50^{th},\ and\ 90^{th}}$ percentiles of HST Class 1+2 optical clusters for 3--5~Myr (dark purple) and greater than 5~Myr (light purple). The dashed vertical line at $CI=1.4$ marks the limit \cite{Whitmore2023} uses to distinguish between stars and clusters in the JWST F200W band.  The median $CI_{\rm F150W}$ is 1.22 for embedded cluster candidates with ages greater than 5~Myr and 1.69 for sources in the 3--5~Myr age bin.}
    \label{fig:CI_plots}
\end{figure*}

Figure~\ref{fig:CI_plots} (left) shows the resulting histograms of $CI_{\rm F150W}$ for each of these populations: embedded cluster candidates (ECC) in cyan, HST Class~1+2 optical clusters (OC) in orange, and AGB stars (AGB) in pink. We include sources with a concentration index greater than zero and an F150W S/N$> 5$, resulting in $CI_{\rm F150W}$ values for 181 embedded cluster candidates, 2308 HST Class~1+2 optical clusters, and 1117 AGB stars. The cyan, orange, and pink vertical lines mark the median of each population, respectively. Each histogram is normalized to the total number of objects and the bin width.

We note a bimodality in our embedded cluster candidate distribution, with one peak consistent with the median of the AGB stars and the second consistent with the median of HST Class~1+2 optical clusters. We examine potential causes of this bimodality in Figure~\ref{fig:CI_plots} (right), where we show F150W flux density versus $CI_{\rm F150W}$. The colors indicate Pa$\alpha$ equivalent-width age bins of 3--5~Myr (light purple) and $\geq5$~Myr (dark purple). For our embedded cluster candidates, there are 108 and 71 sources in each age bin, respectively. There is one source in the 1--3~Myr age bin and one without a good age estimate, which are not included in Figure~\ref{fig:CI_plots} (right).

\cite{Whitmore2023} use $CI=1.4$ in the F200W band to distinguish between stars and stellar clusters in NGC~1365 (vertical black dashed line).  Adopting this limit, we find that 67.6\% of our sources with ages $\geq5$~Myr have a $CI_{\rm F150W}$ less than 1.4, while 66.1\% of sources with ages $<5$~Myr have a CI larger than 1.4. As expected, the HST Class 1+2 clusters do not display the same extent of stratification in CI with age evolution. For Pa$\alpha$ ages $<5$~Myr and ages $\geq5$~Myr, a majority of HST Class 1+2 clusters, 95.4\% and 89.2\%, respectively, with lie beyond this compactness limit. We explore explanations for the compact, F150W bright and faint subpopulations in Section~\ref{sec:CI_populations}.

\subsubsection{Masses}\label{sec:mass_funcs}
To compute embedded cluster candidate masses, we use CIGALE to calculate synthetic F150W fluxes along with a theoretical mass-to-light ratio for the F150W band.  We model a 1~Myr old cluster using \cite{bruzual2003} solar metallicity tracks and a fully sampled Chabrier IMF. Due to the uncertainty in our source ages, we choose to model a 1~Myr cluster rather than using the median age of our sample. Our model assumes an instantaneous star formation history, maximal nebular emission ($f_{\rm{esc}}=0$), and an $A_V$ of 1.\footnote{Similar to our age estimates (Section~\ref{sec:pa-alpha_ew_age_dating}), increasing the escape fraction will decrease nebular emission at a given age and decrease the total F150W emission. For our model age of 1~Myr, this effect is extremely relevant, as nebular emission can contribute up to $\sim25\%$ to the total F150W emission. Moreover, model 1.5\m\ flux will decrease with increasing model cluster age until $\sim 5$~Myr when the most massive stars begin to evolve. Hence, our mass estimates reflect lower limits, as changes to these parameters will decrease the model F150W flux density and increase the model mass-to-light ratio.}
With these model assumptions, the resulting mass-to-light scale at 10~pc is $2.36\times10^6\ \rm{mJy\;M_{\odot}^{-1}}$ or $4.23\times10^{-7}\ \rm{M_{\odot}\;mJy^{-1}}$. The scale at 10~pc is converted to the distances of our sample via
\begin{equation}
    F(D\;[\mathrm{pc}])=2.36\times10^6\;\left[\mathrm{mJy\;M_\odot^{-1}}\right] \div \left(\frac{D}{10\;\mathrm{pc}} \right)^{2}.
\end{equation}
For the distances of our target galaxies, the scale range is $7.81\times10^{5}$--$2.27\times10^{6}$~$\rm{M_{\odot}\;mJy^{-1}}$. Source stellar masses are inferred by converting the observed F150W fluxes using the modeled mass-to-light scale at the respective galaxy distance.

\begin{figure}[!ht]
    \centering
    \includegraphics[width=\columnwidth]{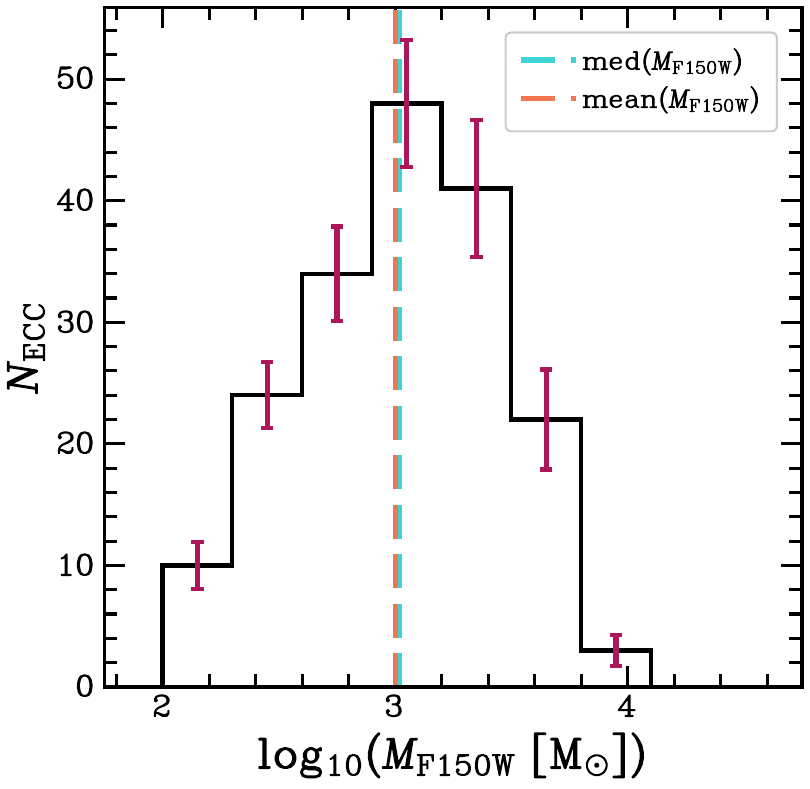}
    \caption{A histogram of the distribution of masses for the subset of 186 embedded cluster candidates with an F150W peak pixel $S/N>5$. There are two sources with mass estimates below $10^2 \ \rm M_{\odot}$ not shown. The maroon error bars show the 1$\sigma$ standard deviations from Monte Carlo simulations for each bin with at least one embedded cluster candidate.}
    \label{fig:MChist}
\end{figure}

Figure~\ref{fig:MChist} shows a histogram of cluster stellar masses in black for the sources with a secure F150W flux ($\mathrm{S/N}>5$). The cyan and orange vertical lines represent the median and mean of our distribution, respectively. The mass distribution of our sample spans a similar range to the emerging young stellar clusters in M83 from \citealt{knutas2025}, but a scarcity of massive clusters ($M>10^{4}\ \rm{M_\odot}$) in comparison to strong 3.3 PAH emitters in \citealt{rodriguez2024}. Section~\ref{sec:ComparisonCatalogs} provides a comprehensive comparison of our derived cluster properties to these catalogs and other similar works.

To estimate measurement-based uncertainties in mass bins, we utilize a Monte Carlo simulation to randomly sample the observed fluxes by their uncertainties. We then convert the resulting fluxes into masses utilizing our previously derived \texttt{CIGALE} mass-to-light ratio. For each real flux, we create 1000 synthetic fluxes. Synthetic fluxes are equal to the original flux plus a random number that is set by a Gaussian, where the standard deviation of that Gaussian is equal to the flux uncertainty. We then convert these synthetic fluxes to masses and bin them, as we binned our sample of embedded cluster candidate masses. Then, we compute the standard deviation of the number of simulated embedded cluster candidates in each bin, and utilize this as the uncertainty on the observed number of embedded cluster candidates. The maroon error bars in Figure~\ref{fig:MChist} display the resulting 1$\sigma$ standard deviations in each mass bin from our Monte Carlo simulations that have at least one count in the embedded cluster candidate mass distribution (black).

As a note, stochastic sampling effects will contribute more heavily to uncertainties in our mass distribution than the uncertainties derived above. The CIGALE model we use to derive masses assumes a fully populated IMF; however, the presence of massive stars, or lack thereof, in our sources will lead to large variability in the composition of the near-infrared light used to derive masses. In particular, the 1.5\m\ emission could be enhanced due to nebular emission contribution (primarily Pa$\beta$ for the JWST F150W filter) for a young cluster, or conversely, for an older cluster, the massive stars being evolved. We revisit this in Section~\ref{sec:age_discussion} and further discuss the impact of stochasticity in the context of our cluster ages. Moreover, the size of our sample is small, meaning it is limited in fully probing the full range of cluster masses, especially at the high-mass end, leading to large uncertainties in our distribution (\cite{Adamo2020}, and references therein). However, the diverse galaxy sample allows for an interesting look at how galaxy environment can influence the mass of clusters formed. Structural features such as bars and nuclear rings are thought to be more effective at producing higher-mass clusters due to differences in cloud properties and dynamics in these regions \citep{Adamo2020, Sun2024}. In our sample, galaxies with one of these features (NGC~1097, NGC~1559, NGC~3059, and NGC~5248) host at least one cluster with a mass greater than 5000~$\rm M_\odot$, and this is often the most massive cluster in the galaxy. The most massive cluster in NGC~2997 is less than 5000~$\rm M_\odot$ but it is located in its central nuclear region. 

\subsection{Machine Learning Analysis}\label{sec:C2_machine_learning}
\subsubsection{Model performance}
\begin{figure*}[!ht]
    \centering
    \includegraphics[width=\columnwidth]{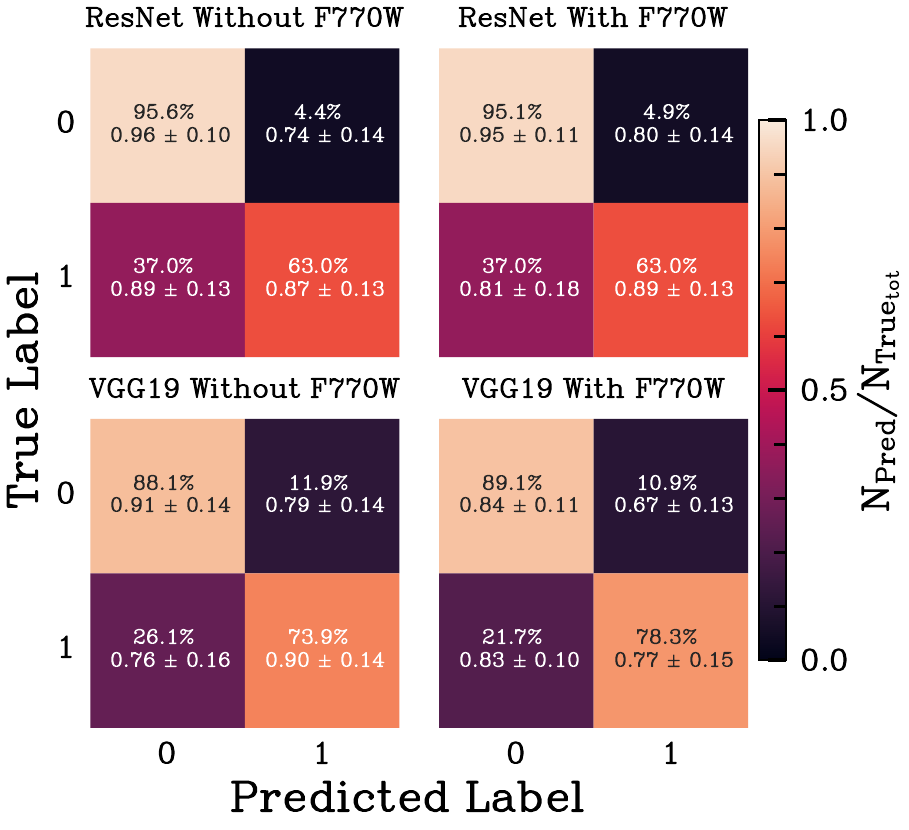}
    \includegraphics[width=\columnwidth]{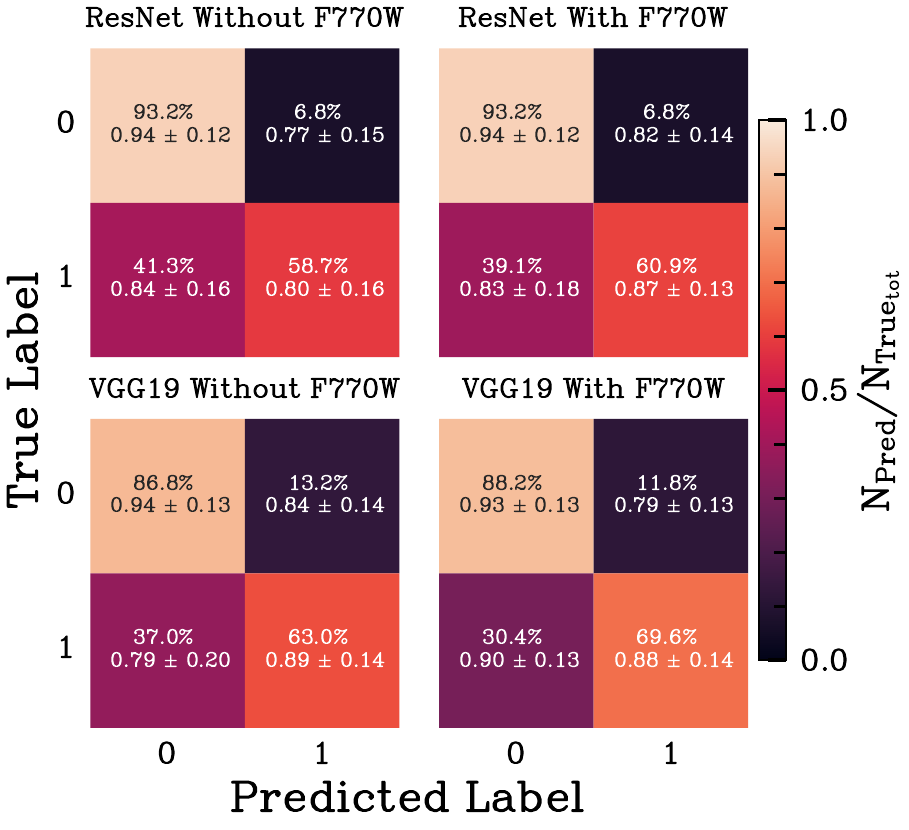}
    \caption{Confusion matrices for all 8 model configurations, normalized to the number of objects tested in each `True Label' category (see Table~\ref{tab:MLobjects} for number of objects in each class). The numbers below the percentages are the average and standard deviation of the model agreement in each quadrant. Left: Confusion matrices for models which include F150W and Pa$\alpha$ data in training. Right: Confusion matrices for models which exclude F150W and Pa$\alpha$ data in training.}
    \label{fig:C2CMs}
\end{figure*}

We assess model performance by labeling objects in the testing sample (20\% of objects from each label) and comparing each object's model label to the true label. For a given model configuration, the mode identification of the 10 models becomes an object's `predicted label', embedded cluster candidate (1 or `positive'), or other PAH `bright' (0 or `negative'). This process is done for each of the eight model configurations (four filter sets with two CNN architectures; reference \S~\ref{sec:ML}), and we present the results as confusion matrices in Figure~\ref{fig:C2CMs}. A confusion matrix compares the model identifications to the known object labels and provides an overview of model performance. For a bimodal labeling system, the confusion matrix contains four quadrants: true negative, false positive, false negative, and true positive. Label~0 and Label~1 objects in our sample are the negative and positive groups, respectively. 

Figure~\ref{fig:C2CMs} (left) displays the confusion matrices for the four configurations that include F150W and Pa$\alpha$ data. Figure~\ref{fig:C2CMs} (right) shows the confusion matrices for the four configurations that exclude these data from training and testing. For all confusion matrices, we normalize each row to the total number of `true label' objects in the row, such that the diagonal values represent the accuracy for each class. The number of testing objects in each class are in Table~\ref{tab:MLobjects}, column (5). The bottom numbers show the average and standard deviation of the ``mode label agreement'' for objects in each quadrant. This metric describes the fraction of models that agree with an object's final predicted label, i.e., the fraction of the 10 models that agree with the mode classification label.

Model accuracies for Label~1 objects reflect expectations for human-to-human variations ($\sim$70\%; \citealt{whitmore2021}), and Label~0 objects reach up to 95\% accuracy. A possible explanation for the difference in accuracy for each class is an imbalance in the training set.  \cite{dablain2022} show that training with an imbalanced sample can produce model inability to generalize classification of the minority class.  Namely, the model can perform well on classifying the training sample of the minority class, but cannot extend that training as effectively to unseen data. A further reflection of this `feature memorization' manifests in the similar features between the false positive classifications in comparison to the true positives \citep{dablain2022}. An additional consideration is that no subclassification of the Label~0 `PAH-bright' catalog is done prior to training and testing. As such, false positive cases could reflect embedded or partially embedded cluster candidates that we miss in our `by-eye' catalog. We explore this further in Section~\ref{sec:ModelTrends}.

Broadly, across all eight model configurations, \texttt{ResNet18} models produce a lower accuracy for Label~1 objects in comparison to \texttt{VGG19-bn} models. However, using accuracy as a metric of performance for two-label systems, especially for samples with label imbalance, is often not the primary metric for assessing model performance \citep{Carvalho2025, Baron2019}. As an additional metric, we compute the $F1$ score for each model as
\begin{equation}
    F1=\frac{2(\texttt{precision}*\texttt{recall})}{(\texttt{precision}+\texttt{recall})}.
\end{equation}
\texttt{Precision} is the ratio of correct Label~1 identifications to the total objects with positive model classifications, while \texttt{recall}, or accuracy, is the fraction of Label~1 objects that the model correctly labels (lower right quadrant in all confusion matrices in Figure \ref{fig:C2CMs}).  Qualitatively, \texttt{precision} describes the quality of the model identifications for a given class, or how well it mitigates spurious positive classifications.  Accuracy, or \texttt{recall}, simply measures the model's ability to correctly label objects from each class. The $F1$ score describes the balance between these two metrics, highlighting the separability between classes, and is often a better measure of performance for imbalanced datasets \citep{Carvalho2025}.

For the confusion matrices that include Pa$\alpha$ and F150W data (Figure~\ref{fig:C2CMs}, left), we find that \texttt{ResNet18} produces higher $F1$ scores than \texttt{VGG19-bn} for both object labels when comparing models with the same data inclusion. For models excluding the F770W data, \texttt{ResNet18} produces $F1$ scores of 0.97 for Label~0 objects and 0.5 for Label~1 objects, whereas \texttt{VGG19-bn} has an $F1$ score of 0.93 and 0.35 for each of these labels, respectively. When we include the F770W data, the $F1$ score for Label~0 objects remained the same for \texttt{ResNet18} models (0.97) and marginally increased for \texttt{VGG19-bn} models (0.94). For Label~1 objects, the \texttt{ResNet18} model performance slightly decreased, from 0.5 to 0.48, while the \texttt{VGG19-bn} model performance slightly increased, from 0.35 to 0.39. In all cases, these models produce $F1$ scores at, or lower than, 0.5 (50\%) for Label~1 objects, meaning poor model performance for this group of objects. Similar trends in $F1$ score are produced by the models that exclude the Pa$\alpha$ and F150W data (Figure~\ref{fig:C2CMs}, right). The average $F1$ score for Label~1 objects is 0.36 and 0.94 for Label~0 objects, with \texttt{ResNet18} performing better than \texttt{VGG19-bn} models for both object labels. We address possible reasons for poorer performance and potential methods of improvement for Label~1 objects later in Section~\ref{sec:MLdiscussion}.

\subsubsection{Trends with Model Identification}
\label{sec:ModelTrends}
To gain insight as to how data inclusion and CNN architecture impact model classification, we examine objects with a consistent positive model classification across each subset of 4 model configurations. We conduct aperture photometry for all training and testing MEFs. To emulate the input data the model uses for classification, the central $50\times50$ pixels for each filter are kept and resized to $299\times299$ (as in \S~\ref{sec:ML}). Since the MEF centers match our source centers, we use the central $x-y$ pixel of the MEF for the aperture center and use a radius of 12 pixels. During training and testing, MEF data are kept in the native units, $\rm{electrons\ s^{-1}}$ for HST broadband data, $\rm{MJy\ sr^{-1}}$ for JWST medium- and wide-band data, $\rm{erg\ s^{-1}\ cm^{-2}}$ for nebular data. For ease of comparison here, broadband photometry is given in units of Jansky and nebular data in $\rm{erg\ s^{-1}\ cm^{-2}}$.

\begin{figure}[!ht]
    \centering
    \includegraphics[width=\columnwidth]{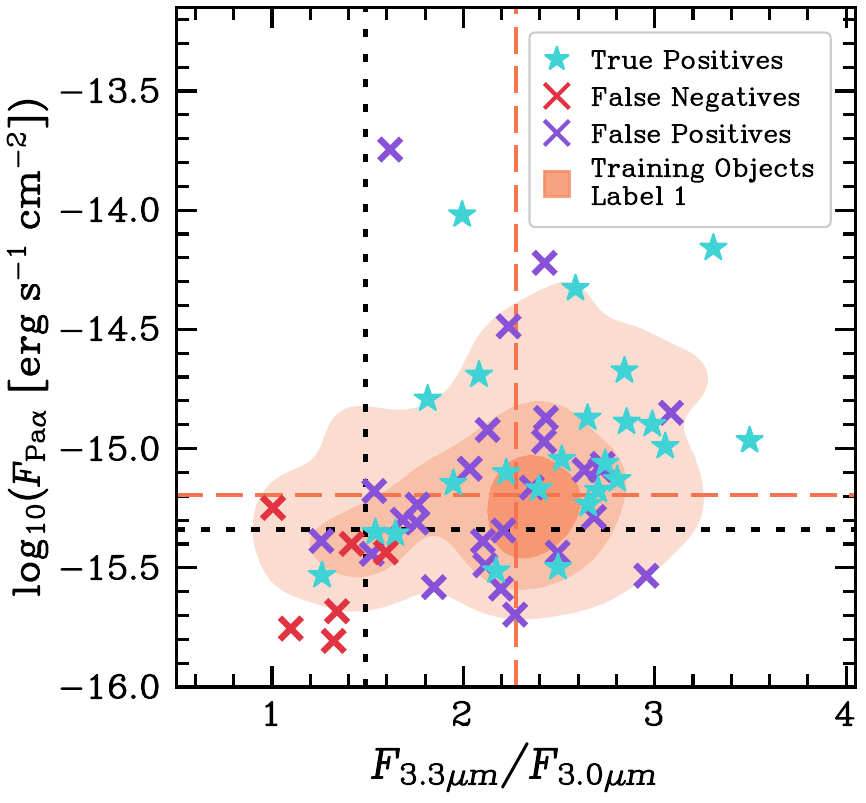}
    \caption{Pa$\alpha$ flux versus F335M/F300M flux ratio for Label~1 objects in Figure~\ref{fig:C2CMs} (left). The dashed lines indicate the median along each axis for the training samples: Label~0 (black) and Label~1 (orange).}
    \label{fig:PaavPAH}
\end{figure}

For models that include Pa$\alpha$ and F150W data, there are 805 true negatives, 27 false positives, 6 false negatives, and 24 true positives consistent across all four model configurations. Figure \ref{fig:PaavPAH} compares the training sample of embedded cluster candidates to objects with a  predicted classification of 1 or true label of 1. The purple and red Xs are the false positive and false negative cases, respectively. The cyan stars show embedded cluster candidate objects that receive a correct model label of 1. The orange contours show the 16th , 50th , and 84th percentiles of the training Label~1 objects, with the vertical and horizontal orange dashed lines denoting the median of the sample along each axis ($F_{\rm{3.3~\mu m}}/F_{\rm{3.0~\mu m}}=2.27$ and $F_{\rm{Pa}\alpha}=6.38\times10^{-16}\ \rm{erg\ s^{-1}\ cm^{-2}}$). For comparison, the dotted black lines show the median Pa$\alpha$ flux and flux ratio of F335M to F300M of the training Label~0 objects ($4.57\times10^{-16}\ \rm{erg\ s^{-1}\ cm^{-2}}$ and 1.49).

As a result of how we build our binary label sample for training, there is feature overlap between the Label~1 and Label~0 training objects. Both feature overlap and imbalance in label groups have been shown to degrade model performance, more so for the minority class \citep{Carvalho2025, Santos2022}. However, investigation of this region allows for important insight into how the model delineates between label groups in regions of feature overlap. In imbalanced neural network training, a lack of diverse features in the minority class can lead to overreliance on fewer features for classifying the minority label in comparison to the majority label \citep{dablain2022}. Notably, this `minimal feature reliance' for Label~1 objects is evident in the color space of Figure \ref{fig:PaavPAH}. All 6 false negatives lie below one or both of the median values of the training Label~0 sample and both medians of the Label~1. In this regime, the false negatives have an average model agreement of 91.3\% and true negative objects $\sim$94\%. All Model Label~1 objects in this same region show lower model agreement than Model Label~0 objects, with 85\% for true positives and 75.9\% for false positives. Inversely, true positive objects above both training Label~0 medians and one or both training Label~1 medians have a slightly higher average model agreement of $\sim$90\%.

The difference in model certainty indicates that the model is heavily reliant on a narrow range of few features for classifying Label~1 objects and is unable to generalize to test Label~1 objects when these decision-making features are at the low end of the training sample. Conversely, the model is able to generalize to the testing sample of Label~0 objects and with more confidence since it uses a broader range of features for decision-making of this class. As a result, small fluctuations in the few key features used for Label~1 classification can lead to drastic shifts in the model's decision in object labeling. In regions of feature overlap between the two groups, this is concretely evident by the lower model certainty for model Label~1 objects in comparison to Label~0 objects. This is consistent with \cite{Hannon2023}, who found that color bias in their training sample of HST Class~1 and Class~2 stellar clusters impacted neural network distinction between these two classes of objects.

For the four model configurations without Pa$\alpha$ and F150W data (Figure~\ref{fig:C2CMs}; right), there are 798 true negative, 37 false positive, 11 false negative, and 24 true positive sources consistent across all four model configurations. Of these, 20 true positives (83.3\%), 18 false positives (48.6\%), and 5 false negatives (45.5\%) receive the same label as all four model configurations where we include Pa$\alpha$ and F150W data. The average model agreement for Model Label~1 objects increases in comparison to Figure \ref{fig:PaavPAH}-- $\sim 87\%$ for false positives and $\sim 88\%$ for true positives--indicating improvement to model generalization of Label~1 objects when we remove the Pa$\alpha$ and F150W data.

\begin{figure}[!ht]
    \centering
    \includegraphics[width=\columnwidth]{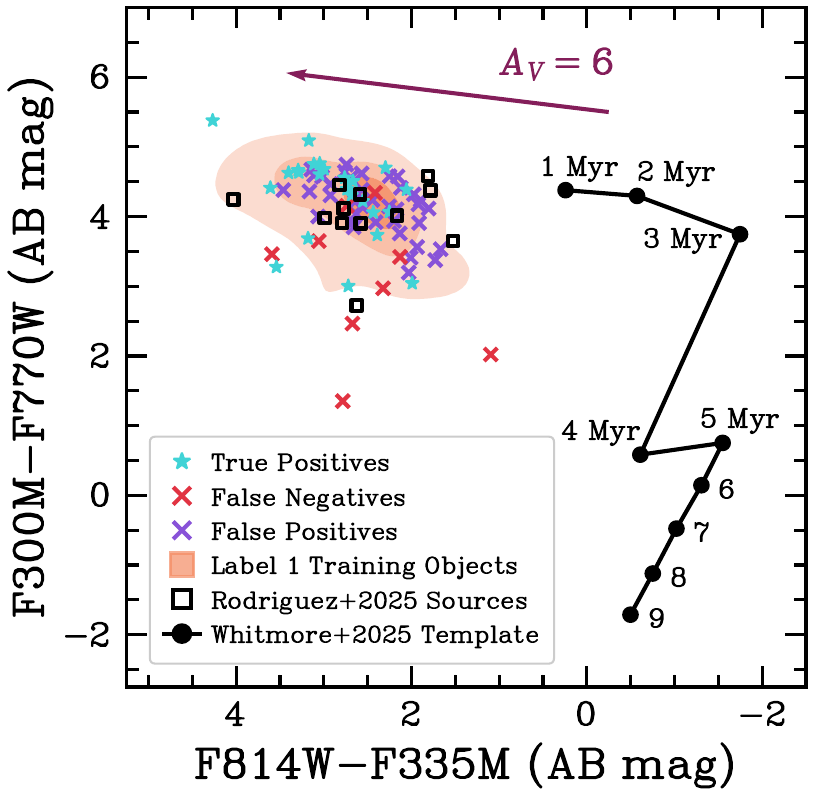}
    \caption{$\mathrm{F300M-F770W}$ versus $\mathrm{F814W-F335M}$ diagram for consistent Label~1 test objects for models without Pa$\alpha$ and F150W data. The black line is the mean empirical template from \cite{whitmore2025} for ages of 1--9 Myr (black dots), as this range spans just beyond the timescale we expect for embedded sources (\cite{rodriguez2024, whitmore2025, Ramambason2025}). The reddening vector (maroon) is for $A_V=6$, representative of the median we derive in Section \ref{sec:cluster_extinctions} for our by-eye sample.}
    \label{fig:Whitmore2025CC}
\end{figure}

These model configurations are made to emulate object classification using data available for the Cycle~1 sample. As such, we compare the results of testing to expectations of other stellar cluster populations in the Cycle~1 galaxy, NGC~628. Figure \ref{fig:Whitmore2025CC} is a color-color diagram of test objects with a consistent Label~1 model classification or True Label of 1 -- as in Figure \ref{fig:PaavPAH} -- across all four models without Pa$\alpha$ and F150W data. The orange contours represent the 16th, 50th, and 84th percentiles of the training Label~1 objects. We select axes to match those in Figure 12 of \cite{whitmore2025} and overlay the mean empirical template from  \cite{whitmore2025} showing the color evolution of clusters from 1--9 Myr (black dots). The maroon arrow indicates the reddening vector for the median line of sight extinction of our by-eye sample (see Section~\ref{sec:cluster_extinctions}). For comparison, we include 12 nearly embedded sources in NGC~628 from \cite{rodriguez2024} (open black squares) using photometry from Table 5 of \cite{whitmore2025}.
 
Overall, we find good agreement between the distribution of our sample of Model Label~1 objects in comparison to the nearly embedded clusters in NGC~628 from \cite{rodriguez2024}. Of the 12 sources, 2 lie within the 18th percentile of our Label~1 training sample and 9 within the 84th percentile. This shows that while these models are subject to overfitting of the Label~1 object features, they may still be able to produce embedded clusters candidate catalogs consistent with observation and increase the number of these objects by up to $\sim40\%$, with closer examination of the 18 consistent false positive objects. Next steps to improve training and generalization of model classification are discussed in Section~\ref{sec:MLdiscussion}

%% file: analysis_summary.tex
\begin{deluxetable}{lccccccc}[!ht]
\tablecaption{Embedded cluster candidates}
\label{tab:phot_summary}
\setlength{\tabcolsep}{2.2pt}
\tablehead{
\colhead{Galaxy} & \colhead{$N_{\mathrm{ECC}}$} & \colhead{X pc} &
\colhead{$N_{\mathrm{F150W}}$} &
\colhead{$\left<CI\right>$}&
\colhead{$\left<{Age}\right>$} &
\colhead{$\left<M_*\right>$}&
\colhead{$\left<A_V\right>$}\\
\colhead{} & \colhead{} & \colhead{$\mathrm{pix^{-1}}$} & \colhead{} & \colhead{F150W} &
\colhead{Myr} &\colhead{$10^3 M_\odot$} &\colhead{mag} 
}
\startdata
NGC~1097$\dagger$ & 13 & 2.04& 12 & 1.48& 4.61 & 2.71 & 6.85\\ %
NGC~1559$\dagger$ & 57 & 2.90& 33 & 1.50& 4.48 & 1.32 & 5.52\\ %
NGC~2775$\dagger$ &  2 & 3.45&  2 & 1.72& 5.17 & 1.07 & 7.99\\ %
NGC~2997          & 32 & 2.10& 27 & 1.49& 4.52 & 1.84 & 7.36\\ %
NGC~3059          & 73 & 3.02& 55 & 1.52& 4.83 & 2.01 & 6.00\\ %
NGC~4298$\dagger$ & 35 & 2.22& 22 & 1.52& 4.77 & 0.88 & 6.26\\ %
NGC~4424          & 11 & 2.41& 10 & 1.49& 5.38 & 2.80 & 7.38\\ %
NGC~4571$\dagger$ &  4 & 2.22&  3 & 1.55& 5.03 & 1.63 & 6.16\\ %
NGC~4694          &  2 & 2.35&  2 & 2.04& 5.72 & 1.91 & 5.38\\ %
NGC~4951          &  5 & 2.24&  4 & 1.52& 5.04 & 1.43 & 5.50\\ %
NGC~5248$\dagger$ & 58 & 2.20& 42 & 1.56& 4.71 & 1.12 & 5.82\\ 
\hline
Total    &292 &     & 212& &     &      &\\
\enddata

\tablenotetext{}{The number of human-identified embedded cluster candidates is  post quality assurance. There are 212 embedded cluster candidates with matching F150W. A fraction 186/212 of these matching sources have peak F150W $S/N> 5$. Mass averages are determined using sources with a peak F150W $S/N> 5$. $\dagger$ indicates galaxies with HST Class~1+2 cluster catalogs available from \citet{maschmann2024}.}
\end{deluxetable}

%% file: 06_discussion.tex
\subsection{Variations in the Number of Sources}\label{sec:Var_Num}
The final number of by-eye detections varied significantly between galaxies, ranging from two (NGC~4694 and NGC~2775) to 73 (NGC~3059). NGC~4694 is a lenticular galaxy and structurally contains a central galactic bulge and a disk with no distinct spiral arms. The locations of the two candidates from this galaxy are consistent with star-forming or composite regions as classified by \citet{williams2024}. Interestingly, a significant amount of our sample is classified as having flocculant spiral arms by \cite{stuber2023}, and exhibits a wide range in the number of sources identified. Two of these galaxies, NGC~2775 and NGC~4571, both have a low number of embedded cluster candidates, 2 and 4, respectively. Whereas NGC~3059, also classified as flocculant by \citet{stuber2023}, had the highest number of candidates identified (73). These differences likely arise from differences in local galaxy environment, which can influence the total gas available for star formation \citep{maschmann2024}, as well as the presence or lack of other internal structures that enhance star formation, such as bars and nuclear rings, as mentioned in Section~\ref{sec:mass_funcs}.

The variation in the number of sources detected can be best explained by the span of global properties of the sample. Variations in inclination result in differing obscuration by the disk of the galaxy and can heavily impact source detection. However, we detected a similar number of sources in our lowest inclined galaxy (NGC~3059) as in one of our highest inclined galaxies (NGC~1559), suggesting inclination is not the primary reason for differences in detection rate. More importantly, we find that the final number of detections scales with global SFR and sSFR. Namely, the amount of new star formation scales with the global level of star formation activity, as was shown in Figure~\ref{fig:Nclus_props}.

\subsection{Derived Ages and Clearing Timescales} \label{sec:age_discussion}
One of our main results is the distribution of ages for the embedded stellar cluster candidates sample. Previous studies in other PHANGS galaxies show that the youngest stellar clusters emerge from their natal clouds in less than $\sim$5~Myr. \cite{henny2025} find that the youngest and most gas-rich clusters in the PHANGS Cycle~1 galaxies have a median age of 3~Myr. Using mid-infrared detected sources in the F1000W and F2100W bands, Hassani et al.\ (2025, in preparation) derive H$\alpha$ equivalent widths consistent with ages of $<5$~Myr in their sample of exposed clusters. They conclude that these young ages, in conjunction with the small percentage of embedded sources in their catalog ($\sim10\%$), imply a short embedded period during star-formation. Likewise, \citet{Ramambason2025} obtain a short transition ($\sim4$~Myr) between the embedded and exposed phase using spatial correlation between the CO, 21~\m\, and H$\alpha$ emission. Similarly, based on an analysis of the spatial correlations in the peaks of CO and H$\alpha$ emission, \citet{kim2023} find that massive stars are embedded within their dusty natal sites for an average of 5.1~Myr, and about half of this timescale, the stars are deeply embedded. Outside of `normal' star-forming galaxies, \citet{linden2024} show that in the luminous infrared galaxy NGC~3256, massive clusters ($M_\odot>10^5$) have rapid dust clearing time scales in the 3--4~Myr range. However, they note that as stellar cluster mass increases, presupernova feedback mechanisms may not completely clear the natal gas and rely on supernovae, resulting in longer emergence times \citep{linden2024, Stanway2023}. Moreover, \citet{McQuaid2024} find longer emergence timescales for their sample of young, Pa$\alpha$ and Pa$\beta$ selected clusters in the dwarf starburst galaxy NGC~4449. Reflective of our sample, they find longer emergence timescales by 1--2~Myr for infrared-selected clusters in comparison to UV/optical selected stellar clusters. This highlights the importance of using infrared-selected cluster catalogs to resolve differences in the process and duration of natal cloud clearing in comparison to UV/optical clusters (see Section~\ref{sec:ComparisonCatalogs} for further discussion).

Our ages are based on comparing measured Pa$\alpha$ equivalent widths to the output of population synthesis models. However, there are several caveats to age dating via the Pa$\alpha$ equivalent width, specifically in relation to the choice of stellar population model. The two primary model assumptions of Starburst99 that impact the resulting model Pa$\alpha$ equivalent widths are the escape fraction and the initial mass function. In Figure \ref{fig:EWage}, we show that the escape fraction assumed has a moderate impact on the age for a given equivalent width. In particular, for a constant equivalent width, we expect to derive younger ages for larger values of $f_{\rm esc}$, with this effect being more prominent at smaller equivalent widths. For example, at an equivalent width of $\sim30$~\AA, the model age for an $f_{\rm esc}$ of 0.5 is approximately 2.2~Myr younger than for an $f_{\rm esc}$ of 0, while at an equivalent width of $\sim1000$~\AA~ this difference reduces to 0.9~Myr. Similarly, in a comparison of seven simple stellar population models, \citet{wofford2016} show that, though the solar metallicity models are in agreement over their hydrogen-ionizing rates at 1~Myr, they can differ by a factor of 3 at ages near 3~Myr. Whether or not stellar rotation is included in the stellar population models will impact the equivalent width (e.g., \cite{maeder2000}).

Assumptions in the IMF, such as the upper-mass limit, also affect the equivalent widths of hydrogen recombination lines. For example, employing an upper-mass cutoff of 100~$\rm M_\odot$ instead of 30~$\rm M_\odot$ will increase the inferred equivalent width by a factor of $\sim$3 \citep{lee2009}. In relation to the IMF, Starburst99 and many other single stellar population synthesis models assume a fully populated IMF. However, all of our source mass estimates produce masses below 10$^4$~M$_\odot$ and are subject to effects of stochastic sampling. \citet{Stanway2023} use Binary Population and Spectral Synthesis (\texttt{BPASS}) to model the effects of stochasticity on cluster properties for cluster masses between 10$^2$ and 10$^7$~M$_\odot$. They find a difference of 34~M$_\odot$ between the average maximum stellar mass produced for a 10$^3$~M$_\odot$ cluster in comparison to a 10$^5$~M$_\odot$ cluster. Moreover, their models show that ionizing-photon output for clusters of 10$^4$~M$_\odot$ can vary by a factor of 10$^6$ at ages less than 10~Myr \citep{Stanway2023}. For equivalent-width derived ages, this means the inferred ages could be artificially older (i.e lower equivalent width) due to a dearth of massive stars in lower-mass clusters rather than the stellar population being evolved.

\subsection{Presence of Two Compact Populations} \label{sec:CI_populations}
In Section~\ref{sec:F150wCI}, we find two compact (CI~$<$~1.4) sub-populations consistent with different age bins in our embedded cluster candidate sample. The first is a near-infrared bright ($\mathrm{log_{10}}(F_{1.5\mu\rm{m}}[\mu \rm Jy])>0$) subpopulation of 44 sources, consistent with ages greater than 5~Myr. The second population is a fainter near-infrared population ($\mathrm{log_{10}}(F_{1.5\mu\rm{m}}[\mu \rm Jy])<-0.5$) of 22 sources with ages between 3 and 5~Myr. Based on the low masses of our sources, a possible explanation is that these sources represent massive main-sequence stars that dominate the emission of their birth cluster and thus the near infrared light ($\lambda\sim1~\mu$m) of the stellar cluster. This is possible either through direct radiation, evidence of the presence of evolved massive stars, or through dust absorption and remission, indicative of a still young, embedded population. Based on the presence of Pa$\alpha$ emission, we infer that the latter scenario is the more likely of the two. In support of this, the embedded source mass distribution (Figure~\ref{fig:MChist}) and stochastic sampling models from \cite{Stanway2023} lead us to infer that most of our sources will not produce many stars with masses exceeding 20~M$_\odot$. Stars less massive than 20~M$_\odot$ will still be on the main-sequence after 5 Myr, and are luminous enough ($\sim10^4$~L$_\odot$) to easily dominate the radiation profile of a small cluster $<10^4 \rm~M_\odot$ (see \citealt{Eldridge2022} and references therein for a full review of modern stellar luminosities and timescales).

Another variable to consider is that the two subpopulations exhibit significantly different $F_{3.3\mu \rm m}$/$F_{3.0\mu \rm m}$ ratios. The brighter and older population has a median ratio of 1.03, while the fainter, younger population has a median ratio of 3.56. \cite{hassani2023} and Hassani et al. (2025, in preparation) implement color cuts to their compact 10$\mu$m and 21$\mu$m sources to classify stellar sources in their sample, primarily using PAH emission as a key indicator of embedded sources in comparison to evolved stars. We direct the reader to these works for an in-depth discussion of candidate star selection at near- and mid-infrared wavelengths. The low ratio of the bright and older population is consistent with the $\mathrm{F300M-F335M}$ color of candidate AGB stars in \cite{rodriguez2024}, but has Pa$\alpha$ equivalent widths spanning 21--245~\AA, inconsistent with what we expect for evolved stars. It could be that these sources are still extremely gas-rich but have pushed away their natal dust \citep{henny2025}. Since we are working with unresolved stellar populations and in a heavily stochastic regime for IMF sampling, we cannot draw robust conclusions regarding the exact stellar composition of the compact embedded sources.

\subsection{Comparison to Other Catalogs}\label{sec:ComparisonCatalogs}
\begin{figure}[ht!]
    \centering
    \includegraphics[width=\columnwidth]{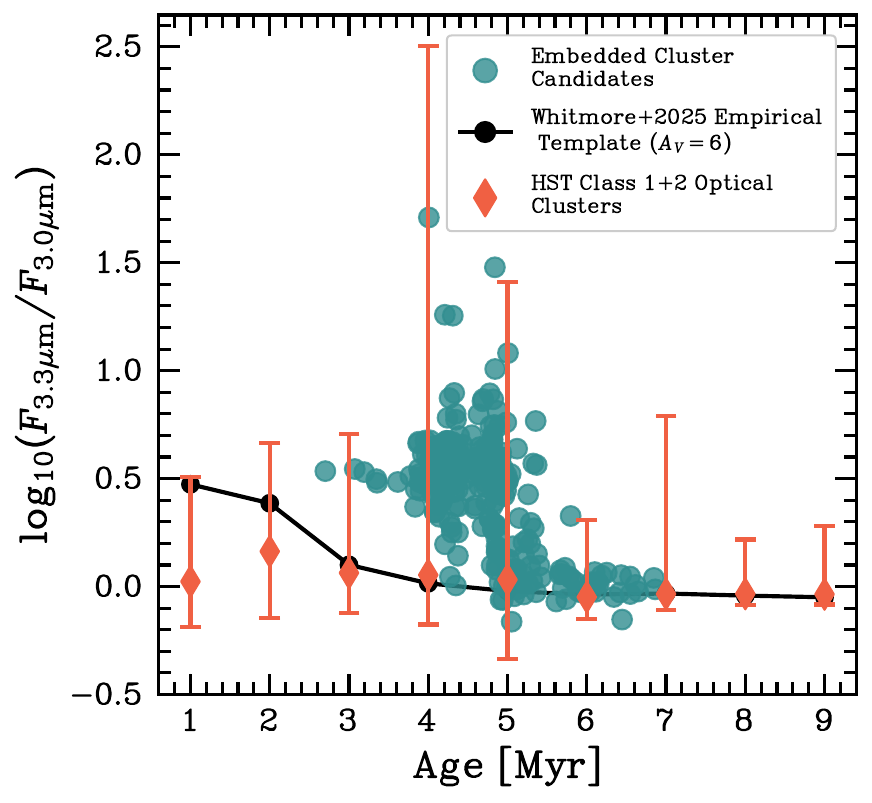}
    \caption{3.3\m/3.0\m \ flux ratio versus stellar population age for our embedded cluster candidate sample (cyan circles) and the HST Class 1+2 optical clusters from \cite{maschmann2024} for five galaxies in our sample (orange diamonds). The vertical orange lines denote the 10$^{\rm th}$ to 90$^{\rm th}$ percentile range for each age bin and the diamond marks the 50$^{\rm th}$ percentile for the for the Class 1+2 clusters. Optical cluster ages are from \cite{thilker2024} and embedded cluster candidate ages are from Section~\ref{sec:pa-alpha_ew_age_dating} of this work. We compare these values to the empirical template from \cite{whitmore2025}, reddened for an $A_V=6$ (black track).}
    \label{fig:catalog_comp}
\end{figure}

Assessing our embedded cluster candidates in the context of similar works is necessary for understanding of the duration of 3.3~\m\ PAH emission and natal cloud clearing timescales. Clearing timescales inferred from optical cluster catalogs are often found to be shorter than those inferred from infrared source catalogs. In Figure~\ref{fig:catalog_comp}, the 3.3~\m/3.~0\m\ ratio versus age is shown for our embedded cluster candidates (cyan circles) compared to the Class 1+2 optical clusters from \cite{maschmann2024} (orange diamonds). The embedded cluster candidate ages are those derived in Section~\ref{sec:pa-alpha_ew_age_dating} and the optical cluster ages are from \cite{thilker2024}. The orange vertical lines indicate the 10th to 90th percentile range for each age and the orange diamond marks the 50th percentile for the Class 1+2 optical clusters. We overlay these values on the empirical template from \cite{whitmore2025}. They derive these templates by combining empirical SEDs and H$\alpha$ morphologies for a subsample of 40 HST Class 1+2 clusters in NGC~628. To limit stochastic effects, the template is built off of sources with SED mass estimates over 3000~$\rm M_\odot$ \citep{whitmore2025}. The black track in Figure~\ref{fig:catalog_comp} is the template from \cite{whitmore2025} reddened for the median $A_V$ of our sample ($A_V=6$). The 3.3~\m\ extinction ($A_{3.3\mu \rm{m}}$) and 3.0~\m\ extinction ($A_{3.0\mu \rm{m}}$) values are determined via
\begin{equation}
    A_\lambda\ [\mathrm{mag}]=\left(\frac{A_\lambda}{A_V}\right)\cdot6\ [\mathrm{mag}]
\end{equation}
where $A_{3.3\mu \rm{m}}/A_V$ and $A_{3.0\mu\rm{m}}/A_V$ are from the Milky Way ($R_V=3.1$) dust models of \citet{draine2003}.
The Figure shows that the embedded cluster candidates exhibit a steep drop in 3.3\m\ PAH emission at $\sim5$~Myr, similar to timeline predictions from \cite{whitmore2025} and \cite{rodriguez2023}. \cite{whitmore2025} suggest that the lack of PAH emission associated with young stellar clusters $\sim$4~Myr after their birth is likely due to a combination of the destruction of the smallest dust grains and the removal of dust grains near the vicinity of clusters due to presupernova feedback, such as stellar winds and radiation pressure. Another finding of interest is that for both catalogs, there is large scatter at this ``turnoff" age of 4--5~Myr. For the embedded cluster candidates, we interpret this as being an effect of stochasticity, as the presence and evolution of the most massive stars will strongly influence the clearing timescale of the natal dust (as discussed in Section~\ref{sec:age_discussion}). This could also provide an explanation for the lower than expected 3.3\m/3.0\m\ ratio for the Class 1+2 clusters between 1 and 2~Myr. These clusters have an average mass of $3.2\times10^4\ \rm M_\odot$, meaning they may be more effective at rapidly clearing the natal cloud in comparison to the lower-mass embedded clusters, as also found in \cite{McQuaid2024}. Similarly, \cite{rodriguez2024} find that their 3.3\m\ emitters not detected in any HST filters, including H$\alpha$, have a larger $\mathrm{F300M-F335M}$ color than HST Class 1+2 optical clusters from \cite{maschmann2024} with ages between 1 and 3~Myr (Figure~18 of \cite{rodriguez2024}). From Figure~\ref{fig:Whitmore2025CC}, we also see that our embedded cluster candidates have ratios consistent with ages of 1--2~Myr using the optically derived template from \cite{whitmore2025}. 
 
\subsection{Imbalanced Training in Machine Learning}\label{sec:MLdiscussion}
\citet{Santos2022} provide a comprehensive review of how imbalanced training and feature overlap in training degrade model performance, especially when the two are jointly acting. This impact was evident in our results from CNN model testing in Section~\ref{sec:C2_machine_learning}. These properties of our catalog led to overfitting of the training data and subsequently, poor translation to the testing data \citep{Carvalho2025}. These conclusions are apparent by the large difference in model accuracy for the minority class (embedded cluster candidates; Label~1 objects) in comparison to the majority class (other 3.3~$\mu$m PAH-bright sources; Label~0 objects) and further evident by the nearly identical features of false positives and true positives in Figure~\ref{fig:Whitmore2025CC}. \cite{Santos2022} describe that the feature overlap, when acting alone, provides a larger source of complexity for model decision-making than class imbalance. As such, a logical first step would be to alter how we choose Label~0 objects such that there is less feature overlap with distinguishing Label~1 features, such as the 3.3 PAH emission.

The ideal mode of improving model performance would be to expand and diversify, specifically, the Label~1 dataset. This could be done in two primary ways: incorporating embedded cluster catalogs from other nearby-galaxy studies \citep{rodriguez2023, rodriguez2024, hassani2023, pedrini2024} or producing artificial sources that match the selection criteria used to produce our by-eye sample. The benefit of incorporating catalogs from other samples is that models would still be trained using human-selected sources, similar to the current study. Moreover, enlarging the sample of galaxies would have the added benefit of larger statistics for dissecting systematics in model performance based on galaxy global properties, such as distance, as in \cite{Hannon2023}, inclination, and global SFR. As a secondary benefit, such an analysis would allow for a comparison of selection methods for embedded cluster candidates to see which criteria could be used to systematically produce catalogs of these objects moving forward.

As a secondary step to altering the training and testing samples, we could harness the available PHANGS-HST optical cluster catalogs to populate the Label~0 object catalog. This would have the benefit of reducing feature overlap immensely, since these catalogs have contrasting selection criteria to the embedded cluster candidates selected here, and there are ample objects to randomly sample from to populate a diverse Label~0 object catalog. We do not pursue this mode of training and testing, as not all galaxies in our sample have HST optical cluster catalogs available. With a larger sample, we could potentially subdivide our Label~1 and Label~0 objects into a multilabel system based on other features, such as 3.3~\m\ PAH morphology; however, model performance can decrease as the complexity of subclasses increases \citep{Ghosh2024}.

This method is limited, however, in that it is necessary for galaxies outside the current sample used here to have the same data available for training and testing models. As such, another option would be to incorporate artificial sources into the training and testing catalogs. The primary limitation of this method would be the loss of surrounding environment complexity. Namely, to retain the environmental diversity of observational data–-point source crowding, surrounding extended structures, varying background levels--the sources would need to be injected into the data prior to making cutouts for training and testing. Otherwise, we would be limited to isolated point-source objects, which do not reflect the same process of human selection of these embedded sources. These alternative pathways are left as an opportunity for future studies. To improve model performance using the same training and testing objects, other methods must be implemented.

One way to offset model overfitting due to imbalanced data is to implement data augmentation and selective sampling of each class \citep{Krizhevsky2012, Carvalho2025}. Data augmentation enlarges the training set and increases feature variability in model training. This helps reduce model overfitting, as the model is exposed to a larger variety of features to classify on a given label group. We implement two forms of data augmentation---rotation and axial flipping---but other forms include slight translations, RGB scaling adjustments, and inclusion of random noise. Selective sampling, such as undersampling the majority class or oversampling the minority class, is another form of data augmentation that can improve model performance for imbalanced training sets \citep{Carvalho2025}, as this assists in balancing the number of objects from each class that the model is exposed to in training.

Fine-tuning the CNN hyperparameters is also shown to significantly improve the robustness of models trained with imbalanced datasets \citep{Soekhoe2016, Pasupa2016}. The hyperparameters of a CNN can be broken down into three categories: the convolutional layers' hyperparameters, the fully connected layers' hyperparameters, and general hyperparameters \citep{Raiaan2024}. The convolutional layers' hyperparameters define the convolution operations and dictate feature extraction of the input data. Parameters include the number and size of kernels and the activation function used for feature recognition. The fully connected layers consist of hyperparameters that dictate the interpretation of the feature vector, extracted in the convolution and sampling layers, into a final object classification. The primary hyperparameter that has been shown to reduce over-fitting in small sample training is dropout \citep{Alzubaidi2021}. Dropout is a form of normalization in the fully connected layers, where a percentage of neurons and their connections are randomly deactivated. As a result, the model relies less heavily on certain neurons and their connections for decision-making. This broadens feature connections and can result in better generalization to unseen data \citep{Srivastava2014}. Lastly, we can adjust general hyperparameters such as the learning rate, the batch size and number of batches, and the optimizer and loss function used. One specific alteration that is shown to improve performance for imbalanced datasets is using cost-sensitive learning \citep{khan2017}, where the loss function is weighted according to the cost of errors for each class. In the case of our sample, rather than the loss being weighted equally for each label group, we would introduce a larger weight for the mislabeling of Label~1 objects. This, in turn, would make the model more sensitive to the correct classification of the underrepresented Label~1 group and put more weight on the accuracy of these objects rather than overall accuracy.

%% file: 07_conclusions.tex
\label{sec:Conclusions}
Using available JWST and HST data, we conduct a multiwavelength search for embedded stellar clusters in 11 targets drawn from the PHANGS sample of nearby galaxies. We catalog 292 embedded stellar cluster candidates via their prominent 3.3~$\mu$m PAH emission, high ratio of Pa$\alpha$/H$\alpha$, and total extinction at visible wavelengths. We find a positive correlation between the number of embedded cluster candidates and both the SFR and sSFR of galaxies in our sample. We derive physical properties for our by-eye sample, including ages, masses, extinctions, and concentration index values, using photometric data. A summary of the key findings is as follows:
\begin{itemize}
    \item Using Pa$\alpha$ equivalent-width age dating, our sources span an age range of 2.7--6.9 Myr, with a majority younger than 5~Myr. These ages are consistent with the SED-derived ages and emergence timescales for embedded stellar clusters from similar works \citep{rodriguez2024, linden2024, hassani2023, pedrini2024, knutas2025}. However, underlying model assumptions and stochastic sampling effects provide substantial sources of uncertainty in our sample ages.
    \item  We derive embedded source masses between  $12\ \rm M_\odot$--$9.89\times10^3\rm M_\odot$ via the F150W mass-to-light ratio. Only two sources have a mass estimate less than $10^2\ \rm M_\odot$. Similar to \cite{pedrini2024} and \cite{knutas2025}, there is a dearth of massive embedded sources ($>10^4\rm \ M_\odot$) in our sample, but masses for our sample are subject to similar caveats as source ages due to stochastic sampling effects.
    \item Our embedded sources exhibit a significant drop in 3.3~\m\ PAH emission after 5~Myr, consistent with 3.3~\m\ PAH emission timescales from the empirical templates of \cite{whitmore2025}. Our timescales are longer than those of \cite{rodriguez2024}, who derive a shorter average timescale of $\sim3$~Myr for their sample of 3.3~\m\ emitters in the PHANGS Cycle~1 galaxies. Discrepancies in clearing timescales may be a product of stochastic sampling effects present within our sample or due to probing different phases of cluster emergence altogether.
\end{itemize}

We also explore the capability of CNNs as a pathway for future classification of these objects in nearby galaxies. We train two deep CNNS--\texttt{ResNet18} and \texttt{VGG19-bn}--on four data configurations  that reflect the data available in the PHANGS-JWST Cycle~1 and Cycle~2 programs. Imbalance and feature overlap between our two-label groups in model training led to poorer model performance for classifying the embedded cluster candidates (Label~1 objects) in comparison to Label~0 objects for all eight model configurations. However, we do find that most false positive objects have features consistent with the Label~1 objects, meaning they could represent objects missed in our `by-eye' identification of sources. We determine that fine-tuning of model hyperparameters and precleaning our training catalog would greatly improve the model performance and potentially lead to robust enough models to classify these sources in other PHANGS galaxies.

%% file: 08_acknowledge.tex
\section{acknowledgments}
We thank the excellent support provided by the staff at the University of Wyoming's Advanced Research Computing Center.  
This work is based on observations made with the NASA/ESA/CSA James Webb and Hubble Space Telescopes. The data were obtained from the Mikulski Archive for Space Telescopes at the Space Telescope Science Institute, which is operated by the Association of Universities for Research in Astronomy, Inc., under NASA contract NAS 5-03127 for JWST and NASA contract NAS 5-26555 for HST. The JWST observations are associated with Program~3707, and those from HST with Program~15454.  
This publication uses data generated via the Zooniverse.org platform, development of which is funded by generous support, including a Global Impact Award from Google, and by a grant from the Alfred P. Sloan Foundation.
D.D. acknowledges support from grant JWST-GO-02107.005-A
M.B. acknowledges support from the ANID BASAL project FB210003. This work was supported by the French government through the France 2030 investment plan managed by the National Research Agency (ANR), as part of the Initiative of Excellence of Université Côte d’Azur under reference number ANR-15-IDEX-01.
K.S. and H.K. acknowledge funding support from grants JWST-GO-02107.006-A and JWST-GO-03707.005-A.

\software{
\texttt{NumPy} \citep{NumPy2020},
\texttt{SciPy} \citep{SciPy2020},
\texttt{Astropy} \citep{AstroPy2022},
\texttt{Matplotlib} \citep{Matplotlib2007},
\texttt{PyTorch} \citep{PyTorch2019},
\texttt{CIGALE} \citep{boquien2019}}

\section*{Data Availability Statement}
Data from the High-Level Science Project PHANGS-HST \citep{lee2022} were obtained from the Mikulski Archive for Space Telescopes (MAST) at the Space Telescope Science Institute \citep{PHANGSHST}. The DR4 stellar cluster catalog and empirical SED templates from the High-Level Science Project PHANGS-CAT \citep{lee2022, thilker2024, maschmann2024, whitmore2025} are also available through MAST at the Space Telescope Science Institute \citep{PHANGSCAT}.

%% file: 09_appendixA.tex
\begin{figure*}[h!]
    \centering
    \includegraphics[width=0.75\linewidth]{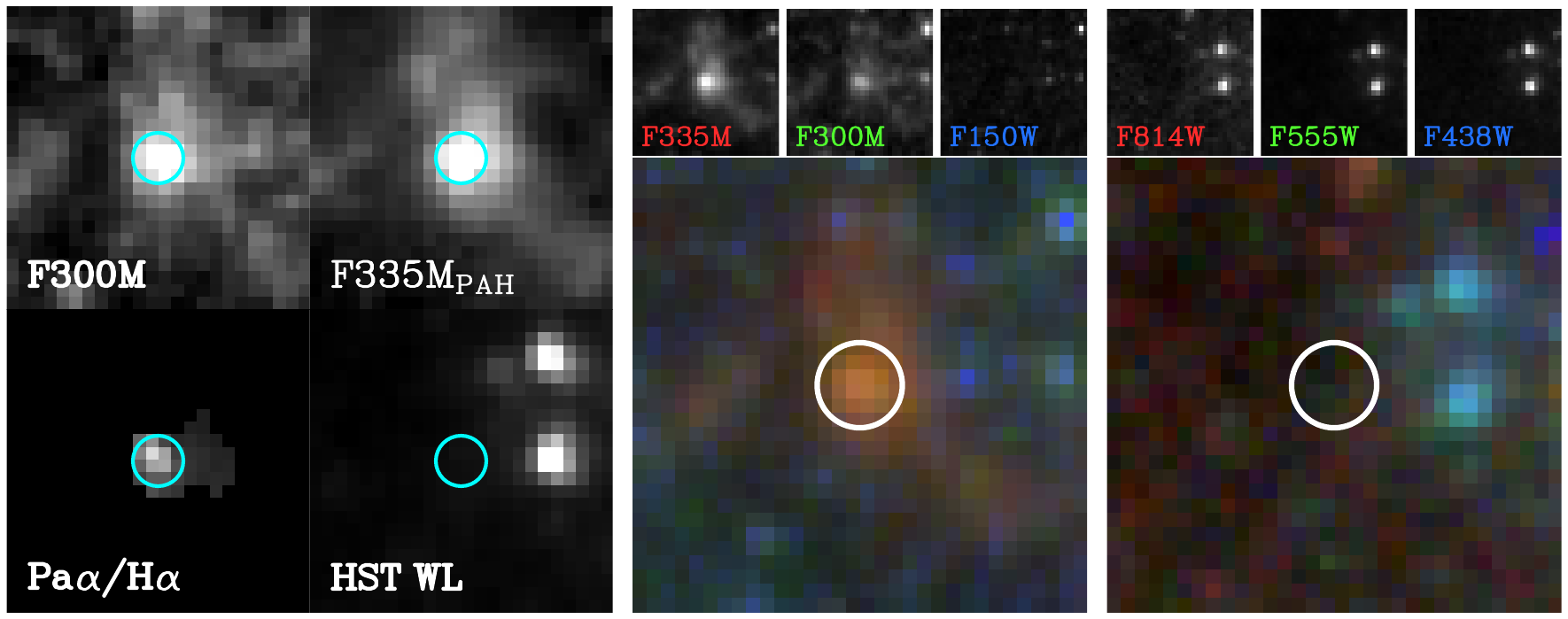}
    \includegraphics[width=0.75\linewidth]{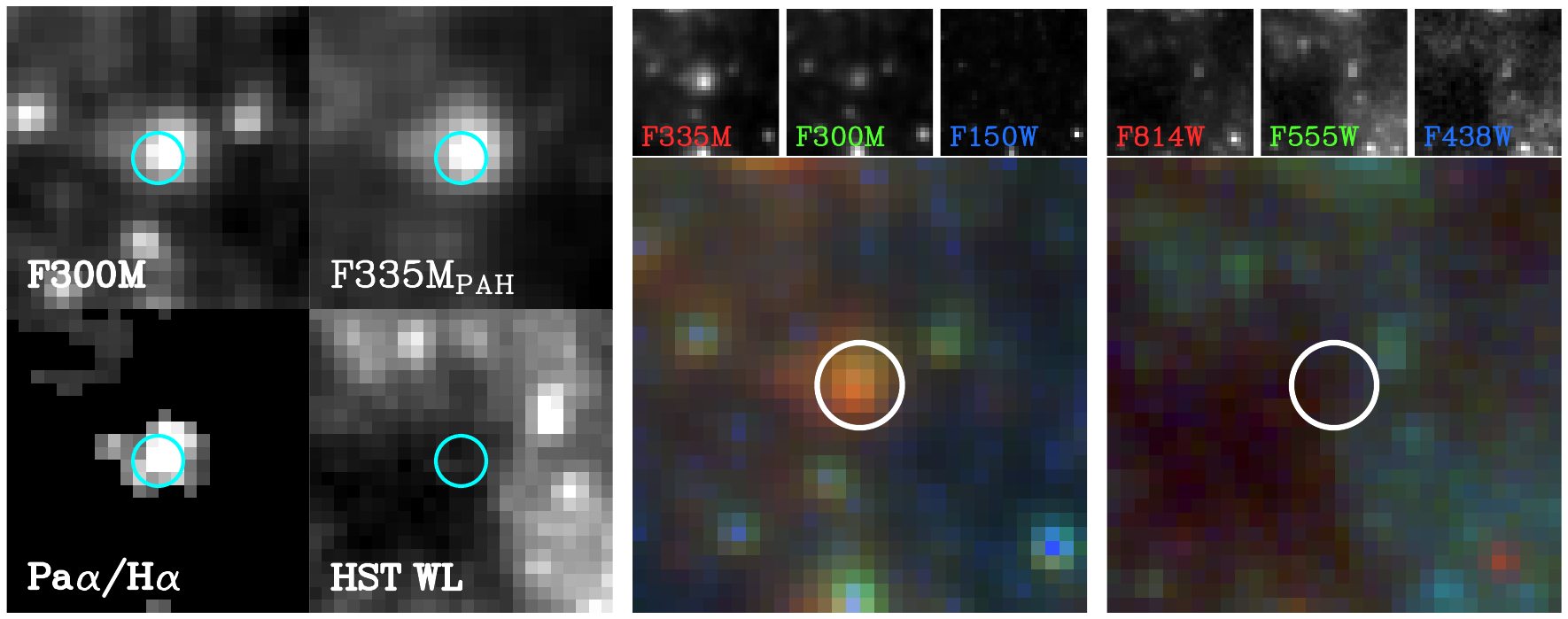}
    \includegraphics[width=0.75\linewidth]{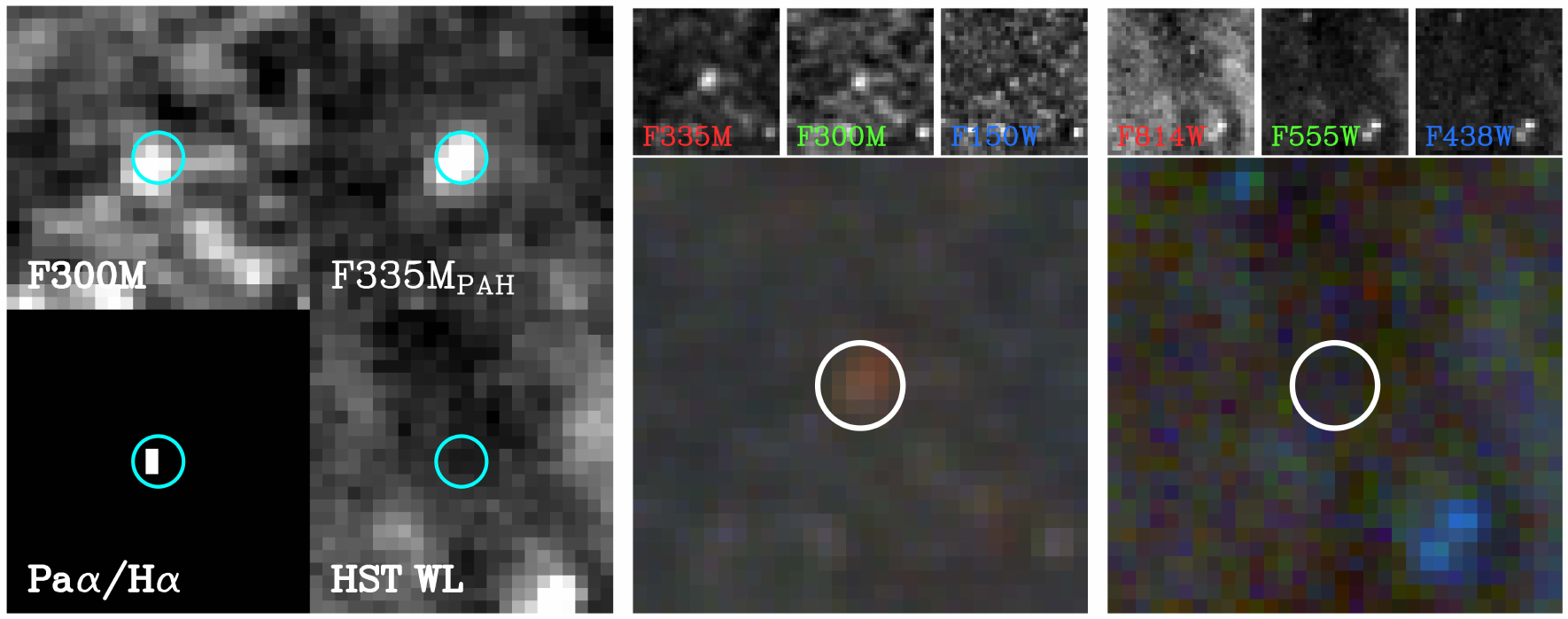}
    \includegraphics[width=0.75\linewidth]{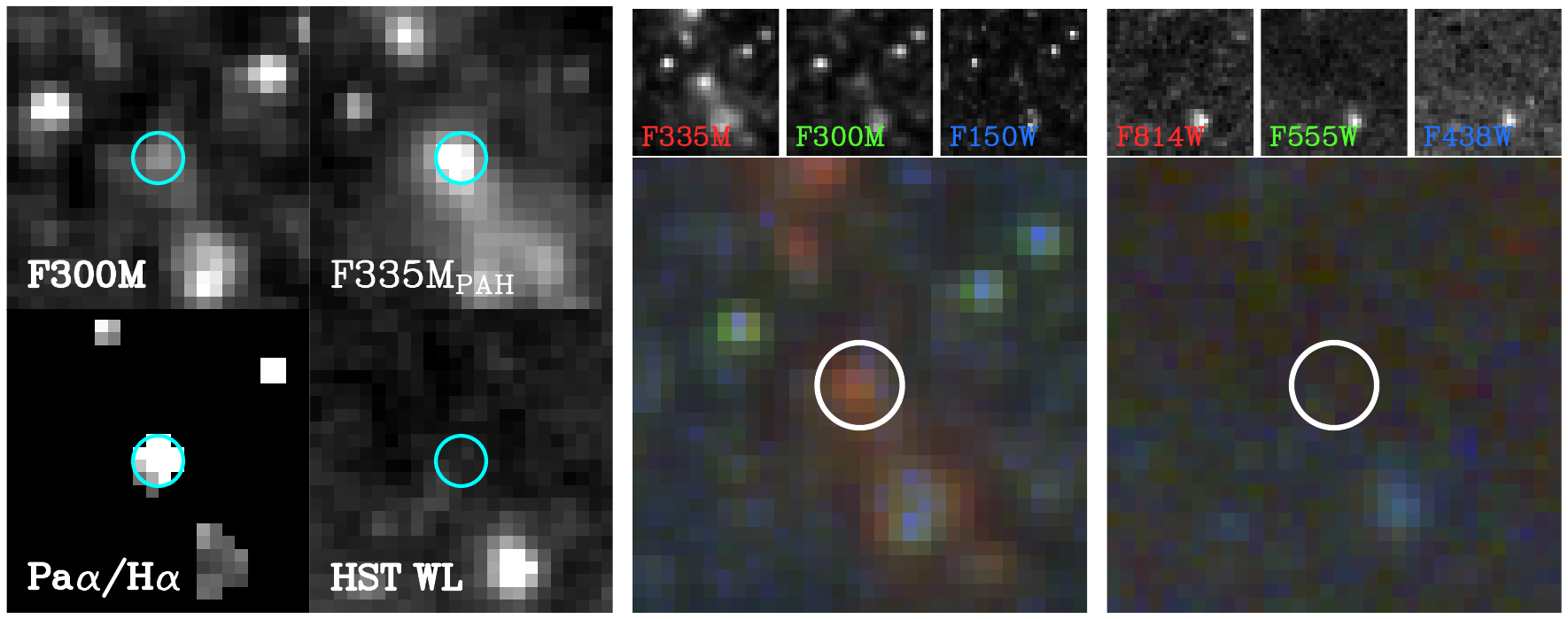}
    \caption{Example ideal cluster candidates as seen in the \textit{Zooniverse} interface (left), a three-color NIRCam image (center), and a three-color HST image (right). From top to bottom, we show one candidate from NGC~1097, NGC~1559, NGC~2775, and NGC~2997. The four cutouts in the left panel are 2\farcs0 cutouts centered on the source using the same images as Figure~\ref{fig:idCase}, with the cyan circle denoting the final F335M source center. The three-color images also display a 2\farcs0 cutout centered on the source. The individual bands are shown in the small boxes above the composite image.}
    \label{fig:Ideal_Sample}
\end{figure*}

Figure~\ref{fig:Ideal_Sample} displays an ideal `by-eye' candidate for each galaxy in our sample. For each object, the first panel shows how the source appears in each image used for identification with \textit{Zooniverse}, with the cyan circle showing the final F335M-centered source. The next two panels show three-color images using JWST NIRCam data (center panel) and HST broadband data (right panel), with the data for each color channel shown in the small panels.

%% file: 10_appendixB.tex
\begin{figure}
    \centering
    \includegraphics[width=0.5\linewidth]{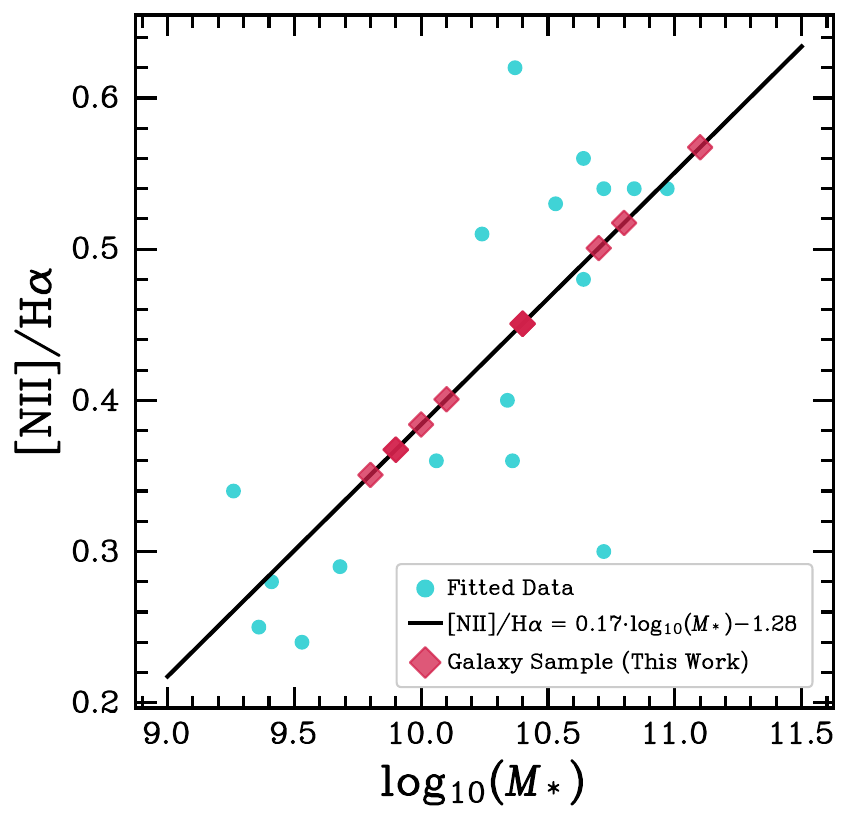}
    \caption{Scatter of \NII/H$\alpha$ vs. $\mathrm{log}_{10}(M_*)$. We select 17 galaxies at a distance of less than 11~Mpc which have $\mathrm{log}_{10}(M_*)$ values available from \cite{leroy2021}, as well as \NII/H$\alpha$ estimates from \cite{kennicutt2008} (cyan circles). The linear fit (black line) to these data are used to determine \NII/H$\alpha$ for the 11 galaxies studied in this work (magenta diamonds).}
    \label{fig:NII_Ha_fit}
\end{figure}

We select a sample of 17 galaxies from \cite{leroy2021} which have \NII/H$\alpha$ values available from \cite{kennicutt2008}. To estimate the contribution of \NII\ to the H$\alpha$ emission, we fit a line of the form
\begingroup
\setlength{\abovedisplayskip}{12pt}
\setlength{\belowdisplayskip}{12pt}
\begin{equation}
    \label{eqn:NII_Ha_fit}
    \mathrm{[NII]/H}\alpha=\gamma\cdot\mathrm{log}_{10}(M_*)+\beta
\end{equation}
\endgroup
where $\gamma$ and $\beta$ are the fit parameters. The stellar mass of each galaxy (log$_{10}(M_*)$) comes from \cite{leroy2021} and the \NII/H$\alpha$ values are from \cite{kennicutt2008}. This sample is limited to galaxies at distances less than 11~Mpc as a result of sample selection in \cite{kennicutt2008}. However, the log$_{10}(M_*)$ values used for fitting encompass the log$_{10}(M_*)$ extent for the sample of galaxies in this work. Figure~\ref{fig:NII_Ha_fit} shows \NII/H$\alpha$ versus $\mathrm{log}_{10}(M_*)$ for the 17 nearby galaxies we selected for fitting, as well as the values we derive for our sample using the fit. The fitted data are shown by cyan circles, and the fit is denoted by the solid black line. We derive a $\gamma$ of 0.17 and a $\beta$ of $-1.28$. Using these fit parameters in Equation~\ref{eqn:NII_Ha_fit}, we calculate [NII]/H$\alpha$ for each of the 11 galaxies in our sample (magenta diamonds) and present them in Table~\ref{tab:sample}.

%% file: main.bbl
\begin{thebibliography}{}
\expandafter\ifx\csname natexlab\endcsname\relax\def\natexlab#1{#1}\fi
\providecommand{\url}[1]{\href{#1}{#1}}

\bibitem[{{Adamo} {et~al.}(2017){Adamo}, {Ryon}, {Messa}, {Kim}, {Grasha}, {Cook}, {Calzetti}, {Lee}, {Whitmore}, {Elmegreen}, {Ubeda}, {Smith}, {Bright}, {Runnholm}, {Andrews}, {Fumagalli}, {Gouliermis}, {Kahre}, {Nair}, {Thilker}, {Walterbos}, {Wofford}, {Aloisi}, {Ashworth}, {Brown}, {Chandar}, {Christian}, {Cignoni}, {Clayton}, {Dale}, {de Mink}, {Dobbs}, {Elmegreen}, {Evans}, {Gallagher}, {Grebel}, {Herrero}, {Hunter}, {Johnson}, {Kennicutt}, {Krumholz}, {Lennon}, {Levay}, {Martin}, {Nota}, {{\"O}stlin}, {Pellerin}, {Prieto}, {Regan}, {Sabbi}, {Sacchi}, {Schaerer}, {Schiminovich}, {Shabani}, {Tosi}, {Van Dyk}, \& {Zackrisson}}]{adamo2017}
{Adamo}, A., {Ryon}, J.~E., {Messa}, M., {et~al.} 2017, \apj, 841, 131

\bibitem[{Adamo {et~al.}(2020)Adamo, Zeidler, Kruijssen, Chevance, Gieles, Calzetti, Charbonnel, Zinnecker, \& Krause}]{Adamo2020}
Adamo, A., Zeidler, P., Kruijssen, J. M.~D., {et~al.} 2020, Space Science Reviews, 216, doi:10.1007/s11214-020-00690-x

\bibitem[{Alzubaidi {et~al.}(2021)Alzubaidi, Zhang, Humaidi, Al-Dujaili, Duan, Al-Shamma, Santamar{\'\i}a, Fadhel, Al-Amidie, \& Farhan}]{Alzubaidi2021}
Alzubaidi, L., Zhang, J., Humaidi, A.~J., {et~al.} 2021, J. Big Data, 8, 53

\bibitem[{{Astropy Collaboration} {et~al.}(2022){Astropy Collaboration}, {Price-Whelan}, {Lim}, {Earl}, {Starkman}, {Bradley}, {Shupe}, {Patil}, {Corrales}, {Brasseur}, {N{\"o}the}, {Donath}, {Tollerud}, {Morris}, {Ginsburg}, {Vaher}, {Weaver}, {Tocknell}, {Jamieson}, {van Kerkwijk}, {Robitaille}, {Merry}, {Bachetti}, {G{\"u}nther}, {Aldcroft}, {Alvarado-Montes}, {Archibald}, {B{\'o}di}, {Bapat}, {Barentsen}, {Baz{\'a}n}, {Biswas}, {Boquien}, {Burke}, {Cara}, {Cara}, {Conroy}, {Conseil}, {Craig}, {Cross}, {Cruz}, {D'Eugenio}, {Dencheva}, {Devillepoix}, {Dietrich}, {Eigenbrot}, {Erben}, {Ferreira}, {Foreman-Mackey}, {Fox}, {Freij}, {Garg}, {Geda}, {Glattly}, {Gondhalekar}, {Gordon}, {Grant}, {Greenfield}, {Groener}, {Guest}, {Gurovich}, {Handberg}, {Hart}, {Hatfield-Dodds}, {Homeier}, {Hosseinzadeh}, {Jenness}, {Jones}, {Joseph}, {Kalmbach}, {Karamehmetoglu}, {Ka{\l}uszy{\'n}ski}, {Kelley}, {Kern}, {Kerzendorf}, {Koch}, {Kulumani}, {Lee}, {Ly}, {Ma}, {MacBride}, {Maljaars}, {Muna}, {Murphy}, {Norman},
  {O'Steen}, {Oman}, {Pacifici}, {Pascual}, {Pascual-Granado}, {Patil}, {Perren}, {Pickering}, {Rastogi}, {Roulston}, {Ryan}, {Rykoff}, {Sabater}, {Sakurikar}, {Salgado}, {Sanghi}, {Saunders}, {Savchenko}, {Schwardt}, {Seifert-Eckert}, {Shih}, {Jain}, {Shukla}, {Sick}, {Simpson}, {Singanamalla}, {Singer}, {Singhal}, {Sinha}, {Sip{\H{o}}cz}, {Spitler}, {Stansby}, {Streicher}, {{\v{S}}umak}, {Swinbank}, {Taranu}, {Tewary}, {Tremblay}, {de Val-Borro}, {Van Kooten}, {Vasovi{\'c}}, {Verma}, {de Miranda Cardoso}, {Williams}, {Wilson}, {Winkel}, {Wood-Vasey}, {Xue}, {Yoachim}, {Zhang}, {Zonca}, \& {Astropy Project Contributors}}]{AstroPy2022}
{Astropy Collaboration}, {Price-Whelan}, A.~M., {Lim}, P.~L., {et~al.} 2022, \apj, 935, 167

\bibitem[{{Barnes} {et~al.}(2025){Barnes}, {Chandar}, \& {Kreckel}}]{barnes2025}
{Barnes}, A., {Chandar}, R., \& {Kreckel}, K. 2025, \apj, submitted

\bibitem[{{Baron}(2019)}]{Baron2019}
{Baron}, D. 2019, arXiv e-prints, arXiv:1904.07248

\bibitem[{{Bastian} \& {Goodwin}(2006)}]{bastian2006}
{Bastian}, N., \& {Goodwin}, S.~P. 2006, \mnras, 369, L9

\bibitem[{{Boquien} {et~al.}(2019){Boquien}, {Burgarella}, {Roehlly}, {Buat}, {Ciesla}, {Corre}, {Inoue}, \& {Salas}}]{boquien2019}
{Boquien}, M., {Burgarella}, D., {Roehlly}, Y., {et~al.} 2019, \aap, 622, A103

\bibitem[{{Bruzual} \& {Charlot}(2003)}]{bruzual2003}
{Bruzual}, G., \& {Charlot}, S. 2003, \mnras, 344, 1000

\bibitem[{{Calzetti} {et~al.}(2025){Calzetti}, {Adamo}, \& {Sandstrom}}]{calzetti2025}
{Calzetti}, D., {Adamo}, A., \& {Sandstrom}, K. 2025, \apj, submitted

\bibitem[{Calzetti {et~al.}(1994)Calzetti, Kinney, \& Storchi-Bergmann}]{Calzetti1994}
Calzetti, D., Kinney, A.~L., \& Storchi-Bergmann, T. 1994, Astrophys. J., 429, 582

\bibitem[{{Calzetti} {et~al.}(2024){Calzetti}, {Adamo}, {Linden}, {Gregg}, {Krumholz}, {Bajaj}, {Bik}, {Cignoni}, {Correnti}, {Elmegreen}, {Faustino Vieira}, {Gallagher}, {Grasha}, {Gutermuth}, {Johnson}, {Messa}, {Melinder}, {{\"O}stlin}, {Pedrini}, {Sabbi}, {Smith}, \& {Tosi}}]{calzetti2024}
{Calzetti}, D., {Adamo}, A., {Linden}, S.~T., {et~al.} 2024, \apj, 971, 118

\bibitem[{Carvalho {et~al.}(2025)Carvalho, Pinho, \& Br{\'a}s}]{Carvalho2025}
Carvalho, M., Pinho, A.~J., \& Br{\'a}s, S. 2025, J. Big Data, 12

\bibitem[{{Chabrier}(2003)}]{chabrier2003}
{Chabrier}, G. 2003, \pasp, 115, 763

\bibitem[{{Chandar} {et~al.}(2025){Chandar}, {Barnes}, {Thilker}, {Caputo}, {Floyd}, {Leroy}, {{\'U}beda}, {Lee}, {Boquien}, {Maschmann}, {Belfiore}, {Kreckel}, {Glover}, {Klessen}, {Groves}, {Dale}, {Schinnerer}, {Emsellem}, {Rosolowsky}, {Bigiel}, {Blanc}, {Chevance}, {Congiu}, {Egorov}, {Faesi}, {Grasha}, {Hannon}, {Larson}, {Lopez}, {Mok}, {Neumann}, {Ostriker}, {Razza}, {S{\'a}nchez-Bl{\'a}zquez}, {Santoro}, {Schruba}, {Sun}, {Usero}, {Watkins}, {Whitmore}, \& {Williams}}]{chandar2025}
{Chandar}, R., {Barnes}, A.~T., {Thilker}, D.~A., {et~al.} 2025, \aj, 169, 150

\bibitem[{{Chown} {et~al.}(2025){Chown}, {Leroy}, {Sandstrom}, {Chastenet}, {Sutter}, {Koch}, {Koziol}, {Neumann}, {Sun}, {Williams}, {Baron}, {Anand}, {Barnes}, {Bazzi}, {Belfiore}, {Bigiel}, {Bolatto}, {Boquien}, {Cao}, {Chevance}, {Colombo}, {Dale}, {den Brok}, {Egorov}, {Eibensteiner}, {Emsellem}, {Hassani}, {Henshaw}, {He}, {Kim}, {Klessen}, {Kreckel}, {Larson}, {Lee}, {Meidt}, {Murphy}, {Oakes}, {Ostriker}, {Pan}, {Pathak}, {Rosolowsky}, {Sarbadhicary}, {Schinnerer}, {Teng}, {Thilker}, {Weinbeck}, \& {Watkins}}]{chown2025}
{Chown}, R., {Leroy}, A.~K., {Sandstrom}, K., {et~al.} 2025, \apj, 983, 64

\bibitem[{{Corbelli} {et~al.}(2017){Corbelli}, {Braine}, {Bandiera}, {Brouillet}, {Combes}, {Druard}, {Gratier}, {Mata}, {Schuster}, {Xilouris}, \& {Palla}}]{Corbelli2014}
{Corbelli}, E., {Braine}, J., {Bandiera}, R., {et~al.} 2017, \aap, 601, A146

\bibitem[{Dablain {et~al.}(2022)Dablain, Jacobson, Bellinger, Roberts, \& Chawla}]{dablain2022}
Dablain, D., Jacobson, K.~N., Bellinger, C., Roberts, M., \& Chawla, N. 2022, Understanding CNN Fragility When Learning With Imbalanced Data, , , arXiv:2210.09465

\bibitem[{Deng {et~al.}(2009)Deng, Dong, Socher, Li, Li, \& Fei-Fei}]{Deng2015}
Deng, J., Dong, W., Socher, R., {et~al.} 2009, in 2009 IEEE Conference on Computer Vision and Pattern Recognition, 248--255

\bibitem[{{Deshmukh} {et~al.}(2024){Deshmukh}, {Linden}, {Calzetti}, {Adamo}, {Messa}, {Grasha}, {Sabbi}, {Smith}, \& {Johnson}}]{deshmukh2024}
{Deshmukh}, S., {Linden}, S.~T., {Calzetti}, D., {et~al.} 2024, arXiv e-prints, arXiv:2409.08994

\bibitem[{{Draine}(2003)}]{draine2003}
{Draine}, B.~T. 2003, \araa, 41, 241

\bibitem[{{Draine} {et~al.}(2021){Draine}, {Li}, {Hensley}, {Hunt}, {Sandstrom}, \& {Smith}}]{draine2021}
{Draine}, B.~T., {Li}, A., {Hensley}, B.~S., {et~al.} 2021, \apj, 917, 3

\bibitem[{Eldridge \& Stanway(2022)}]{Eldridge2022}
Eldridge, J.~J., \& Stanway, E.~R. 2022, Annual Review of Astronomy and Astrophysics, 60, 455, first published as a Review in Advance on June 09, 2022

\bibitem[{{Emsellem} {et~al.}(2022){Emsellem}, {Schinnerer}, {Santoro}, {Belfiore}, {Pessa}, {McElroy}, {Blanc}, {Congiu}, {Groves}, {Ho}, {Kreckel}, {Razza}, {Sanchez-Blazquez}, {Egorov}, {Faesi}, {Klessen}, {Leroy}, {Meidt}, {Querejeta}, {Rosolowsky}, {Scheuermann}, {Anand}, {Barnes}, {Be{\v{s}}li{\'c}}, {Bigiel}, {Boquien}, {Cao}, {Chevance}, {Dale}, {Eibensteiner}, {Glover}, {Grasha}, {Henshaw}, {Hughes}, {Koch}, {Kruijssen}, {Lee}, {Liu}, {Pan}, {Pety}, {Saito}, {Sandstrom}, {Schruba}, {Sun}, {Thilker}, {Usero}, {Watkins}, \& {Williams}}]{emsellem2022}
{Emsellem}, E., {Schinnerer}, E., {Santoro}, F., {et~al.} 2022, \aap, 659, A191

\bibitem[{Fall {et~al.}(2009)Fall, Chandar, \& Whitmore}]{Fall2009}
Fall, S.~M., Chandar, R., \& Whitmore, B.~C. 2009, The Astrophysical Journal, 704, 453

\bibitem[{{Ferland} {et~al.}(2017){Ferland}, {Chatzikos}, {Guzm{\'a}n}, {Lykins}, {van Hoof}, {Williams}, {Abel}, {Badnell}, {Keenan}, {Porter}, \& {Stancil}}]{ferland2017}
{Ferland}, G.~J., {Chatzikos}, M., {Guzm{\'a}n}, F., {et~al.} 2017, \rmxaa, 53, 385

\bibitem[{{Frissell} {et~al.}(2024){Frissell}, {Chandler}, {Oldroyd}, {Trujillo}, {Sedaghat}, {Burris}, {Kueny}, {DeSpain}, {Farrell}, {Hsieh}, {Magbanua}, {Sheppard}, {Mazzucato}, {Bosch}, {Shaw-Diaz}, {Gonano}, {Lamperti}, {da Silva Campos}, {Goodwin}, {Terentev}, \& {Dukes}}]{Frissell2024}
{Frissell}, M.~K., {Chandler}, C.~O., {Oldroyd}, W.~J., {et~al.} 2024, Research Notes of the American Astronomical Society, 8, 225

\bibitem[{Ghosh {et~al.}(2024)Ghosh, Bellinger, Corizzo, Branco, Krawczyk, \& Japkowicz}]{Ghosh2024}
Ghosh, K., Bellinger, C., Corizzo, R., {et~al.} 2024, Mach. Learn., 113, 4845

\bibitem[{{Gregg} {et~al.}(2024){Gregg}, {Calzetti}, {Adamo}, {Bajaj}, {Ryon}, {Linden}, {Correnti}, {Cignoni}, {Messa}, {Sabbi}, {Gallagher}, {Grasha}, {Pedrini}, {Gutermuth}, {Melinder}, {Kotulla}, {P{\'e}rez}, {Krumholz}, {Bik}, {{\"O}stlin}, {Johnson}, {Bortolini}, {Smith}, {Tosi}, {Maji}, \& {Faustino Vieira}}]{Gregg2024}
{Gregg}, B., {Calzetti}, D., {Adamo}, A., {et~al.} 2024, \apj, 971, 115

\bibitem[{Hannon {et~al.}(2023)Hannon, Whitmore, Lee, Thilker, Deger, Huerta, Wei, Mobasher, Klessen, Boquien, Dale, Chevance, Grasha, Sanchez-Blazquez, Williams, Scheuermann, Groves, Kim, Kruijssen, \& the PHANGS-HST~Team}]{Hannon2023}
Hannon, S., Whitmore, B.~C., Lee, J.~C., {et~al.} 2023, Monthly Notices of the Royal Astronomical Society, 526, 2991

\bibitem[{Harris {et~al.}(2020)Harris, Millman, van~der Walt, Gommers, Virtanen, Cournapeau, Wieser, Taylor, Berg, Smith, Kern, Picus, Hoyer, van Kerkwijk, Brett, Haldane, del R{\'{i}}o, Wiebe, Peterson, G{\'{e}}rard-Marchant, Sheppard, Reddy, Weckesser, Abbasi, Gohlke, \& Oliphant}]{NumPy2020}
Harris, C.~R., Millman, K.~J., van~der Walt, S.~J., {et~al.} 2020, Nature, 585, 357

\bibitem[{{Hassani} {et~al.}(2023){Hassani}, {Rosolowsky}, {Leroy}, {Boquien}, {Lee}, {Barnes}, {Belfiore}, {Bigiel}, {Cao}, {Chevance}, {Dale}, {Egorov}, {Emsellem}, {Faesi}, {Grasha}, {Kim}, {Klessen}, {Kreckel}, {Kruijssen}, {Larson}, {Meidt}, {Sandstrom}, {Schinnerer}, {Thilker}, {Watkins}, {Whitmore}, \& {Williams}}]{hassani2023}
{Hassani}, H., {Rosolowsky}, E., {Leroy}, A.~K., {et~al.} 2023, \apjl, 944, L21

\bibitem[{He {et~al.}(2016)He, Zhang, Ren, \& Sun}]{he2015}
He, K., Zhang, X., Ren, S., \& Sun, J. 2016, in 2016 IEEE Conference on Computer Vision and Pattern Recognition (CVPR), 770--778

\bibitem[{{Henny} {et~al.}(2025){Henny}, {Dale}, \& {Chandar}}]{henny2025}
{Henny}, K., {Dale}, D., \& {Chandar}, R. 2025, \apj, submitted

\bibitem[{{Hoyer} {et~al.}(2023){Hoyer}, {Pinna}, {Kamlah}, {Nogueras-Lara}, {Feldmeier-Krause}, {Neumayer}, {Sormani}, {Boquien}, {Emsellem}, {Seth}, {Klessen}, {Williams}, {Schinnerer}, {Barnes}, {Leroy}, {Bonoli}, {Kruijssen}, {Neumann}, {S{\'a}nchez-Bl{\'a}zquez}, {Dale}, {Watkins}, {Thilker}, {Rosolowsky}, {Bigiel}, {Grasha}, {Egorov}, {Liu}, {Sandstrom}, {Larson}, {Blanc}, \& {Hassani}}]{hoyer2023}
{Hoyer}, N., {Pinna}, F., {Kamlah}, A. W.~H., {et~al.} 2023, \apjl, 944, L25

\bibitem[{Hunter(2007)}]{Matplotlib2007}
Hunter, J.~D. 2007, Computing in Science \& Engineering, 9, 90

\bibitem[{Johnson {et~al.}(2015)Johnson, Seth, Dalcanton, Wallace, Simpson, Lintott, Kapadia, Skillman, Caldwell, Fouesneau, Weisz, Williams, Beerman, Gouliermis, \& Sarajedini}]{Johnson2015}
Johnson, L.~C., Seth, A.~C., Dalcanton, J.~J., {et~al.} 2015, The Astrophysical Journal, 802, 127

\bibitem[{{Jones} {et~al.}(2017){Jones}, {Woods}, {Kemper}, {Kraemer}, {Sloan}, {Srinivasan}, {Oliveira}, {van Loon}, {Boyer}, {Sargent}, {McDonald}, {Meixner}, {Zijlstra}, {Ruffle}, {Lagadec}, {Pauly}, {Sewi{\l}o}, {Clayton}, \& {Volk}}]{Jones2017}
{Jones}, O.~C., {Woods}, P.~M., {Kemper}, F., {et~al.} 2017, \mnras, 470, 3250

\bibitem[{{Kennicutt} {et~al.}(2008){Kennicutt}, {Lee}, {Funes}, {J.}, {Sakai}, \& {Akiyama}}]{kennicutt2008}
{Kennicutt}, Robert~C., J., {Lee}, J.~C., {Funes}, J.~G., {et~al.} 2008, \apjs, 178, 247

\bibitem[{Khan {et~al.}(2017)Khan, Hayat, Bennamoun, Sohel, \& Togneri}]{khan2017}
Khan, S.~H., Hayat, M., Bennamoun, M., Sohel, F., \& Togneri, R. 2017, Cost Sensitive Learning of Deep Feature Representations from Imbalanced Data, , , arXiv:1508.03422

\bibitem[{{Kim} {et~al.}(2023){Kim}, {Chevance}, {Kruijssen}, {Barnes}, {Bigiel}, {Blanc}, {Boquien}, {Cao}, {Congiu}, {Dale}, {Egorov}, {Faesi}, {Glover}, {Grasha}, {Groves}, {Hassani}, {Hughes}, {Klessen}, {Kreckel}, {Larson}, {Lee}, {Leroy}, {Liu}, {Longmore}, {Meidt}, {Pan}, {Pety}, {Querejeta}, {Rosolowsky}, {Saito}, {Sandstrom}, {Schinnerer}, {Smith}, {Usero}, {Watkins}, \& {Williams}}]{kim2023}
{Kim}, J., {Chevance}, M., {Kruijssen}, J.~M.~D., {et~al.} 2023, \apjl, 944, L20

\bibitem[{Kingma \& Ba(2015)}]{kingma2017}
Kingma, D.~P., \& Ba, J. 2015, in 3rd International Conference on Learning Representations, {ICLR} 2015, San Diego, CA, USA, May 7-9, 2015, Conference Track Proceedings, ed. Y.~Bengio \& Y.~LeCun

\bibitem[{Klessen \& Glover(2016)}]{klessen2016}
Klessen, R.~S., \& Glover, S. C.~O. 2016, Physical Processes in the Interstellar Medium, ed. Y.~Revaz, P.~Jablonka, R.~Teyssier, \& L.~Mayer (Berlin, Heidelberg: Springer Berlin Heidelberg), 85--249

\bibitem[{Knutas {et~al.}(2025)Knutas, Adamo, Pedrini, Linden, Bajaj, Ryon, Gregg, Ali, Andersson, Bik, Bortolini, Buckner, Calzetti, Duarte-Cabral, Elmegreen, Vieira, Gallagher, Grasha, Johnson, Lai, Lapeer, Messa, Östlin, Sabbi, Smith, \& Tosi}]{knutas2025}
Knutas, A., Adamo, A., Pedrini, A., {et~al.} 2025, FEAST: JWST uncovers the emerging timescales of young star clusters in M83, , , arXiv:2505.08874

\bibitem[{Krizhevsky {et~al.}(2012)Krizhevsky, Sutskever, \& Hinton}]{Krizhevsky2012}
Krizhevsky, A., Sutskever, I., \& Hinton, G.~E. 2012, in Advances in Neural Information Processing Systems, ed. F.~Pereira, C.~Burges, L.~Bottou, \& K.~Weinberger, Vol.~25 (Curran Associates, Inc.)

\bibitem[{{Kroupa}(2002)}]{Kroupa2002}
{Kroupa}, P. 2002, Science, 295, 82

\bibitem[{Kruijssen(2012)}]{Kruijssen2012}
Kruijssen, J. M.~D. 2012, Monthly Notices of the Royal Astronomical Society, 426, 3008

\bibitem[{Krumholz {et~al.}(2019)Krumholz, McKee, \& Bland-Hawthorn}]{Krumholtz2019}
Krumholz, M.~R., McKee, C.~F., \& Bland-Hawthorn, J. 2019, Annual Review of Astronomy and Astrophysics, 57, 227

\bibitem[{Lada \& Lada(2003)}]{Lada&Lada2003}
Lada, C.~J., \& Lada, E.~A. 2003, Annual Review of Astronomy and Astrophysics, 41, 57

\bibitem[{{Lee} {et~al.}(2009){Lee}, {Gil de Paz}, {Tremonti}, {Kennicutt}, {Salim}, {Bothwell}, {Calzetti}, {Dalcanton}, {Dale}, {Engelbracht}, {Funes}, {Johnson}, {Sakai}, {Skillman}, {van Zee}, {Walter}, \& {Weisz}}]{lee2009}
{Lee}, J.~C., {Gil de Paz}, A., {Tremonti}, C., {et~al.} 2009, \apj, 706, 599

\bibitem[{{Lee} {et~al.}(2022){Lee}, {Whitmore}, {Thilker}, {Deger}, {Larson}, {Ubeda}, {Anand}, {Boquien}, {Chandar}, {Dale}, {Emsellem}, {Leroy}, {Rosolowsky}, {Schinnerer}, {Schmidt}, {Lilly}, {Turner}, {Van Dyk}, {White}, {Barnes}, {Belfiore}, {Bigiel}, {Blanc}, {Cao}, {Chevance}, {Congiu}, {Egorov}, {Glover}, {Grasha}, {Groves}, {Henshaw}, {Hughes}, {Klessen}, {Koch}, {Kreckel}, {Kruijssen}, {Liu}, {Lopez}, {Mayker}, {Meidt}, {Murphy}, {Pan}, {Pety}, {Querejeta}, {Razza}, {Saito}, {S{\'a}nchez-Bl{\'a}zquez}, {Santoro}, {Sardone}, {Scheuermann}, {Schruba}, {Sun}, {Usero}, {Watkins}, \& {Williams}}]{lee2022}
{Lee}, J.~C., {Whitmore}, B.~C., {Thilker}, D.~A., {et~al.} 2022, \apjs, 258, 10

\bibitem[{{Lee} {et~al.}(2023){Lee}, {Sandstrom}, {Leroy}, {Thilker}, {Schinnerer}, {Rosolowsky}, {Larson}, {Egorov}, {Williams}, {Schmidt}, {Emsellem}, {Anand}, {Barnes}, {Belfiore}, {Be{\v{s}}li{\'c}}, {Bigiel}, {Blanc}, {Bolatto}, {Boquien}, {den Brok}, {Cao}, {Chandar}, {Chastenet}, {Chevance}, {Chiang}, {Congiu}, {Dale}, {Deger}, {Eibensteiner}, {Faesi}, {Glover}, {Grasha}, {Groves}, {Hassani}, {Henny}, {Henshaw}, {Hoyer}, {Hughes}, {Jeffreson}, {Jim{\'e}nez-Donaire}, {Kim}, {Kim}, {Klessen}, {Koch}, {Kreckel}, {Kruijssen}, {Li}, {Liu}, {Lopez}, {Maschmann}, {Chen}, {Meidt}, {Murphy}, {Neumann}, {Neumayer}, {Pan}, {Pessa}, {Pety}, {Querejeta}, {Pinna}, {Rodr{\'\i}guez}, {Saito}, {S{\'a}nchez-Bl{\'a}zquez}, {Santoro}, {Sardone}, {Smith}, {Sormani}, {Scheuermann}, {Stuber}, {Sutter}, {Sun}, {Teng}, {Tre{\ss}}, {Usero}, {Watkins}, {Whitmore}, \& {Razza}}]{lee2023}
{Lee}, J.~C., {Sandstrom}, K.~M., {Leroy}, A.~K., {et~al.} 2023, \apjl, 944, L17

\bibitem[{{Leitherer} {et~al.}(2014){Leitherer}, {Ekstr{\"o}m}, {Meynet}, {Schaerer}, {Agienko}, \& {Levesque}}]{leitherer2014}
{Leitherer}, C., {Ekstr{\"o}m}, S., {Meynet}, G., {et~al.} 2014, \apjs, 212, 14

\bibitem[{{Leitherer} {et~al.}(1999){Leitherer}, {Schaerer}, {Goldader}, {Delgado}, {Robert}, {Kune}, {de Mello}, {Devost}, \& {Heckman}}]{leitherer1999}
{Leitherer}, C., {Schaerer}, D., {Goldader}, J.~D., {et~al.} 1999, \apjs, 123, 3

\bibitem[{{Leroy} {et~al.}(2021){Leroy}, {Schinnerer}, {Hughes}, {Rosolowsky}, {Pety}, {Schruba}, {Usero}, {Blanc}, {Chevance}, {Emsellem}, {Faesi}, {Herrera}, {Liu}, {Meidt}, {Querejeta}, {Saito}, {Sandstrom}, {Sun}, {Williams}, {Anand}, {Barnes}, {Behrens}, {Belfiore}, {Benincasa}, {Be{\v{s}}li{\'c}}, {Bigiel}, {Bolatto}, {den Brok}, {Cao}, {Chandar}, {Chastenet}, {Chiang}, {Congiu}, {Dale}, {Deger}, {Eibensteiner}, {Egorov}, {Garc{\'\i}a-Rodr{\'\i}guez}, {Glover}, {Grasha}, {Henshaw}, {Ho}, {Kepley}, {Kim}, {Klessen}, {Kreckel}, {Koch}, {Kruijssen}, {Larson}, {Lee}, {Lopez}, {Machado}, {Mayker}, {McElroy}, {Murphy}, {Ostriker}, {Pan}, {Pessa}, {Puschnig}, {Razza}, {S{\'a}nchez-Bl{\'a}zquez}, {Santoro}, {Sardone}, {Scheuermann}, {Sliwa}, {Sormani}, {Stuber}, {Thilker}, {Turner}, {Utomo}, {Watkins}, \& {Whitmore}}]{leroy2021}
{Leroy}, A.~K., {Schinnerer}, E., {Hughes}, A., {et~al.} 2021, \apjs, 257, 43

\bibitem[{{Levy} {et~al.}(2024){Levy}, {Bolatto}, {Mayya}, {Cuevas-Otahola}, {Tarantino}, {Boyer}, {Boogaard}, {B{\"o}ker}, {Cronin}, {Dale}, {Donaghue}, {Emig}, {Fisher}, {Glover}, {Herrera-Camus}, {Jim{\'e}nez-Donaire}, {Klessen}, {Lenki{\'c}}, {Leroy}, {De Looze}, {Meier}, {Mills}, {Ott}, {Rela{\~n}o}, {Veilleux}, {Villanueva}, {Walter}, \& {van der Werf}}]{levy2024}
{Levy}, R.~C., {Bolatto}, A.~D., {Mayya}, D., {et~al.} 2024, \apjl, 973, L55

\bibitem[{{Linden} {et~al.}(2023){Linden}, {Evans}, {Armus}, {Rich}, {Larson}, {Lai}, {Privon}, {U}, {Inami}, {Bohn}, {Song}, {Barcos-Mu{\~n}oz}, {Charmandaris}, {Medling}, {Stierwalt}, {Diaz-Santos}, {B{\"o}ker}, {van der Werf}, {Aalto}, {Appleton}, {Brown}, {Hayward}, {Howell}, {Iwasawa}, {Kemper}, {Frayer}, {Law}, {Malkan}, {Marshall}, {Mazzarella}, {Murphy}, {Sanders}, \& {Surace}}]{linden2023}
{Linden}, S.~T., {Evans}, A.~S., {Armus}, L., {et~al.} 2023, \apjl, 944, L55

\bibitem[{{Linden} {et~al.}(2024){Linden}, {Lai}, {Evans}, {Armus}, {Larson}, {Rich}, {U}, {Privon}, {Inami}, {Song}, {Bianchin}, {Bohn}, {Buiten}, {Sanchez-Garc{\'\i}a}, {Kader}, {Lenki{\'c}}, {Medling}, {B{\"o}ker}, {D{\'\i}az-Santos}, {Charmandaris}, {Barcos-Mu{\~n}oz}, {van der Werf}, {Stierwalt}, {Aalto}, {Appleton}, {Hayward}, {Howell}, {Malkan}, {Mazzarella}, {Murphy}, \& {Surace}}]{linden2024}
{Linden}, S.~T., {Lai}, T., {Evans}, A.~S., {et~al.} 2024, \apjl, 974, L27

\bibitem[{{Maeder} \& {Meynet}(2000)}]{maeder2000}
{Maeder}, A., \& {Meynet}, G. 2000, \araa, 38, 143

\bibitem[{{Mahabal} {et~al.}(2019){Mahabal}, {Rebbapragada}, {Walters}, {Masci}, {Blagorodnova}, {van Roestel}, {Ye}, {Biswas}, {Burdge}, {Chang}, {Duev}, {Golkhou}, {Miller}, {Nordin}, {Ward}, {Adams}, {Bellm}, {Branton}, {Bue}, {Cannella}, {Connolly}, {Dekany}, {Feindt}, {Hung}, {Fortson}, {Frederick}, {Fremling}, {Gezari}, {Graham}, {Groom}, {Kasliwal}, {Kulkarni}, {Kupfer}, {Lin}, {Lintott}, {Lunnan}, {Parejko}, {Prince}, {Riddle}, {Rusholme}, {Saunders}, {Sedaghat}, {Shupe}, {Singer}, {Soumagnac}, {Szkody}, {Tachibana}, {Tirumala}, {van Velzen}, \& {Wright}}]{Mahabal2019}
{Mahabal}, A., {Rebbapragada}, U., {Walters}, R., {et~al.} 2019, \pasp, 131, 038002

\bibitem[{{Maschmann} {et~al.}(2024){Maschmann}, {Lee}, {Thilker}, {Whitmore}, {Deger}, {Boquien}, {Chandar}, {Dale}, {Wofford}, {Hannon}, {Larson}, {Leroy}, {Schinnerer}, {Rosolowsky}, {{\'U}beda}, {Barnes}, {Emsellem}, {Grasha}, {Groves}, {Indebetouw}, {Kim}, {Klessen}, {Kreckel}, {Levy}, {Pinna}, {Rodr{\'\i}guez}, {Tian}, \& {Williams}}]{maschmann2024}
{Maschmann}, D., {Lee}, J.~C., {Thilker}, D.~A., {et~al.} 2024, \apjs, 273, 14

\bibitem[{{McQuaid} {et~al.}(2024){McQuaid}, {Calzetti}, {Linden}, {Messa}, {Adamo}, {Elmegreen}, {Grasha}, {Johnson}, {Smith}, \& {Bajaj}}]{McQuaid2024}
{McQuaid}, T., {Calzetti}, D., {Linden}, S.~T., {et~al.} 2024, \apj, 967, 102

\bibitem[{{Messa} {et~al.}(2021){Messa}, {Calzetti}, {Adamo}, {Grasha}, {Johnson}, {Sabbi}, {Smith}, {Bajaj}, {Finn}, \& {Lin}}]{Messa2021}
{Messa}, M., {Calzetti}, D., {Adamo}, A., {et~al.} 2021, \apj, 909, 121

\bibitem[{{Meynet} {et~al.}(1994){Meynet}, {Maeder}, {Schaller}, {Schaerer}, \& {Charbonnel}}]{Meynet1994}
{Meynet}, G., {Maeder}, A., {Schaller}, G., {Schaerer}, D., \& {Charbonnel}, C. 1994, \aaps, 103, 97

\bibitem[{Pasupa \& Sunhem(2016)}]{Pasupa2016}
Pasupa, K., \& Sunhem, W. 2016, in 2016 8th International Conference on Information Technology and Electrical Engineering (ICITEE), 1--6

\bibitem[{Paszke {et~al.}(2019{\natexlab{a}})Paszke, Gross, Massa, Lerer, Bradbury, Chanan, Killeen, Lin, Gimelshein, Antiga, Desmaison, Kopf, Yang, DeVito, Raison, Tejani, Chilamkurthy, Steiner, Fang, Bai, \& Chintala}]{Paszke2019}
Paszke, A., Gross, S., Massa, F., {et~al.} 2019{\natexlab{a}}, in Advances in Neural Information Processing Systems, ed. H.~Wallach, H.~Larochelle, A.~Beygelzimer, F.~d\textquotesingle Alch\'{e}-Buc, E.~Fox, \& R.~Garnett, Vol.~32 (Curran Associates, Inc.)

\bibitem[{Paszke {et~al.}(2019{\natexlab{b}})Paszke, Gross, Massa, Lerer, Bradbury, Chanan, Killeen, Lin, Gimelshein, Antiga, Desmaison, K{\"o}pf, Yang, DeVito, Raison, Tejani, Chilamkurthy, Steiner, Fang, Bai, \& Chintala}]{PyTorch2019}
Paszke, A., Gross, S., Massa, F., {et~al.} 2019{\natexlab{b}}, 1912.01703

\bibitem[{{Pedrini} {et~al.}(2024){Pedrini}, {Adamo}, {Calzetti}, {Bik}, {Gregg}, {Linden}, {Bajaj}, {Ryon}, {Ali}, {Bortolini}, {Correnti}, {Elmegreen}, {Elmegreen}, {Gallagher}, {Grasha}, {Gutermuth}, {Johnson}, {Melinder}, {Messa}, {{\"O}stlin}, {Sabbi}, {Smith}, {Tosi}, \& {Faustino Vieira}}]{pedrini2024}
{Pedrini}, A., {Adamo}, A., {Calzetti}, D., {et~al.} 2024, \apj, 971, 32

\bibitem[{{Prescott} {et~al.}(2007){Prescott}, {Kennicutt}, {Bendo}, {Buckalew}, {Calzetti}, {Engelbracht}, {Gordon}, {Hollenbach}, {Lee}, {Moustakas}, {Dale}, {Helou}, {Jarrett}, {Murphy}, {Smith}, {Akiyama}, \& {Sosey}}]{prescott2007}
{Prescott}, M. K.~M., {Kennicutt}, Robert~C., J., {Bendo}, G.~J., {et~al.} 2007, \apj, 668, 182

\bibitem[{Raiaan {et~al.}(2024)Raiaan, Sakib, Fahad, Mamun, Rahman, Shatabda, \& Mukta}]{Raiaan2024}
Raiaan, M. A.~K., Sakib, S., Fahad, N.~M., {et~al.} 2024, Decision Analytics Journal, 11, 100470

\bibitem[{{Ramambason} {et~al.}(2025){Ramambason}, {Chevance}, {Kim}, {Belfiore}, {Kruijssen}, {Romanelli}, {Amiri}, {Boquien}, {Chown}, {Dale}, {Dlamini}, {Egorov}, {Gerasimov}, {Glover}, {Grasha}, {Hassani}, {Kim}, {Kreckel}, {Koziol}, {Leroy}, {M{\'e}ndez-Delgado}, {Neumann}, {Neumann}, {Pan}, {Pathak}, {Sandstrom}, {Sarbadhicary}, {Schinnerer}, {Sun}, {Sutter}, {Thilker}, {Ubeda}, {Weinbeck}, {Williams}, \& {Whitmore}}]{Ramambason2025}
{Ramambason}, L., {Chevance}, M., {Kim}, J., {et~al.} 2025, arXiv e-prints, arXiv:2507.01508

\bibitem[{{Reddy} {et~al.}(2023){Reddy}, {Topping}, {Sanders}, {Shapley}, \& {Brammer}}]{reddy2023}
{Reddy}, N.~A., {Topping}, M.~W., {Sanders}, R.~L., {Shapley}, A.~E., \& {Brammer}, G. 2023, \apj, 948, 83

\bibitem[{{Rodr{\'\i}guez} {et~al.}(2023){Rodr{\'\i}guez}, {Lee}, {Whitmore}, {Thilker}, {Maschmann}, {Chandar}, {Deger}, {Boquien}, {Dale}, {Larson}, {Williams}, {Kim}, {Schinnerer}, {Rosolowsky}, {Leroy}, {Emsellem}, {Sandstrom}, {Kruijssen}, {Grasha}, {Watkins}, {Barnes}, {Sormani}, {Kim}, {Anand}, {Chevance}, {Bigiel}, {Klessen}, {Hassani}, {Liu}, {Faesi}, {Cao}, {Belfiore}, {Pessa}, {Kreckel}, {Groves}, {Pety}, {Indebetouw}, {Egorov}, {Blanc}, {Saito}, \& {Hughes}}]{rodriguez2023}
{Rodr{\'\i}guez}, M.~J., {Lee}, J.~C., {Whitmore}, B.~C., {et~al.} 2023, \apjl, 944, L26

\bibitem[{{Rodr{\'\i}guez} {et~al.}(2024){Rodr{\'\i}guez}, {Lee}, {Indebetouw}, {Whitmore}, {Maschmann}, {Williams}, {Chandar}, {Barnes}, {Gnedin}, {Sandstrom}, {Rosolowsky}, {Sun}, {Klessen}, {Groves}, {Wofford}, {Boquien}, {Dale}, {Leroy}, {Thilker}, {Kim}, {Levy}, {Sarbadhicary}, {Ubeda}, {Larson}, {Johnson}, {Bigiel}, {Hassani}, \& {Grasha}}]{rodriguez2024}
{Rodr{\'\i}guez}, M.~J., {Lee}, J.~C., {Indebetouw}, R., {et~al.} 2024, arXiv e-prints, arXiv:2412.07862

\bibitem[{{Sandstrom} {et~al.}(2023){Sandstrom}, {Chastenet}, {Sutter}, {Leroy}, {Egorov}, {Williams}, {Bolatto}, {Boquien}, {Cao}, {Dale}, {Lee}, {Rosolowsky}, {Schinnerer}, {Barnes}, {Belfiore}, {Bigiel}, {Chevance}, {Grasha}, {Groves}, {Hassani}, {Hughes}, {Klessen}, {Kruijssen}, {Larson}, {Liu}, {Lopez}, {Meidt}, {Murphy}, {Sormani}, {Thilker}, \& {Watkins}}]{sandstrom2023}
{Sandstrom}, K.~M., {Chastenet}, J., {Sutter}, J., {et~al.} 2023, \apjl, 944, L7

\bibitem[{Santos {et~al.}(2022)Santos, Abreu, Japkowicz, Fern{\'a}ndez, Soares, Wilk, \& Santos}]{Santos2022}
Santos, M.~S., Abreu, P.~H., Japkowicz, N., {et~al.} 2022, Artif. Intell. Rev., 55, 6207

\bibitem[{{Schinnerer} \& {Leroy}(2024)}]{schinnerer2024}
{Schinnerer}, E., \& {Leroy}, A.~K. 2024, \araa, 62, 369

\bibitem[{Simonyan \& Zisserman(2015)}]{Simonyan&Zisserman2015}
Simonyan, K., \& Zisserman, A. 2015, Very Deep Convolutional Networks for Large-Scale Image Recognition, , , arXiv:1409.1556

\bibitem[{Soekhoe {et~al.}(2016)Soekhoe, van~der Putten, \& Plaat}]{Soekhoe2016}
Soekhoe, D., van~der Putten, P., \& Plaat, A. 2016, in Advances in Intelligent Data Analysis XV, ed. H.~Bostr{\"o}m, A.~Knobbe, C.~Soares, \& P.~Papapetrou (Cham: Springer International Publishing), 50--60

\bibitem[{Srivastava {et~al.}(2014)Srivastava, Hinton, Krizhevsky, Sutskever, \& Salakhutdinov}]{Srivastava2014}
Srivastava, N., Hinton, G., Krizhevsky, A., Sutskever, I., \& Salakhutdinov, R. 2014, Journal of Machine Learning Research, 15, 1929

\bibitem[{Stanway \& Eldridge(2023)}]{Stanway2023}
Stanway, E.~R., \& Eldridge, J.~J. 2023, Monthly Notices of the Royal Astronomical Society, 522, 4430

\bibitem[{{Stuber} {et~al.}(2023){Stuber}, {Schinnerer}, {Williams}, {Querejeta}, {Meidt}, {Emsellem}, {Barnes}, {Klessen}, {Leroy}, {Neumann}, {Sormani}, {Bigiel}, {Chevance}, {Dale}, {Faesi}, {Glover}, {Grasha}, {Kruijssen}, {Liu}, {Pan}, {Pety}, {Pinna}, {Saito}, {Usero}, \& {Watkins}}]{stuber2023}
{Stuber}, S.~K., {Schinnerer}, E., {Williams}, T.~G., {et~al.} 2023, \aap, 676, A113

\bibitem[{{Sun} {et~al.}(2024){Sun}, {He}, {Batschkun}, {Levy}, {Emig}, {Rodr{\'\i}guez}, {Hassani}, {Leroy}, {Schinnerer}, {Ostriker}, {Wilson}, {Bolatto}, {Mills}, {Rosolowsky}, {Lee}, {Dale}, {Larson}, {Thilker}, {Ubeda}, {Whitmore}, {Williams}, {Barnes}, {Bigiel}, {Chevance}, {Glover}, {Grasha}, {Groves}, {Henshaw}, {Indebetouw}, {Jim{\'e}nez-Donaire}, {Klessen}, {Koch}, {Liu}, {Mathur}, {Meidt}, {Menon}, {Neumann}, {Pinna}, {Querejeta}, {Sormani}, \& {Tress}}]{Sun2024}
{Sun}, J., {He}, H., {Batschkun}, K., {et~al.} 2024, \apj, 967, 133

\bibitem[{Tammina(2019)}]{Tammina2019}
Tammina, S. 2019, International Journal of Scientific and Research Publications (IJSRP), 9

\bibitem[{{Thilker} {et~al.}(2024){Thilker}, {Leroy}, \& {Sandstrom}}]{thilker2024}
{Thilker}, D., {Leroy}, A., \& {Sandstrom}, K. 2024, \apj

\bibitem[{Thilker {et~al.}(2022)Thilker, Whitmore, Lee, Deger, Maschman, Larson, Hannon, Wofford, \& Lilly}]{PHANGSCAT}
Thilker, D., Whitmore, B., Lee, J.~C., {et~al.} 2022, Physics at High Angular resolution in Nearby GalaxieS - Catalogs ("PHANGS-CAT"),  STScI/MAST, doi:10.17909/JRAY-9798

\bibitem[{{Thilker} {et~al.}(2022){Thilker}, {Whitmore}, {Lee}, {Deger}, {Chandar}, {Larson}, {Hannon}, {Ubeda}, {Dale}, {Glover}, {Grasha}, {Klessen}, {Kruijssen}, {Rosolowsky}, {Schruba}, {White}, \& {Williams}}]{thilker2022}
{Thilker}, D.~A., {Whitmore}, B.~C., {Lee}, J.~C., {et~al.} 2022, \mnras, 509, 4094

\bibitem[{Ubeda {et~al.}(2021)Ubeda, Whitmore, Thilker, Lee, White, Anand, \& Hannon}]{PHANGSHST}
Ubeda, L., Whitmore, B., Thilker, D., {et~al.} 2021, Physics at High Angular resolution in Nearby GalaxieS - HST ("PHANGS-HST"),  STScI/MAST, doi:10.17909/T9-R08F-DQ31

\bibitem[{{V{\'e}ron-Cetty} \& {V{\'e}ron}(2010)}]{veroncetty2010}
{V{\'e}ron-Cetty}, M.~P., \& {V{\'e}ron}, P. 2010, \aap, 518, A10

\bibitem[{Virtanen {et~al.}(2020)Virtanen, Gommers, Oliphant, Haberland, Reddy, Cournapeau, Burovski, Peterson, Weckesser, Bright, {van der Walt}, Brett, Wilson, Millman, Mayorov, Nelson, Jones, Kern, Larson, Carey, Polat, Feng, Moore, {VanderPlas}, Laxalde, Perktold, Cimrman, Henriksen, Quintero, Harris, Archibald, Ribeiro, Pedregosa, {van Mulbregt}, \& {SciPy 1.0 Contributors}}]{SciPy2020}
Virtanen, P., Gommers, R., Oliphant, T.~E., {et~al.} 2020, Nature Methods, 17, 261

\bibitem[{Wei {et~al.}(2020)Wei, Huerta, Whitmore, Lee, Hannon, Chandar, Dale, Larson, Thilker, Ubeda, Boquien, Chevance, Kruijssen, Schruba, Blanc, \& Congiu}]{Wei2020}
Wei, W., Huerta, E.~A., Whitmore, B.~C., {et~al.} 2020, Monthly Notices of the Royal Astronomical Society, 493, 3178

\bibitem[{{Whitmore} {et~al.}(2014){Whitmore}, {Chandar}, {Bowers}, {Larsen}, {Lindsay}, {Ansari}, \& {Evans}}]{Whitmore2014}
{Whitmore}, B.~C., {Chandar}, R., {Bowers}, A.~S., {et~al.} 2014, \aj, 147, 78

\bibitem[{Whitmore {et~al.}(2007)Whitmore, Chandar, \& Fall}]{Whitmore2007}
Whitmore, B.~C., Chandar, R., \& Fall, S.~M. 2007, Astron. J., 133, 1067

\bibitem[{{Whitmore} {et~al.}(2021){Whitmore}, {Lee}, {Chandar}, {Thilker}, {Hannon}, {Wei}, {Huerta}, {Bigiel}, {Boquien}, {Chevance}, {Dale}, {Deger}, {Grasha}, {Klessen}, {Kruijssen}, {Larson}, {Mok}, {Rosolowsky}, {Schinnerer}, {Schruba}, {Ubeda}, {Van Dyk}, {Watkins}, \& {Williams}}]{whitmore2021}
{Whitmore}, B.~C., {Lee}, J.~C., {Chandar}, R., {et~al.} 2021, \mnras, 506, 5294

\bibitem[{{Whitmore} {et~al.}(2023){Whitmore}, {Chandar}, {Rodr{\'\i}guez}, {Lee}, {Emsellem}, {Floyd}, {Kim}, {Kruijssen}, {Mok}, {Sormani}, {Boquien}, {Dale}, {Faesi}, {Henny}, {Hannon}, {Thilker}, {White}, {Barnes}, {Bigiel}, {Chevance}, {Henshaw}, {Klessen}, {Leroy}, {Liu}, {Maschmann}, {Meidt}, {Rosolowsky}, {Schinnerer}, {Sun}, {Watkins}, \& {Williams}}]{Whitmore2023}
{Whitmore}, B.~C., {Chandar}, R., {Rodr{\'\i}guez}, M.~J., {et~al.} 2023, \apjl, 944, L14

\bibitem[{{Whitmore} {et~al.}(2025){Whitmore}, {Chandar}, {Lee}, {Henny}, {Rodr{\'\i}guez}, {Baron}, {Bigiel}, {Boquien}, {Chevance}, {Chown}, {Dale}, {Floyd}, {Grasha}, {Glover}, {Gnedin}, {Hassani}, {Indebetouw}, {Kapoor}, {Larson}, {Leroy}, {Maschmann}, {Scheuermann}, {Sutter}, {Schinnerer}, {Sarbadhicary}, {Thilker}, {Williams}, \& {Wofford}}]{whitmore2025}
{Whitmore}, B.~C., {Chandar}, R., {Lee}, J.~C., {et~al.} 2025, \apj, 982, 50

\bibitem[{{Williams} {et~al.}(2024){Williams}, {Lee}, {Larson}, {Leroy}, {Sandstrom}, {Schinnerer}, {Thilker}, {Belfiore}, {Egorov}, {Rosolowsky}, {Sutter}, {DePasquale}, {Pagan}, {Berger}, {Anand}, {Barnes}, {Bigiel}, {Boquien}, {Cao}, {Chastenet}, {Chevance}, {Chown}, {Dale}, {Deger}, {Eibensteiner}, {Emsellem}, {Faesi}, {Glover}, {Grasha}, {Hannon}, {Hassani}, {Henshaw}, {Jim{\'e}nez-Donaire}, {Kim}, {Klessen}, {Koch}, {Li}, {Liu}, {Meidt}, {M{\'e}ndez-Delgado}, {Murphy}, {Neumann}, {Neumann}, {Neumayer}, {Oakes}, {Pathak}, {Pety}, {Pinna}, {Querejeta}, {Ramambason}, {Romanelli}, {Sormani}, {Stuber}, {Sun}, {Teng}, {Usero}, {Watkins}, \& {Weinbeck}}]{williams2024}
{Williams}, T.~G., {Lee}, J.~C., {Larson}, K.~L., {et~al.} 2024, \apjs, 273, 13

\bibitem[{{Wofford} {et~al.}(2016){Wofford}, {Charlot}, {Bruzual}, {Eldridge}, {Calzetti}, {Adamo}, {Cignoni}, {de Mink}, {Gouliermis}, {Grasha}, {Grebel}, {Lee}, {{\"O}stlin}, {Smith}, {Ubeda}, \& {Zackrisson}}]{wofford2016}
{Wofford}, A., {Charlot}, S., {Bruzual}, G., {et~al.} 2016, \mnras, 457, 4296

\bibitem[{{Zevin} {et~al.}(2017){Zevin}, {Coughlin}, {Bahaadini}, {Besler}, {Rohani}, {Allen}, {Cabero}, {Crowston}, {Katsaggelos}, {Larson}, {Lee}, {Lintott}, {Littenberg}, {Lundgren}, {{\O}sterlund}, {Smith}, {Trouille}, \& {Kalogera}}]{Zevin2017}
{Zevin}, M., {Coughlin}, S., {Bahaadini}, S., {et~al.} 2017, Classical and Quantum Gravity, 34, 064003

\end{thebibliography}
